\documentclass[graybox]{svmult}

\usepackage{graphicx}
\usepackage{amsmath}

\usepackage{mathptmx}       % selects Times Roman as basic font
\usepackage{helvet}         % selects Helvetica as sans-serif font
\usepackage{courier}        % selects Courier as typewriter font
\usepackage{type1cm}        % activate if the above 3 fonts are
\usepackage{makeidx}         % allows index generation
\usepackage{graphicx}        % standard LaTeX graphics tool
\usepackage{multicol}        % used for the two-column index
\usepackage[bottom]{footmisc}% places footnotes at page bottom
\usepackage{aas_macros,natbib,amsmath,amssymb}   % omit 'round' option if you prefer square brackets

\def \curlB {\vec{\nabla}\times (e^\nu \vec{B})}

\def\src{SGR\,0418$+$5729}

\def\3xmm{3XMM\,J1852$+$0033}
\def\lowbb{Swift\,J1822.3$-$1606}

\newcommand{\bc}{\begin{center}}
\newcommand{\ec}{\end{center}}
\def\ltsima{$\; \buildrel < \over \sim \;$}
\def\lsim{\lower.5ex\hbox{\ltsima}}
\def\loe{\lower.5ex\hbox{\ltsima}}
\def\gtsima{$\; \buildrel > \over \sim \;$}
\def\gsim{\lower.5ex\hbox{\gtsima}}
\def\goe{\lower.5ex\hbox{\gtsima}}

\def\ltsima{$\; \buildrel < \over \sim \;$}
\def\lsim{\lower.5ex\hbox{\ltsima}}
\def\loe{\lower.5ex\hbox{\ltsima}}
\def\gtsima{$\; \buildrel > \over \sim \;$}
\def\gsim{\lower.5ex\hbox{\gtsima}}
\def\goe{\lower.5ex\hbox{\gtsima}}

\def\ergscm2 {erg\,s$^{-1}$cm$^{-2}$}

\def\cm2 {cm$^{-2}$}

\makeindex             % used for the subject index
\begin{document}

\title*{Magnetars: a short review and some sparse considerations}
% Use \titlerunning{Short Title} for an abbreviated version of
% your contribution title if the original one is too long
\author{Paolo Esposito, Nanda Rea and Gian Luca Israel}
% Use \authorrunning{Short Title} for an abbreviated version of
% your contribution title if the original one is too long
\institute{Paolo Esposito \at Anton Pannekoek Institute for Astronomy, University of Amsterdam, Science Park 904, Postbus 94249, 1090\,GE Amsterdam, The Netherlands; \email{P.Esposito@uva.nl}
\and Nanda Rea \at Institute of Space Sciences (ICE, CSIC), Campus UAB, Carrer de Can Magrans s/n, 08193, Barcelona, Spain;\\ Institut d'Estudis Espacials de Catalunya (IEEC), Gran Capit\`a 2-4,
08034, Barcelona, Spain;\\ Anton Pannekoek Institute for Astronomy, University of Amsterdam, Science Park 904, Postbus 94249, 1090\,GE Amsterdam, The Netherlands; \email{rea@ice.csic.es}
\and Gian Luca Israel \at Osservatorio Astronomico di Roma, INAF, via Frascati 33, 00040, Monteporzio Catone (Rome), Italy; \email{gianluca@oa-roma.inaf.it}}
%
% Use the package "url.sty" to avoid
% problems with special characters
% used in your e-mail or web address
%
\maketitle

\abstract{We currently know about 30 magnetars: seemingly isolated neutron stars whose properties can be (in part) comprehended only acknowledging that they are endowed with magnetic fields of complex morphology and exceptional intensity---at least in some components of the field structure. Although magnetars represent only a small percentage of the known isolated neutron stars, there are almost certainly many more of them, since most magnetars were discovered in transitory phases called \emph{outbursts}, during which they are particularly noticeable. In outburst, in fact, a magnetar can be brighter in X-rays by orders of magnitude and usually emit powerful bursts of hard-X/soft-gamma-ray photons that can be detected almost everywhere in the Galaxy with all-sky monitors such as those on board the Fermi satellite or the Neil Gehrels Swift Observatory. Magnetars command great attention because the large progress that has been made in their understanding is proving fundamental to fathom the whole population of isolated neutron stars, and because, due to their extreme properties, they are relevant for a vast range of different astrophysical topics, from the study of gamma-ray bursts and superluminous supernovae, to ultraluminous X-ray sources, fast radio bursts, and even to sources of gravitational waves. Several excellent reviews with different focuses were published on magnetars in the last few years: among others, \citet{israel11,rea11,turolla13,mereghetti15,turolla15,kaspi17}. Here, we quickly recall the history of these sources and travel through the main observational facts, trying to touch some recent and sometimes little-discussed ramifications of magnetars.}

\section{Historical overview}
\label{sec:1}

The event at the birth of magnetar studies was the landmark giant flare observed in 1979 from SGR\,0526--66 \citep{mazets79,cline80}. Nothing alike had been observed before from a pulsar, and the flare was also different from and much brighter than any other gamma-ray burst. The extreme properties of the event forced the astronomer to conceive unusual scenarios. The most appealing one was the existence of a class of neutron stars with super-strong
magnetic fields of  $10^{14}$--$10^{15}$\,G \citep{duncan92,paczynski92}. In particular, the word \emph{magnetar} was introduced to designate these objects by \citet{duncan92}. The giant flare and the more common emission of short fainter bursts of these sources were explained by impulsive releases of magnetic energy stored in the neutron star, which could be triggered by fractures in the magnetically stressed crust, perhaps associated to sudden magnetic reconnections in the star's magnetosphere \citep{duncan92,thompson95,thompson96}. The handfuls of sources associated to these recurrent hard X-/gamma-ray transients were dubbed soft gamma repeaters (SGRs), to distinguish them from the gamma-ray bursts.

In the same years, another class of X-ray pulsars defying any easy pigeonholing was emerging: the anomalous X-ray pulsars (AXPs; \citealt{mereghetti95,vanparadijs95}). They were characterised as persistent X-ray pulsars with periods of a few seconds and owed the adjective `anomalous' to the fact that their X-ray luminosity exceeded that available from spin-down energy loss (while the accretion was ruled out by the lack of any trace of a stellar companion). \citet{thompson96} noted that except for the emission of short bursts, which had never been observed in AXPs, these sources were very similar to the recently-discovered persistent soft X-rays counterparts of SGRs \citep{rothschild94,vasisht94,murakami94}. It was suggested that AXPs could be evolved SGRs, which ended or drastically reduced their explosive activity after having largely depleted their reservoir of magnetic energy, and the opposite possibility, too, was considered: that AXPs might be SGR progenitors in which the magnetic field decay has just begun  (e.g. \citealt{gavriil02}).

The magnetar hypothesis was arguably the most successful in explaining the salient features of SGRs and AXPs, but some skepticism about the scenario persisted for long, in particular for the AXPs, for which several alternative models could not be excluded \citep{chatterjee00,mci02}. The magnetar model really started to become the mainstream when the period derivative of an SGR was measured for the first time \citep{kouveliotou98}. Using RossiXTE, it was established that SGR\,1806--20 was pulsating at $P=7.47$\,s and the period was increasing at the rate $\dot{P}=2.6\times10^{-3}$\,s\,yr$^{-1}$: with the standard assumption of a magnetic dipole rotating in vacuum routinely used for standard pulsar, the values correspond to a surface magnetic field of $B \simeq 3.2\times10^{19}\sqrt{P\dot{P}}= 8\times10^{14}$\,G at the neutron star's equator. Moreover, the release of magnetic energy was necessary to power the X-ray emission of SGR\,1806--20, since the spin-down energy loss of the neutron star was two orders of magnitude lower than the observed X-ray luminosity.

Few years later, the detection of SGR-like bursts from the AXP 1E\,1048-1--5937 \citep{gavriil02} confirmed the suspect that also AXPs could harbour magnetars. Since then, with the discovery of many new magnetars showing a variegated phenomenology and observations of strong bursting activity and powerful flares from AXPs, as well as prolonged periods of quiescence in once-active SGRs, the AXP--SGR dichotomy seems completely obsolete.  

\section{Observational characteristics}

\subsection{Persistent emission}

\subsubsection{X-ray emission}

\begin{figure*}[h]
\centering
\sidecaption
\resizebox{.8\hsize}{!}{\includegraphics[angle=0]{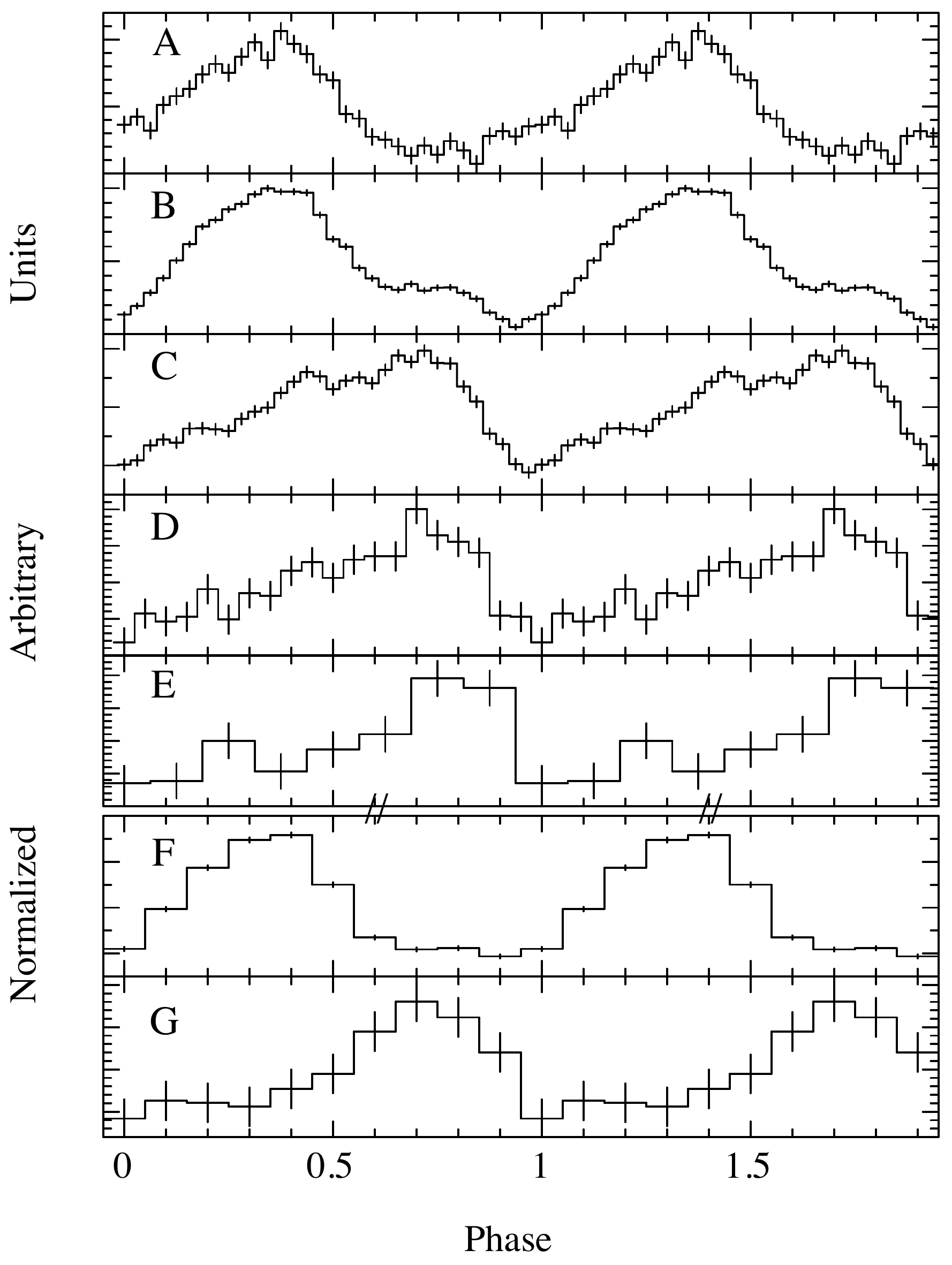}}
\caption{Pulse profiles of 1RXS\,J170849.0--400910, obtained with RossiXTE/PCA (2004 data; panel A: 2.5--4\,keV, B: 4--8\,keV, C: 8--16\,keV, D: 16--32\,keV) and INTEGRAL/IBIS (2004 data; panel E, 20--200\,keV). Panels F and G show the BeppoSAX MECS (1--10\,keV) and PDS (20--200\,keV) pulse profiles obtained during a single pointing in 2001. Note that data in panels A to E were folded using an RossiXTE timing solution, while panels F and G using the period measured in the MECS data: the profiles in the two groups are phase-aligned between themselves but not each other. (From \citealt{gri07}.)}
\label{timing1708}       % Give a unique label
\end{figure*}

The X-ray emission from magnetars is gently modulated at the pulsar spin period, with generally one or two broad sine components and substantial pulsed fractions (the fraction of the flux that changes along the rotation cycle) of 10--30\%. The pulse profiles may be energy dependent and may display dramatic changes with time, especially in connection with strong bursting/outbursting activity (see Sect.\,\ref{transient_activity}). As an example, Fig.\,\ref{timing1708} shows a multi-epoch and multi-instrument pulse profile of 1RXS\,J170849.0--400910 \citep[see][for details]{gri07}.
%%%%%%%%%%%%%%%%%%%%%%%%%%%%%%%%%%%%%%%%%%%%%%
%%%%%%%%%%%%%%%%%%%%%%%%%%%%%%%%%%%%%%%%%%%%%%%

Part of the X-ray luminosity of magnetars in quiescence has a thermal origin and can be fit by a blackbody with temperature $kT\approx0.3$--1\,keV, much higher than the typical values for rotation-powered pulsars; magnetar also tend to be more luminous than rotation-powered pulsars of similar characteristic age. Indeed, in magnetars the neutron star surface is believed to be particularly hot because of the extra-heating by the magnetic field decay (e.g. \citealt{aguilera08}).

The size of the region of the thermal emission inferred from a blackbody fit is generally much smaller than the surface of the star, possibly suggesting that a single-temperature blackbody is an oversimplification. Substantial anisotropies are indeed expected in the presence of a strong magnetic field in the crust and, since most magnetars are located in the Galactic plane, their spectra are generally heavily absorbed: it is possible that only `hot spots' are detectable in the available X-ray spectra. Small and hot regions are also envisaged to result, rather than from internal heat transfer, from particle bombardment and heat deposition from magnetospheric currents induced by the globally twisted external magnetic field and/or by localised  twists. In any case, the thermal emission in magnetars is expected to be significantly distorted by a magnetised atmosphere and also by magnetospheric effects, most likely resonant cyclotron scattering onto magnetospheric charges. Since the charged particles populate vast regions of the magnetosphere, with different magnetic field intensities, the scattering produces a hard tail instead than a narrow line or a set of distinct lines and harmonics.

Phenomenologically, in the 0.5--10\,keV band magnetar spectra are always well described by the already mentioned blackbody and often one or more additionally harder blackbody or power-law components, the latter with photon index generally in the range $\Gamma\sim2$--4 \citep{kb10,olausen14}. Since in the spectral modelling the (usually large) interstellar absorption, the power-law slope and the blackbody temperature(s) are covariant, it is often difficult to disentangle the components or to tell whether a blackbody or a power-law component is to be preferred. For the magnetar with the lowest absorbing column, CXOU\,J0100--7211 in the Small Magellanic Cloud, a double-blackbody model provides a much better fit to the data than a blackbody-plus-power-law model \citep{tiengo08}. On the other hand, a non-thermal component is certainly present at least in the magnetars detected in the hard-X-ray range.

Given the many complications and uncertainties, it is evident that prudence is necessary when drawing physical inferences from the spectral parameters. It is however worth noticing that physical models based on resonant cyclotron scattering (likely, repeated scatterings) of seed thermal photons on mildly relativistic electrons are quite successful in reproducing the general thermal-plus-power-law shape of the continuum and fit the spectra of most magnetars. We refer to \citet{turolla15} for an overview of the state of the art of these models, their application, and the main open problems.

\subsubsection{Hard-X-ray emission}

\begin{figure*}[h]
\centering
\sidecaption
% Use the relevant command for your figure-insertion program
% to insert the figure file.
% For example, with the graphicx style use
\resizebox{\hsize}{!}{\includegraphics[angle=0]{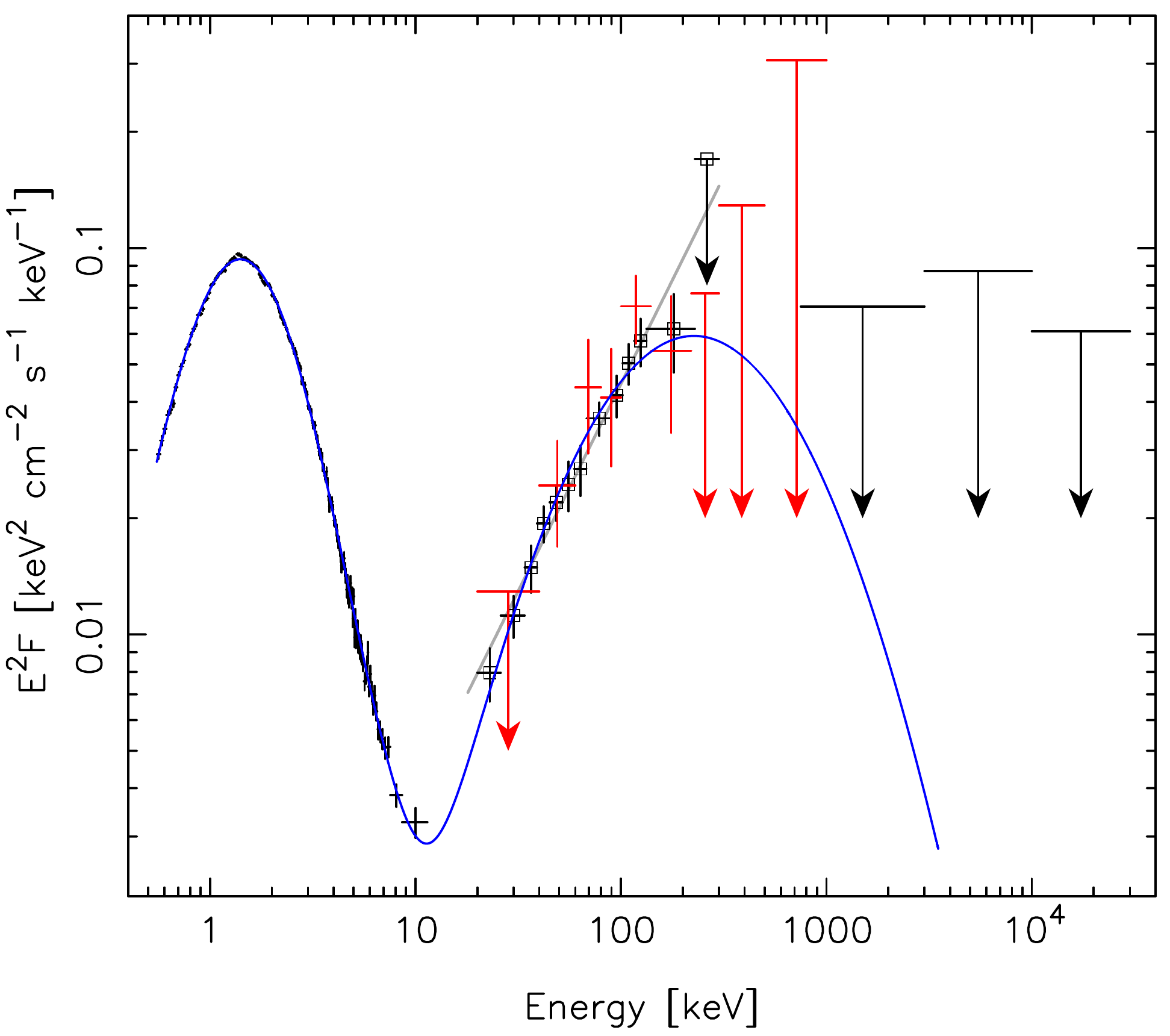}}
%
% If no graphics program available, insert a blank space i.e. use
%\picplace{5cm}{2cm} % Give the correct figure height and width in cm
%
\caption{Unabsorbed soft-to-hard-X-ray spectral energy distribution of 4U\,0142+61 as observed with
  XMM--Newton (in black), INTEGRAL/ISGRI (black open squares), INTEGRAL/SPI (red), and CGRO/COMPTEL
  (black). Down arrows indicate upper limits. See \citet[][from which the figure was taken]{denhartog08} for details on the spectral modelling.}
\label{sed_4u0142}       % Give a unique label
\end{figure*}

In several magnetars, hard X-ray tails with power-law spectra with photon index $\Gamma\approx0.5$--2 (flatter than the soft non thermal components) have been detected with BeppoSAX, RossiXTE, INTEGRAL, Suzaku and NuSTAR extending beyond $\approx$150\,keV \citep{kuiper04,kuiper06,gotz06,gri07,denhartog08,dkh08,an14,vogel14,tendulkar15,enoto17}. Upper limits in the hundreds of keV and MeV regions obtained with CGRO and Fermi indicate that the tails do not extend above $\approx$500\,keV \citep[e.g.][]{denhartog08,li17}. Fig\,\ref{sed_4u0142} shows the soft-to-hard-X-ray spectrum of 4U\,0142+61. Also searches in the GeV and Tev energy bands gave negative results \citep{li17,aleksic13}. 
%%%%%%%%%%%%%%%%%%%%%%%%%%%%%%%%%%%%%%%%%%%%%%
%%%%%%%%%%%%%%%%%%%%%%%%%%%%%%%%%%%%%%%%%%%%%%

The hard components can be variable, but in general their luminosity is comparable with or larger than that measured below 10\,keV; also in the sources in which the hard X-ray emission has not been detected, the upper limits do not exclude a substantial contribution to the total luminosity. A peculiar feature of the hard tails is that the pulsed emission has a harder spectrum than the averaged hard X-ray one and also shows phase dependent variations (also morphological changes  in the pulse profiles and peak shifts with energy are observed; e.g. \citealt{dkh08,gri07}, see also Fig.\,\ref{timing1708}). 

The origin of the hard tails of magnetars is still poorly understood but in general the mechanism suggested, similarly to what proposed for the soft spectra, is the up-scattering on ultra-relativistic electrons with Lorentz factor $\gamma\gg1$ \citep{baring07,fernandez07,wadiasingh18}, possibly associated to relativistic outflows near the neutron star \citet{beloborodov13,b13}. 

\subsubsection{Optical or infrared emission}

Optical or infrared counterparts have been found for about one-third of the known magnetars (e.g. \citealt{israel04,mignani11}). The search is complicated by the intrinsic faintness of magnetars at that wavelengths and by their location in crowded and heavily absorbed regions in the Galactic plane, but in most cases the associations are strengthened by the detection of long-term variability. In three cases, the association is firm because the spin modulation has been detected also in the optical band (4U\,0142+61, \citealt{kern02,dhillon05}; 1E\,1048.1--5937, \citealt{dhillon09}; SGR\,J0501+4516, \citealt{dhillon11}). As anticipated, magnetars are variable also in the infrared/optical range, but it is unclear (possibly because of the lack of adequate multi-wavelength campaigns) whether the changes trace the X-ray flux evolution, as also cases of anti-correlated or simply erratic variations have been reported \citep{tam04,dk05,ccr07,testa08,dhillon11}.  

The magnetar 4U\,0142+61 has been detected in both infrared and optical bands and is the one for which the greatest wealth of data is available at these wavelengths \citep{hulleman00,hulleman04}. In near infrared, it shows an excess with respect to the blackbody that fits the optical data; indeed, a multi-temperature (700--1,200\,K) thermal model provides a better fit to the data. \citet{wang06} suggested that the infrared component arises from an extended disk (possibly from supernova fallback)  illuminated from the star's X-rays and passively heated. This interpretation is supported by the correlation observed in this source  between the X-ray and infrared emissions \citep{tam04}. On the other hand, infrared/optical emission is expected from the inner magnetosphere, a pair-dominated region where the curvature radiation should be able to produce the observed infrared/optical luminosity \citep{zane11}. Moreover, a magnetospheric origin would account more easily for the observed optical pulsations, with profiles nearly aligned with those observed at X-rays and displaying similarly broad modulation and large pulsed fraction (20--50\%). It is also possible that the infrared and optical emissions have different origins or that the infrared excess in 4U\,0142+61 is not well understood. A handful of magnetar has been detected also as pulsating sources at longer wavelength, in the radio band. We give an overview of the properties of magnetars at radio frequencies in Sect.\,\ref{outbursts}.

\subsection{Transient activity}\label{transient_activity}

Magnetars are certainly characterised by an extremely rich observational phenomenology, but this is particularly true when speaking of their transient activity: They display unpredictable and dramatic variations in their emission and timing properties in all the wavelengths at which they are detected, on time scales from milliseconds to months or years, and often with a dynamic range unparalleled by any other embodiment of isolated neutron stars. Their transient radiative events are usually outlined in two main categories: short-duration (ms--minutes) explosive events (giant flares and bursts) and outbursts, in which the X-ray luminosity rises to up to $\sim$$10^3$ times the quiescent level and then decays in weeks to months/years. Perhaps, an outburst more than an event could be considered a `syndrome', in the sense that the flux enhancement is generally accompanied by bursts, spectral changes and timing anomalies, including glitches.  

\subsubsection{Giant flares}\label{giantflares}
Giant flares are the rarest and most energetic events associated with magnetars. They are also the most important, at least historically, as it was the first giant flare, from SGR\,0526--66 in the Large Magellanic Cloud on 1979 March 5 \citep{mazets79,cline80}, that brought magnetars on the astrophysical scene, provided clear-cut evidence of their neutron-star nature and propensity to produce multiple events (at variance with the gamma-ray bursts discovered by the Vela satellites), and prompted---among a multitude of different models (see e.g. \citealt{norris91,woods06})---the concept of super-magnetic neutron stars \citep{paczynski92,duncan92,thompson95}. Moreover, they still provide some of the most compelling clues for the presence of magnetic fields of $10^{14}$\,G close to the neutron star surface in magnetars.

A total of three giant flares have been observed. In 1979 from SGR\,0526--66, on 1998 August 27 from SGR\,1900+14 \citep{hurley99} and on 2004 December 27 from SGR\,1806--20 \citep{hurley05,palmer05}. It is worth noticing that the three giant flares were emitted from the three first SGRs discovered (but only SGR\,0526--66 was discovered because of the event).
Since a mere coincidence seems unlikely and a giant flare makes a significant dent in the total magnetic energy pool of a magnetar, a natural explanation would be that those three sources are at the magnetic-activity pinnacle of their life and their frequent bursting activity got them noticed earlier than other magnetars (see also \citealt{perna11,vigano13}).

All three giant flares started with a short ($\sim$0.1--0.2 s) flash of hard X-rays with peak luminosity $\gtrsim$$10^{44}$--$10^{45}$\,erg s$^{-1}$ ($\gtrsim$$10^{47}$\,erg s$^{-1}$ in the case of SGR\,1806--20; \citealt{hurley05}). The spectrum of this impulsive blaze extends at least to the MeV range and can be described by a blackbody, with initial $kT$ from $\approx$30 to $200$\,keV (for SGR\,0526--66 and SGR\,1806--20, respectively). These sudden releases of an immense amount of energy affected in a measurable way (at least for the two most recent events) the Earth's magnetic field \citep{mandea06} and ionosphere \citep{inan99,inan07}. \citet{hurley05} argued that an extragalactic giant flare as bright as that of SGR\,1806--20 could appear at Earth as a short gamma-ray burst up to a distance of several tens of Mpc an therefore magnetar flares might represent a non-negligible fraction of the population of these transients. 

\begin{figure*}[h]
\centering
\sidecaption
% Use the relevant command for your figure-insertion program
% to insert the figure file.
% For example, with the graphicx style use
\resizebox{\hsize}{!}{\includegraphics[angle=0]{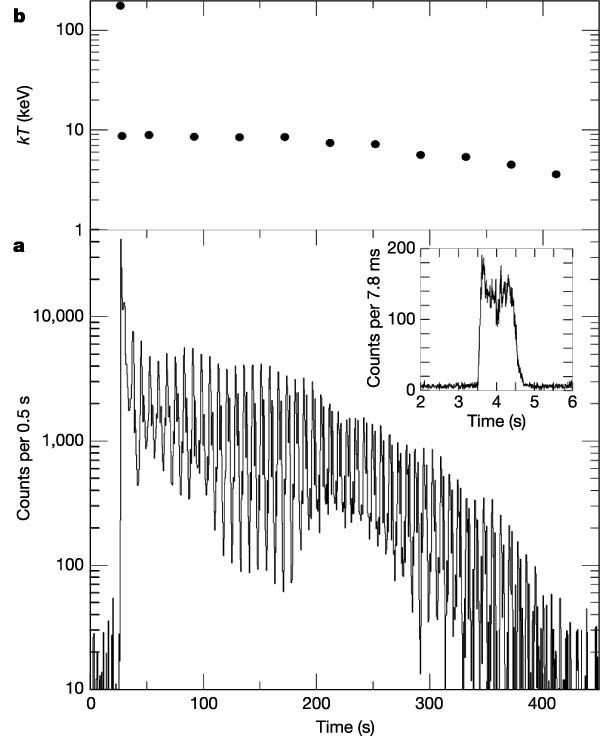}}
%
% If no graphics program available, insert a blank space i.e. use
%\picplace{5cm}{2cm} % Give the correct figure height and width in cm
%
\caption{The 2004 December 27 giant flare of SGR\,1806--20 as observed by RHESSI (from \citealt{hurley05}). Panel a shows the 20--100-keV light curve. The (saturated) spike is at $\sim$30\,s and the inset shows the profile of a bright burst the preceded the flash by $\sim$2\,minutes. The modulation at the spin period of SGR\,1806--20 (7.5\,s) is apparent. Panel b shows the temporal evolution of the spectral temperature.}
\label{sgr1806gf}       % Give a unique label
\end{figure*}

After the initial spikes, followed afterglows that were clearly modulated at the rotational period of the neutron stars. The afterglows were much softer than the flash and over a few minutes gradually further softened and faded (Fig.\,\ref{sgr1806gf}). It is extremely interesting that while the luminosity of the three peaks spans 2--3 orders of magnitude, the total energy of the oscillating tail was similar ($\approx$$10^{44}$\,erg) in all the events. Since the afterglow is believed to arise from a cloud of photon--pair plasma confined by the star's magnetic field that cools as the radiation gradually leaks out, this clearly indicates a similar value of the magnetic field in the three magnetars. To trap the fireball, the magnetic pressure must exceed that of the radiation and pairs at the external boundary of the cloud: $B_{\mathrm{dip}}>2\times10^{14} (E_{\mathrm{fireball}}/(10^{44}\,\mathrm{erg}))^{1/2}(\Delta R/(10\,\mathrm{km}))^{-3/2}((1+\Delta R/R)/2)^3$\,G, where $R$ is the stellar radius and $\Delta R$ the characteristic size of the fireball \citep{thompson95}. 
%%%%%%%%%%%%%%%%%%%%%%%%%%%%%%%%%%%%%%%%%%%%%%
%%%%%%%%%%%%%%%%%%%%%%%%%%%%%%%%%%%%%%%%%%%%%%%

In the SGR\,1900+14 and SGR\,1806--20's giant flares, observations of transient nebular radio emissions provided evidence for outflows \citep{frail99,gaensler05,gelfand05}. For the best-studied case of SGR\,1806--20, the minimum energy in the extended radio emission was estimated at $\approx$$10^{43}$\,erg, which seems too much to be consistent with pair plasma leaked form the fireball. Indeed, the structure, which was observable for more than a year, is better explained in terms of an baryon-rich mildly-relativistic ejection interacting with matter surrounding the star \citep{gelfand05,granot06}.

The detection of quasi-periodic oscillations (QPOs) in the tails of the giant flares from SGR\,0526--66 (with detectors aboard the Prognoz\,7 satellite and the Venera 11 and 12 space probes; \citealt{barat83}), SGR\,1900+14 \citep{sw05} and SGR\,1806--20 (with RossiXTE and RHESSI; \citealt{ibs05,strohmayer06,watts06}), likely associated to seismic vibrations excited by the powerful explosion, started the field of asteroseismologyÁ for neutron stars (see gray box QPOs) and offered a new clue of the presence of a magnetic field $\gtrsim$$10^{14}$\,G in magnetars.
\citet{vietri07} observed that for any source there is a maximum rate of variation of the luminosity ($\Delta L$) on a certain timescale ($\Delta t$): $\Delta L / \Delta t < \eta (2.8\times10^{18})/\sigma_{\mathrm{T}}$\,erg\,s$^{-2}$, where $\sigma_{\mathrm{T}}$ is the Thomson cross section and $\eta$ the energy extraction efficiency \citep{cavallo78,fabian79}. The 1840-Hz QPO detected in SGR\,1806--20 implies a  $\Delta L / \Delta t$ exceeding this limit by a factor larger than 10/$\eta$. However, a strong magnetic field suppresses the electron-scattering cross section (for one photon polarization mode) below the Thomson's value by a factor $\propto$$B^2$ \citep[][see also \citealt{vanputten13}]{herold79}. The presence of a magnetic field $B\gtrsim2\times10^{15}$\,G at the surface of SGR\,1806--20 would reconcile the QPOs with the luminosity variability limit \citep{vietri07}.

\begin{svgraybox}\textbf{Quasi periodic oscillations and seismology of magnetars} \\

The tail of the 2004 giant flare from SGR\,1806--20 displayed clear QPO signals at about 18, 30, 93, 150, 625 and 1840~Hz \citep{ibs05,watts06}. QPOs around frequencies of 28, 54, 84 and 155 Hz were detected in the tail of the 1998 giant flare of SGR\,1900+14 \citep{sw05}, while hints for a signal at $\sim$43 Hz were found in the 1979 event from SGR\,0526--66 \citep{barat83}. Some QPOs were excited simultaneously, others were detected only once in a very narrow time interval, some faded and were re-excited several times. All the detected QPOs are dependent on the phase of the spin period and show large variations of the amplitude with time. Their similarities suggest that the production mechanism is the same and the most obvious responsible are seismic  vibrations induced by the giant flares. This is very exciting, as the QPOs provide a window on the neutron-star and magnetic field structures, and even on the dense matter equation of state.

The QPOs, in accordance with early theoretical suggestions \citep{duncan98}, were initially interpreted in terms of torsional shear modes of the neutron star crust. However, it did not take long to realize that neutron stars can sustain many 
types of oscillation and that the identification of oscillatory modes of magnetars is conceptually (and computationally) extremely challenging because of the magnetic coupling between the crust and the core \citep{glampedakis06,levin06}. Indeed, at the moment, the potential of magnetar asteroseismology is dampened by the high degeneracy because of the many uncertainty associated with the magnetic field and the superfluid state of matter (see for example \citealt{levin07,levin11,glampedakis14} for detailed discussions and \citealt{turolla15} for an overview of the current understanding of the field), but also by the paucity of new data. This motivated searches for QPOs in the more bountiful short bursts. Only one signal, an unusually broad but strong peak at 260\,Hz, was identified in a burst, a 0.5-s-long event from 1E\,1547.0--5408, while several candidates  from $\sim$60 to 130\,Hz were found in data sets combining many short busts \citep{huppenkothen13,huppenkothen14,hhw14}. Unfortunately, after the end of the RossiXTE mission and the non selection of LOFT \citep{feroci16} by ESA, the possibility of collecting a large number of photons with good time resolution and without saturation problems in case of exceptional events in the near future seems rather remote.  \end{svgraybox}

\subsubsection{Short bursts}

The SGR/magnetar short bursts are the hallmark of magnetars and thousands of them have been recorded and studied, both individually and as samples \citep{gogus01,aptekar01,gmm06,israel08,vanderhorst12,huppenkothen15}. In the last few years, and in particular since the launch of Swift (which pairs a sensitive hard X-ray instrument with large field of view to an X-ray telescope that provides $\sim$arcsec localisation), the short bursts have become the primary channel for the discovery of new magnetars and of the onset of magnetic activity from known sources. The bursting activity is unpredictable: Magnetars usually go through long stretches of quiescence (even decades) that can be interrupted by sparse bursts or by paroxysmal activity, during which hundreds or thousands of events are clustered in days. Or by everything in between.
 
\begin{figure*}[h]
\centering
\sidecaption
\resizebox{\hsize}{!}{\includegraphics[angle=0]{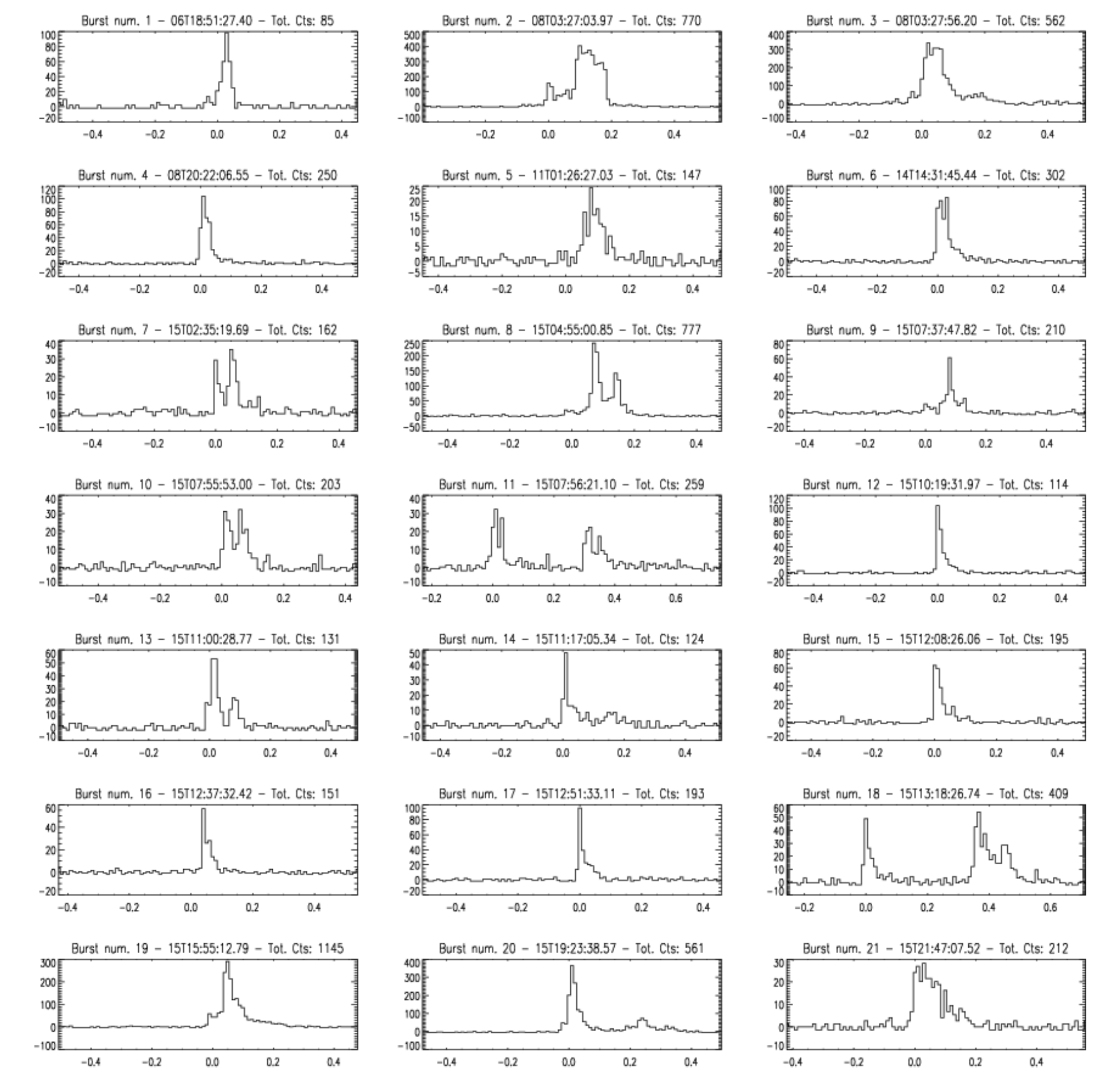}}
\caption{Light curves of bursts observed from SGR\,1806--20 with INTEGRAL in the 15--100\,keV range during an active period in 2003. The units of the axes are seconds for the time (x-axis) and counts per bin in the IBIS/ISGRI instrument for the intensity (y-axis); the time bin is 10 ms (from \citealt{gotz04}).}
\label{gotz_bursts}       
\end{figure*}

The impulsive emission of X\,/\,gamma ray photons lasts from milliseconds to a few seconds and the peak luminosity is typically in the range $\approx$$10^{36}$--$10^{43}$~erg s$^{-1}$. Their profile is usually single-peaked and asymmetric, with a faster rise than decay. Two- or multi-peak bursts are not rare, but since series of bursts can be very rapid, the distinction between single events and multi-peak bursts may be tenuous. Figure\,\ref{gotz_bursts} shows the light curves of 21 bursts collected from SGR\,1806--20 with INTEGRAL in 2003 (from \citealt{gotz04}). 
%%%%%%%%%%%%%%%%%%%%%%%%%%%%%%%%%%%%%%%%%%%%%%
%%%%%%%%%%%%%%%%%%%%%%%%%%%%%%%%%%%%%%%%%%%%%%

Some bursts, in general the brightest (`intermediate flares'), can be followed by X-ray tails surviving from minutes to hours. These events resemble the overall shape of the giant flares and their afterglows, except that in some instances the energy in the tail exceed that emitted in the spike (see e.g. \citealt{pintore17}). Sometimes these tails show flux modulation at the neutron star spin period, linking directly the afterglow to the neutron star (e.g. to a region of the star surface that was heated by the burst), but a significant fraction of them may be due to (or receive a large contribution from) dust-scattering of the burst emission, in which a fraction of the photons of the burst is re-emitted after reprocessing by clouds/layers of interstellar dust between us and the source \citep{lenters03,esposito07,gwk11,pintore17}. In general, it is possible to disentangle the contribution of the delayed dust scattering from that intrinsic to the neutron star only when data sets with a large number of photons and good spatial resolution are available, but sometimes the phenomenon is truly spectacular: see \citet{tiengo10} for the study of a scattering halo surrounding the magnetar 1E\,1547.0--5408 that took the shape of at least three expanding symmetric rings.

A number of models have been used to model the spectral emission of short magnetar bursts, but a single-blackbody model, or a double-blackbody model when broad band data are available (e.g. 1--200 keV), are usually a safe bet \citep{israel08,israel11}. The blackbody temperature $kT$ is typically in the range from $\sim$2 to 12 keV and when the double-blackbody decomposition is viable, a bimodal distribution of radii end temperatures takes shape, with a soft blackbody and a hotter ($\sim$12\,keV) blackbody with smaller surface. 

The fluence ($S$) distribution for number ($N$) of bursts usually follows a power-law function $N(>S)\propto S^{-\alpha}$ over several orders of magnitude, with $\alpha\sim0.6$--0.9, depending on the sources and the instruments \citep{gogus99,gogus00,aptekar01,gmm06}. It has been pointed out many times that this behaviour is similar to what observed for earthquakes. The truth is that such distribution is rather ubiquitous in nature and also in artificial and human-influenced systems: classic examples are the sizes of landslides and avalanches, solar flares, forest fires, lunar craters and cities, or the frequency of use of words in human languages (e.g. \citealt{newman05}). As observed by \citet{paizis14} for the distribution of hard-X luminosity of supergiant fast X-ray transient, the power-law distribution is characteristic (but not exclusive) of the self-organized criticality systems \citep{aschwanden13}, which inherently and perpetually evolves into a critical state where a minor event can start a chain reaction leading to a catastrophe \citep{bak91}.

\subsubsection{Outbursts}\label{outbursts}

In the magnetar world, the term `outburst' is usually used to denote a large enhancement ($\sim$10--1000) of the flux that eventually fades away over the course of months or even years (e.g. \citealt{rea11,cotizelati18}). A few magnetars have never been observed in outburst, others displayed a single outburst in decades, and some  have gone through multiple events.
Outbursts, and their onsets in particular, are generally associated to one or more short bursts. It is not clear, however, whether the bursts actually start the outbursts. In fact, the soft X-ray observations that catch a source in an enhanced-flux phase are generally carried out in response of the detection of a burst; on the other hand, for the outbursts discovered serendipitously it is not possible to pinpoint their exact start or exclude that bursts were missed. At any rate, the few cases in which observations were performed fortuitously shortly prior to a burst--outburst combination, indicate that the flux changes are rapid and happen close to the explosive activity, within $\sim$1--2 days \citep{eiz08,israel07,kennea13,younes17}. 

\begin{figure*}[h]
\centering
\sidecaption
\resizebox{\hsize}{!}{\includegraphics[angle=90]{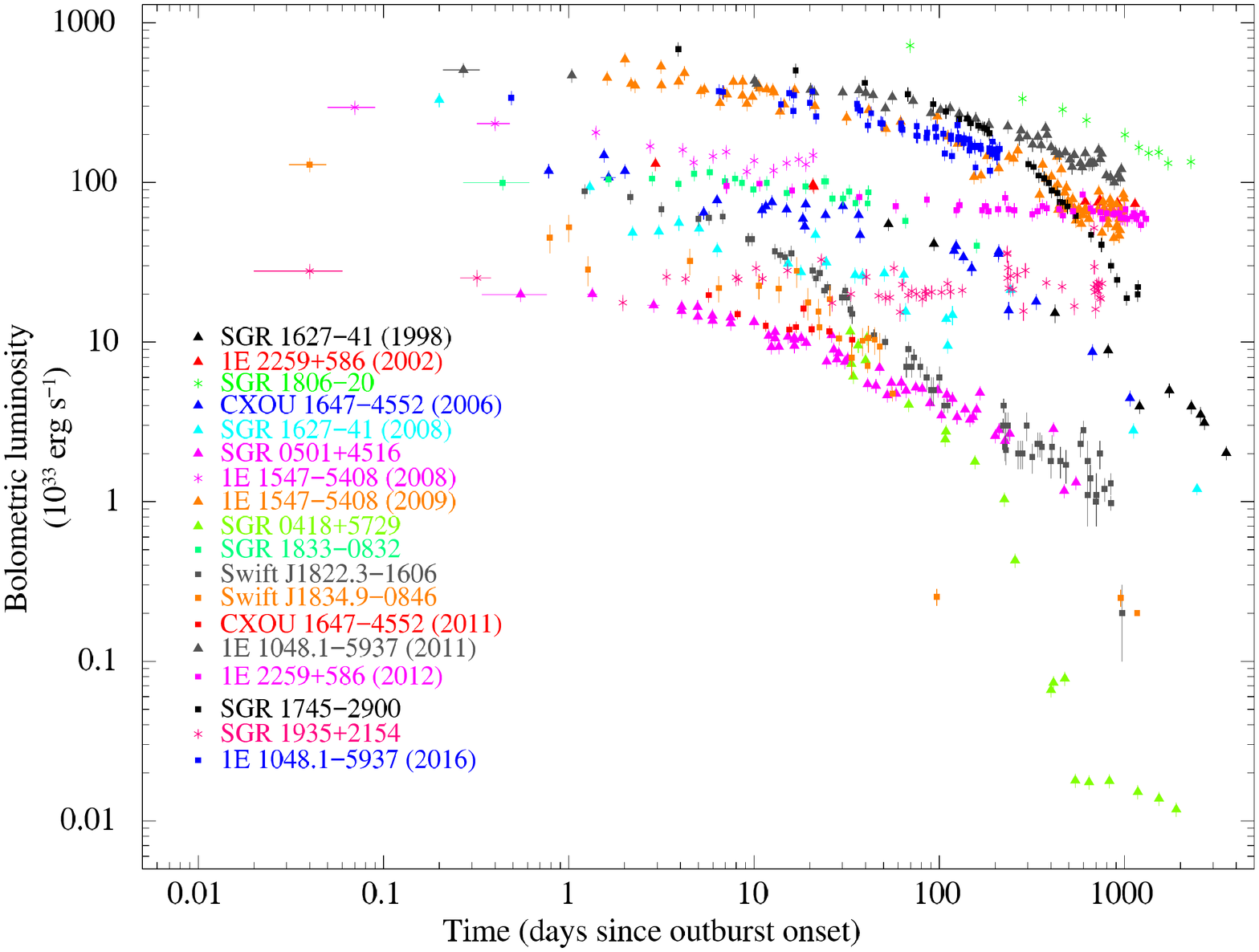}}
\caption{Light curves of all the outbursts with good observed coverage (from \citealt{cotizelati18}). The luminosities are bolometric (obtained from an extrapolation of the best-fit spectral models in the 0.01--100\,keV range).}
\label{outbursts_curves}       
\end{figure*}

The decay pattern is usually complicated, but it often includes an initial rapid decay ($\lesssim$1\,day; e.g. \citealt{woods04,eiz08}) and a more extended phase that can be described by power-law or exponential functions. Sudden flux drops and periods of flux stability have also been observed. Figure\,\ref{outbursts_curves} shows the long-term light curves of all the outbursts discovered up to the end of 2017 and followed with intense and long coverage, mainly using imaging instruments. 
%%%%%%%%%%%%%%%%%%%%%%%%%%%%%%%%%%%%%%%%%%%%%%
%%%%%%%%%%%%%%%%%%%%%%%%%%%%%%%%%%%%%%%%%%%%%%%

\begin{figure*}[h]
\centering
\sidecaption
\resizebox{.75\hsize}{!}{\includegraphics[angle=0]{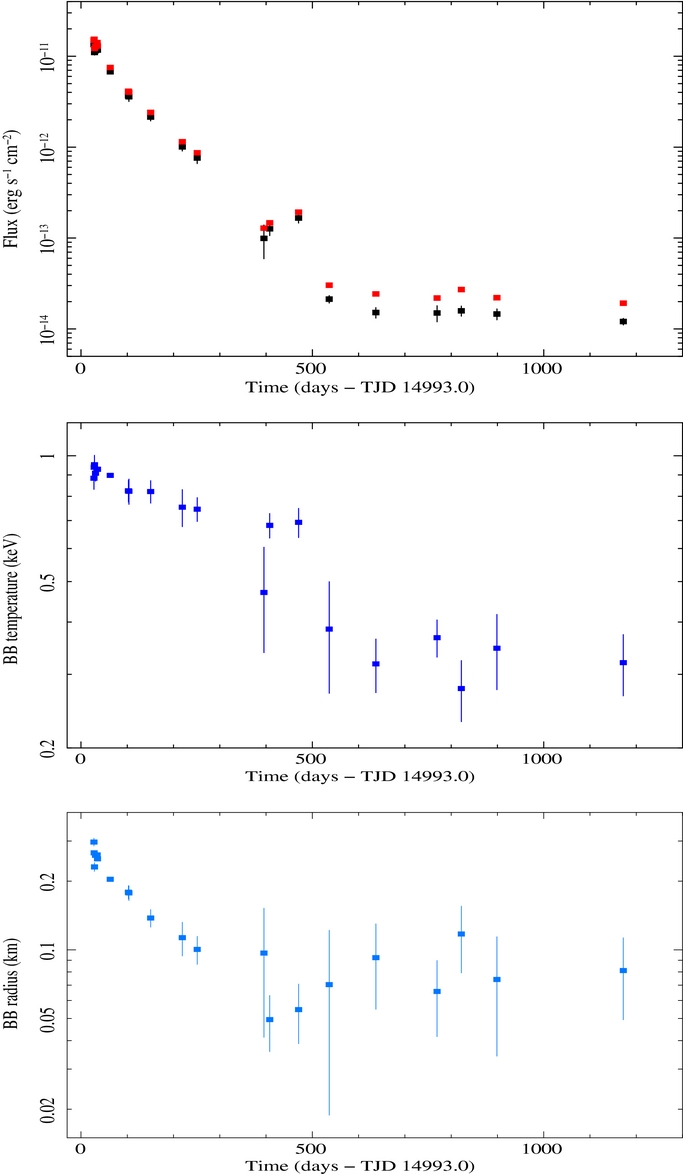}}
\caption{Spectral evolution of SGR\,0418+5729 with time during the outburst started in 2009 June (the time of the detection of the first burst is MJD $54987.9\equiv$ TJD 14987.9). Top panel: flux evolution for the absorbed 0.5--10\,keV flux (black), and for the bolometric unabsorbed flux (red). Middle and bottom panels: evolution of the blackbody temperature and radius, calculated at infinity and assuming a distance of 2\,kpc (from \citealt{rea13}).}
\label{sgr0418_decay}       
\end{figure*}

Despite the great variety of behaviours, all outbursts have some common features: At the beginning of the outburst, the X-ray spectrum is harder than in quiescence, and gradually softens as the flux decreases (Fig.\,\ref{sgr0418_decay}); The larger the luminosity at the outburst peak, the larger the total energy released during the entire episode (generally in the range $\approx$$10^{41}$--$10^{43}$\,erg); The larger the total energy of the outburst, the longer the time scale for the relaxation \citet{cotizelati18}. There is also an anticorrelation between the quiescent X-ray luminosity of a magnetar ($L_{\mathrm{X,q}}$) and the dynamical range of its outbursts; \citet{cotizelati18} found that $L_{\mathrm{X,peak}}/L_{\mathrm{X,q}}  \propto L_{\mathrm{X,q}} ^{-0.7}$, where $L_{\mathrm{X,peak}}$ is the maximum X-ray luminosity achieved during the event (see also \citealt{pons12}). Indeed, several magnetars that could be studied in detail when they were undergoing outbursts, are completely unnoticeable and hardly recognisable as magnetars while in quiescence (or even nondetectable without deep and targeted observations). For this reason, the strategy of the Swift mission \citep{gehrels04} of slewing and pointing its X-ray telescope as soon as possible towards the transient events detected and localized by its wide-field coded-mask detector sensitive to hard X-rays, has proven tremendously successful in discovering new magnetars.
%%%%%%%%%%%%%%%%%%%%%%%%%%%%%%%%%%%%%%%%%%%%%%
%%%%%%%%%%%%%%%%%%%%%%%%%%%%%%%%%%%%%%%%%%%%%%

Outbursts are usually accompanied also by changes in the timing properties of the magnetar. The morphology of the pulse profiles, which generally are broad and with one or two main peaks per cycle, can vary dramatically during an outburst, both in shape (as a rule, becoming more complicate when the activity is high) and in pulsed fraction (e.g. \citealt{woods01,israel10,rodriguez14,esposito10,dib14}). During outbursts and in general in period of enhanced activity, magnetars show a more efficient spin-down torque, with variations of a factor up to $\approx$10 (e.g. \citealt{mte05,olausen14}); the spin-down evolution, however, does not trace or correlate well with the radiative or bursting behaviours \citep{woods07,ybk17}. Among isolated pulsars, magnetars are particularly noisy rotators, and also this aspect is amplified during outbursts \citep{esposito11,dib14}.

In young radio pulsars ($\lesssim$$10^5$ yr), timing noise has been often suggested to be linked to recovery from glitch events \citep{hobbs10}. Magnetars are rather prolific glitchers, comparable to the most frequently glitching radio pulsars, and glitches often happen during outbursts (although in coincidence of some glitches, no X-ray flux enhancements were detected); also, while the amplitude distribution of their glitches peaks on larger $\Delta \nu/\nu$ with respect to the `normal' pulsars, the values observed are in the same range ($\Delta \nu/\nu\sim10^{-9}$--$10^{-5}$; \citealt{dib08,dib14}). Magnetars glitching behaviour appears to be different in the recovery, which is typically very strong, often resulting in an over-recovery, and in the fact that also \emph{anti-glitches} (that is, episodes of sudden spin down) have been reported. The most eminent anti-glitch candidate was reported for 1E\,2259+586 \citep{archibald13}, where in less than 4 days, a spin-down of $\Delta \nu/\nu\sim-10^{-7}$ was achieved; the sudden variation was accompanied by a simultaneous short burst and by a small (factor $\sim$2) but long-lived
(months) flux increase. A large `braking glitch' could also have occurred in SGR\,1900+14 in an 80-days interval including the epoch of its giant flare \citep{wkvp99_3}; the observations were however too sparse to tell whether the abnormal  increase of the period resulted from a sudden event or from a prolonged period of enhanced spin down.

Another magnetar activity associated to X-ray outbursts is the transient pulsed radio emission observed in a few of them. Until the first detection of radio pulses during the outburst of XTE\,J1810--197 \citep{camilo06}, magnetars were (rather staunchly) believed to be radio quiet. Ironically, at the time of its radio activation, XTE\,J1810--197 was the brightest pulsar of the radio sky, with individual pulses reaching flux density of 10\,Jy or more. A few other magnetars were subsequently detected in radio as pulsars: 1E\,1547.0--5408 \citep{camilo07}, PSR\,J1622--4950 (the only magnetar discovered at radio wavelengths so far; \citealt{levin10}), and SGR\,J1745--2900 \citep{eatough13,rep13}. All the detections happened during an X-ray outburst but, interestingly, at least in some cases the radio emission outlived the X-ray flux enhancement \citep{anderson12,camilo16,scholz17}. 
Also, even in periods in which they are overall active in radio, magnetars seem to switch suddenly on and off \citep{burgay09}.\footnote{In this respect, it is interesting to notice that when the radio pulsar PSR\,J1119--6127 emitted a number of SGR-like bursts and entered an outburst, initially it disappeared as a radio pulsar \citep{burgay16}. Furthermore, after the radio reactivation but still during the outburst, \citet{archibald17} using simultaneous radio and X-ray observations observed that the radio emission shut down in coincidence with the X-ray bursts, with a recovery time of $\sim$70\,s.}

\begin{figure*}[h]
\centering
\sidecaption
\resizebox{\hsize}{!}{\includegraphics[angle=0]{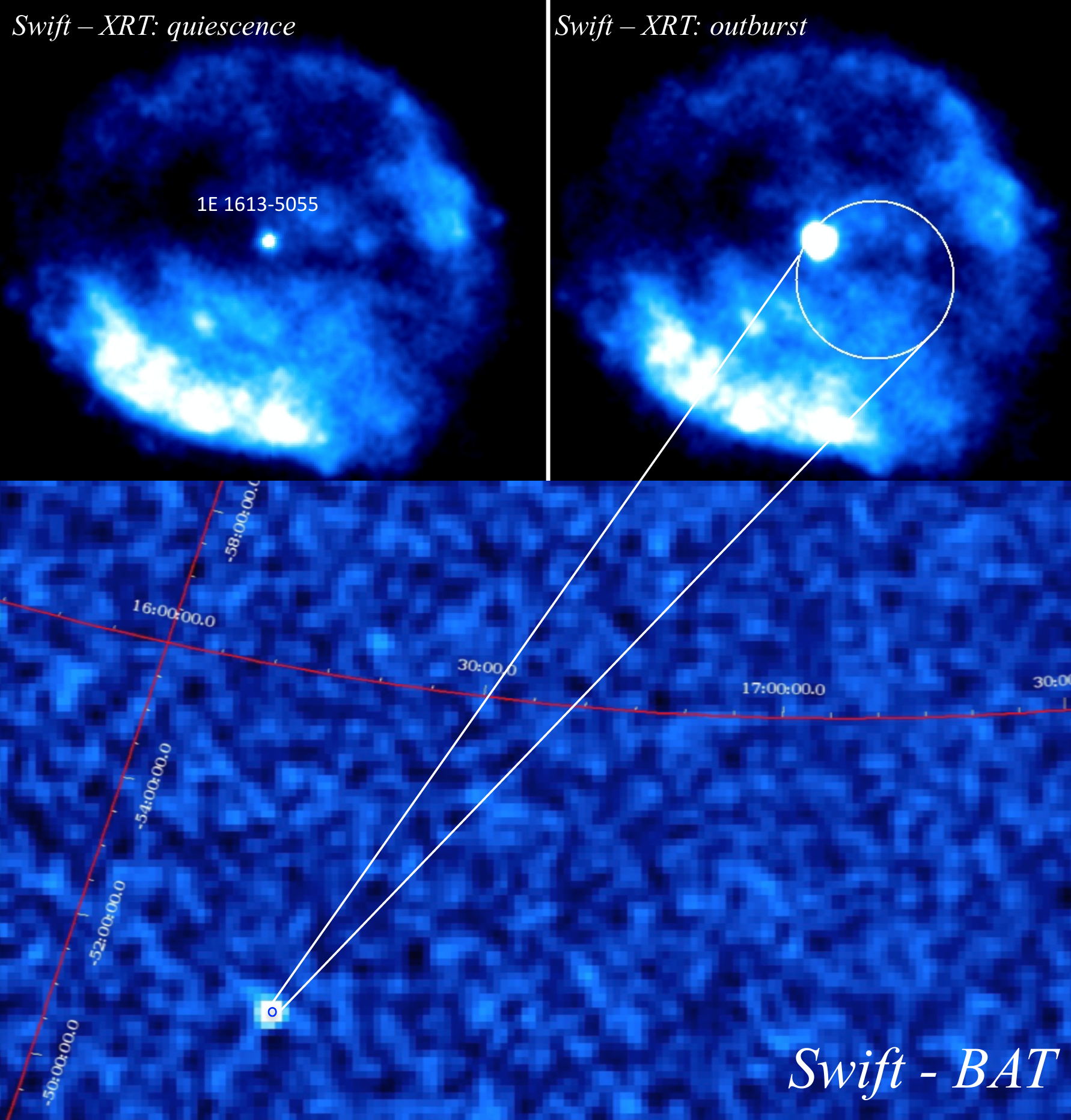}}
\caption{Two Swift-XRT co-added 1--10 keV images of the supernova remnant RCW\,103 during the quiescence state of 1E\,161348--5055 (from 2011 April 18 to 2016 May 16; exposure time
$\sim$33\,ks; top left) and in outburst (from 2016 June 22 to 2016 July 20; exposure time $\sim$67\,ks; top right). The white circle is the positional accuracy of the SGR-like burst detected by BAT (bottom), which has a radius of 1.5\,arcmin (from \citealt{rea16}).}
\label{rcw103image}       
\end{figure*}

Apart from its temporary nature, the pulsed radio emission from magnetars shows in all four sources some clear differences with respect from that of ordinary rotation-powered pulsars \citep{kramer07,camilo08,levin12,shannon13}. The radio emission has a very hard spectrum: $S\propto\nu^{-0.5}$ or flatter (where $S$ is the flux density and $\nu$ is the frequency), while the typical spectral index of radio pulsars is approximately $-1.8$ (e.g \citealt{seiradakis04}). Another peculiar characteristic of magnetar's pulsed radio emission is the instability of the pulse shape. Most radio pulsars show some pulse-by-pulse variability, but the addition of a few hundred pulses is generally sufficient to attain a stable pulse profile; more rarely, radio pulsars switches on time scales from minutes to hours between a small number (usually two) of different pulse profiles \citep{kramer06,lyne10}. The individual pulses of magnetars have a spiky appearance and stable pulse profiles were never observed (e.g. \citealt{kramer07,camilo16}); it is therefore unclear whether the profile stabilization would require for magnetars an unprecedented number of pulses or they simply do not have a stable characteristic pulse profile. Finally,  high degrees of linear polarization have been measured in magnetars, up to very high frequencies \citep{camilo08,torne17}.
\begin{svgraybox}
\textbf{1E\,161348--5055 in RCW\,103: The CCO that was not}\\ 

\noindent Central compact objects in supernova remnants (CCOs) are a small set of isolated neutron stars observed close to centres of young non-plerionic supernova remnants (see \citealt{deluca17} for a review). CCOs are fairly steady X-ray sources with  thermal-like spectra and no counterparts detected at other wavelengths. They owe their redundant designation to the fact that in the years the class emerged, there was not absolute certainty of their neutron-star nature, since no periodic modulations were find in their emission and also searches at radio frequencies failed to detect a pulsar. After numerous deep observations, there is now little doubt that CCOs are indeed neutron stars, and spin periods between 0.1 and 0.5 s and their derivative have been measured in three of them. Interesting, for these sources the inferred dipolar magnetic fields are rather low: $\sim$$10^{10}$--$10^{11}$\,G. This prompted for the CCOs an unifying scenario in which they are either born with weak magnetic fields or with a normal field that has been `buried' beneath the neutron-star surface by a post-supernova stage of hypercritical accretion of fallback matter \citep{ho11,vigano12,gotthelf13}. In the latter case, CCOs could in principle have magnetic field in the magnetar range \citep{vigano12}.

1E\,161348--5055 in the 2-kyr-old supernova remnant RCW\,103 was one of the CCO prototypes. However, observations of large flux variations (about 2 orders of magnitude) and an unusual spin period of 6.7\,h\ set it apart from CCOs or any other class of isolated pulsars \citep{deluca06}. No information or meaningful limit on its magnetic field are available from its rotational parameters \citep{etdl11}, but it is interesting to notice that  \citet{deluca06} discussed as a possible mechanism to slow-down in $\sim$2-kyr a pulsar born with a normal spin period to the rotation rate of 1E\,161348--5055 the propeller interaction between an ultra-magnetised neutron star ($B\sim10^{14}$--$10^{15}$\,G) and a surrounding supernova fallback debris disk.

A major breakthrough was when, on 2016 June 22, the Swift's Burst Alert Telescope detected an X-ray burst (see Fig.\,\ref{rcw103image}) resembling in all respects those of magnetars from the direction of 1E\,161348--5055 \citep{dai16,rea16}. Its duration was $\sim$10\,ms, its luminosity $\sim$$2\times10^{39}$ erg s$^{-1}$ (15--150\,keV), and the spectrum was well described by a blackbody with $kT\sim9$\,keV. Subsequent follow-up observations with Swift, Chandra and NuSTAR showed that 1E\,161348--5055 was undergoing a magnetar-like outburst: Its luminosity was $\sim$100 times higher than the level the source had maintained for several years and up at least to the last observation carried out before the burst (about one month earlier); The pulse profile from single- became double-peaked; A hard power-law component was observed up to $\sim$30\,keV for the first time in its  energy spectrum, superimposed to its usual thermal emission \citep{rea16}. In the first year from the onset of the outburst, the overall energy emitted was $\sim$$3\times10^{42}$ erg \citep{cotizelati18}. Moreover, Hubble observations carried out in the summer of 2016 unveiled at the position of 1E\,161348--5055 a faint infrared source that was not detected in older observations (implying a minimum brightening of 1.3 mag) and therefore can be assumed with a high degree of confidence to be the counterpart of the CCO \citep{tendulkar17}.

While the features exhibited by 1E\,161348--5055 during its outburst match precisely the distinguishing features of magnetars, its 6.7-h period remains puzzling. The infrared observations definitely ruled out any doubt about a binary system but could not confirm of exclude the presence of a fallback disk. Recent modelling of neutron star--debris disk interaction by \citet[see also \citealt{tong16}]{ho17} has shown that a disk with mass of $\approx$$10^{-9}$ M$_{\odot}$ could slow the neutron star period from milliseconds down to hours in $\sim$1--3\,kyr, if the dipole magnetic field is of $5\times10^{15}$\,G.
Since the formation and survival of a supernova fallback disk are unclear \citep{perna14}, theoretical work is ongoing also on the possibility that the source was slowed down in a relatively short-lived phase of propeller during fallback accretion. 
\end{svgraybox}
%%%%%%%%%%%%%%%%%%%%%%%%%%%%%%%%%%%%%%%%%%%%%%
%%%%%%%%%%%%%%%%%%%%%%%%%%%%%%%%%%%%%%%%%%%%%%

\begin{svgraybox}
\textbf{SGR\,J1745--2900: The Galactic Centre magnetar}\\

\noindent When on 24 April 2013 Swift detected with its X-ray telescope a large flare from the region of Sgr\,A* \citep{drm13}, many minds went to the much anticipated pericenter passage of the object G2, which at the time was expected around mid-2013, and its possible tidal disruption by the Milky Way's $4\times10^6$-M$_\odot$ black hole \citep{gillessen12,gillessen13}.  

Two days later, however, a magnetar-like short burst was detected by the Swift's coded mask instrument and also a typical magnetar period of 3.8\,s was detected, using NuSTAR \citep{kennea13,mori13}. The situation was definitely settled when an observation carried out with Chandra (the only X-ray telescope with sufficient angular resolution) showed that a 3.8-s magnetar in outburst, SGR\,J1745--2900, very close to the position of Sgr\,A* (angular separation of $(2.4\pm0.3)$\,arcsec) was responsible for all the X-ray luminosity increase measured by Swift ($\approx$$2\times10^{35}$\,erg s$^{-1}$ at 8.3\,kpc, while Sgr\,A* was not detected in the same exposure; \citealt{rep13}). 
(The closest approach of G2 to Sgr\,A* actually took place in early 2014; G2 survived and no flaring activity clearly associated to the event was observed; \citealt{phifer13,pfuhl15,ponti15,plewa17}.)
On 28 April 2013, the new magnetar was also detected as a radio pulsar, the one with the highest dispersion measure and rotation measure, suggesting that the source is
embedded in the dense and magnetized plasma of the Galactic center \citep{eatough13,rep13,bower15,lynch15,torne15,pennucci15}.

The angular separation between the magnetar and Sgr\,A* corresponds to a projected distance of only 0.1\,pc, and \citet{rep13}  estimated that if SGR\,J1745--2900 was born within 1\,pc of Sgr\,A*, its probability of being in a bound orbit around the black hole is of $\sim$90\%. \citet{bower15} measured a transverse velocity of the source relative to Sgr\,A* of ($236\pm11$)\,km s$^{-1}$ and provided further support to the possibility that the magnetar is bound to Sgr\,A*. 
\citet{rep13} also noted that the high-energy emission produced by the past activity of the magnetar, passing through the molecular clouds surrounding the Galactic center region, might be responsible for a substantial fraction of the light echoes observed in the Fe fluorescence features.

SGR\,J1745--2900 is proving to be an important probe for the compact object population and the interstellar medium in the Galactic center, but is also exhibiting an interesting behaviour as a magnetar. About 3.5 years after the outset of the outburst, it has not reached the quiescent/pre-outburst luminosity level yet \citep{cotizelati17}. Its spectral evolution is difficult to reconcile with crustal cooling models, while a continuous particle bombardment from returning currents of the neutron star surface better explain the data. In this hypothesis, both temperature and size of the region at the footprint point of the bundle of current-carrying field lines decrease, as the magnetospheric twist gradually dissipates and the rate of particles impacting the surface consequently declines \citep{cotizelati15,cotizelati17}.

\end{svgraybox}

\subsection{Magnetar formation}

The generation of magnetar-like magnetic fields from the progenitor star is still a debated and relatively open problem.  All along, preliminary calculations have shown that the effects of a turbulent dynamo amplification occurring in a newly born neutron stars can indeed result in a magnetic field of up to a few $10^{17}$\,G. This dynamo effect is expected to operate only in the first $\sim$10\,s after the supernova explosion of the massive progenitor, and if the proto-neutron star is born with sufficiently small rotational periods (of the order of 1--2\,ms).
The resulting amplified magnetic fields are expected to have a strong multipolar structure and toroidal component (Duncan \& Thompson 1992, Thompson \& Duncan 1993). This formation scenario predicts two main observational consequences: (a) magnetars should have large kick velocities, of the order of $10^{3}$\,km\,s$^{-1}$ and (b) their associated supernovae should be more energetic than ordinary core collapse-supernovae, because of the additional rotational energy loss of such fast spinning proto-neutron star. 

However, this additional energy loss  is not observed in the supernova remnants surrounding magnetars \citep{vink06,martin14}, nor a large kick velocity is observed in the few cases where this could be measured. In particular, measured magnetar proper motions are $v = 212\pm35$\,km\,s$^{-1}$ for XTE\,J1810--197 \citep{helfand07}, $v = 280\pm130$\,km\,s$^{-1}$ for 1E\,1547--5408 \citep{deller12}, $v=157\pm17$\,km\,s$^{-1}$ for 1E\,2259$+$586 and $v=102\pm26$\,km\,s$^{-1}$ for 4U\,0142$+$61 \citep{tendulkar13}, while candidate proper motion velocities are $v = 350\pm100$\,km\,s$^{-1}$ for SGR\,1806--20 and $v = 130\pm30$\,km\,s$^{-1}$ for SGR\,1900+14 \citep{tendulkar12}. All these values are well within the typical radio pulsar distribution (see also the gray box on the Galactic Centre magnetar SGR\,1745--2900).

The fact that the former predictions did not seem to be fulfilled is however not sufficient to dismiss the dynamo formation mechanism (for example, about the lack of evidence for a particularly energetic supernova, \citealt{dallosso09} noted that most of the rotational energy of a proto-neutron star with  internal toroidal field $\approx$$10^{16}$\,G should be released through gravitational waves, without supplying substantial additional energy to the ejecta), but it has lent some support to other formation scenarios. One alternative theory is based on magnetic flux conservation arguments and postulates that the distribution of field strengths in neutron stars simply reflects that of their progenitors. In this fossil field scenario,  magnetars would be the descendant of the massive stars with the highest magnetic fields \citep{ferrario06}.
Counter-arguments have been put forward also in this scenario by \citet{spruit08}, who argued that the number of highly magnetic massive stars with $B\gtrsim$1\,kG is not sufficient to explain the magnetar population. Furthermore, even assuming that most of the magnetic flux is indeed conserved, magnetic fields higher than $10^{14}$\,G seem unattainable.

Recent surveys of very massive stars have shown how in our Galaxy massive stars tend to be in binaries \citep{sana06}. Furthermore,  a detailed radial velocity survey of Westerlund\,1, an open cluster of very massive stars which contains a magnetar, CXOU\,J164710.2--455216, have discovered the possible companion massive star that might have resulted by the disruption of a massive binary progenitor (\citealt{clark14}; see Section\,\ref{binaries} for more details). All these results are pointing to a further element in magnetar formation: the evolution in a binary system of massive stars. The binary scenario might overcome the problem of the spin down by the core--envelope coupling. In particular, both mass transfer and stellar merger in compact binaries may lead to substantial spin-up of the mass-gainer (or of the remnant of the merger), favoring the amplification of the magnetic field via dynamo effects \citep{langer12}. Recent simulations have shown that gamma-ray bursts and hyper-luminous supernovae can indeed be powered by recently formed millisecond magnetar \citep{metzger11}, although no direct or sound observational evidence of the existence of such fast spinning and strongly magnetic neutron stars has been collected thus far.

\subsection{Magnetic field evolution and the neutron star bestiary}

The evolution of magnetic fields in neutron stars has been extensively studied by a number of authors in the past. In the neutron star solid crust, the field evolves under the influence of the Lorentz force (causing the Hall drift) and the Joule effect (responsible for Ohmic dissipation). The evolution in the liquid core is very uncertain. In the core, soon after the neutron star birth (from hours to days) protons undergo a transition to a type-II superconducting phase \citep{baym69}, in which the magnetic  field is confined to tiny flux tubes surrounded by nonmagnetized matter. The dynamics of those flux tubes, likely coupled to the motion of superfluid neutron vortices, is a complex problem that makes the magnetic field evolution in the core formally difficult to tackle (see \citealt{elfritz16}). Most works  \citep[e.g.][]{goldreich92,geppert02,geppert06,pons07,pons09,gonzalez10,vigano13} considered mainly the magnetic evolution in the solid crust, a $\sim$1-km-thick lattice of ions, where the electrical conduction is governed by electrons. 

 The magnetic field evolution in a neutron star is strictly coupled to its thermal evolution. In fact, the magnetic field influences the heating rate and, secondarily, affects the rate of a few neutrino processes; on the other side, the conduction of heat becomes anisotropic in the presence of  a strong magnetic field. The simultaneous study of the magnetic and temperature evolutions (\emph{magneto-thermal evolution}) was started by \citet{pg07} and \citet{aguilera08} with simplifying assumptions, and later implemented in two-dimensional simulations of the fully coupled magneto-thermal evolution in \citet{pons09}, but including only the  Ohmic dissipation. More recently, \citet{vigano12} presented the first two-dimensional magneto-thermal code able to manage arbitrarily large magnetic field intensities while self-consistently including the Hall term throughout the entire evolution. 

The magneto-thermal evolution in the lifetime of the neutron star is governed by the Hall induction equation, for the magnetic evolution:
\begin{description}[Type 1]
\item[ ]{$$\frac{\partial \vec{B}}{\partial t} = - \vec{\nabla}\times \left\{ \frac{c^2}{4\pi\sigma} \left[ \curlB - \omega_B \tau_{e} \left(  \curlB \right) \times \vec{B}/B  \right] \right\}\,, $$ where $\sigma$ is the electrical conductivity,  $e^\nu$ is the lapse function that accounts for redshift corrections, and $\omega_B\tau_{e}=B\sigma/en_ec$ is the magnetization parameter ($\omega_B=eB/m^*_ec$ is the gyration frequency of electrons, $n_e$ is the electron number density, and $\tau_e$ and $m^*_e$ are the relaxation time and effective mass of electrons),}
\item[and the cooling, or energy-balance, equation for the thermal evolution of the crust:]{$$c_v\frac{\partial T}{\partial t} + \vec{\nabla}\cdot(-\hat{\kappa}\cdot\vec{\nabla}T) = -Q_\nu + Q_j\,,$$ where $c_v(T,\rho)$ is the specific heat  (mainly driven by the neutrons), $\hat{\kappa}(T,\rho)$ is the thermal and electrical conductivity, and  $Q_\nu(T,\rho)$ are the neutrino emissivities, and all these quantities depend on the temperature ($T$) and density ($\rho$).}
\end{description}
For low temperatures ($T\lesssim 10^8$ K) or strong magnetic fields, $\omega_B\tau_e\gg 1$, the evolution is Hall-dominated. For high temperatures (large resistivity) or weak fields, \mbox{$\omega_B\tau_e\ll 1$}, we have instead what is called the {\it dissipative regime}.\\ 

\begin{figure*}[h]
\centering
%\vbox{
\includegraphics[width=8cm]{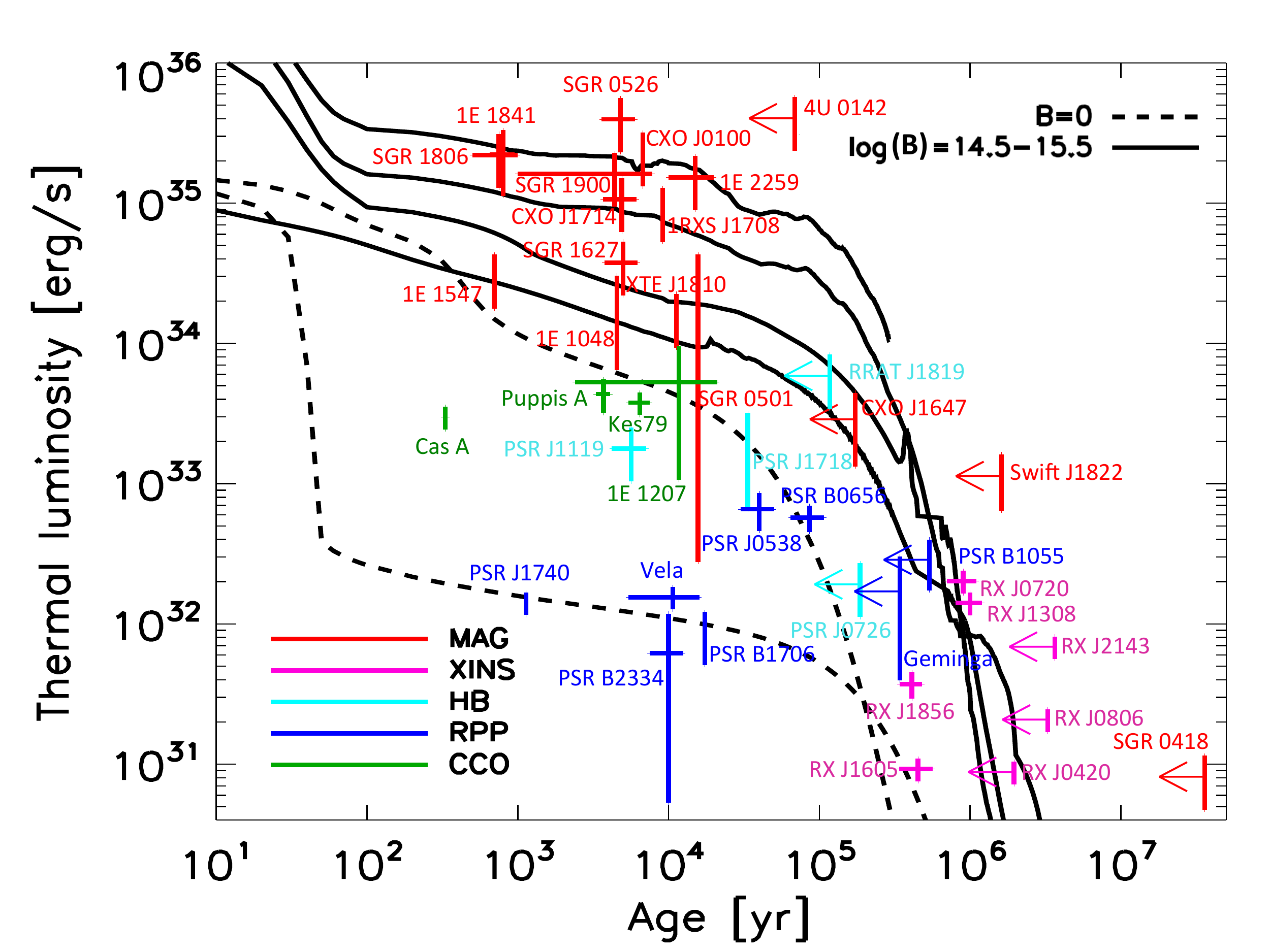} 
\includegraphics[width=8cm]{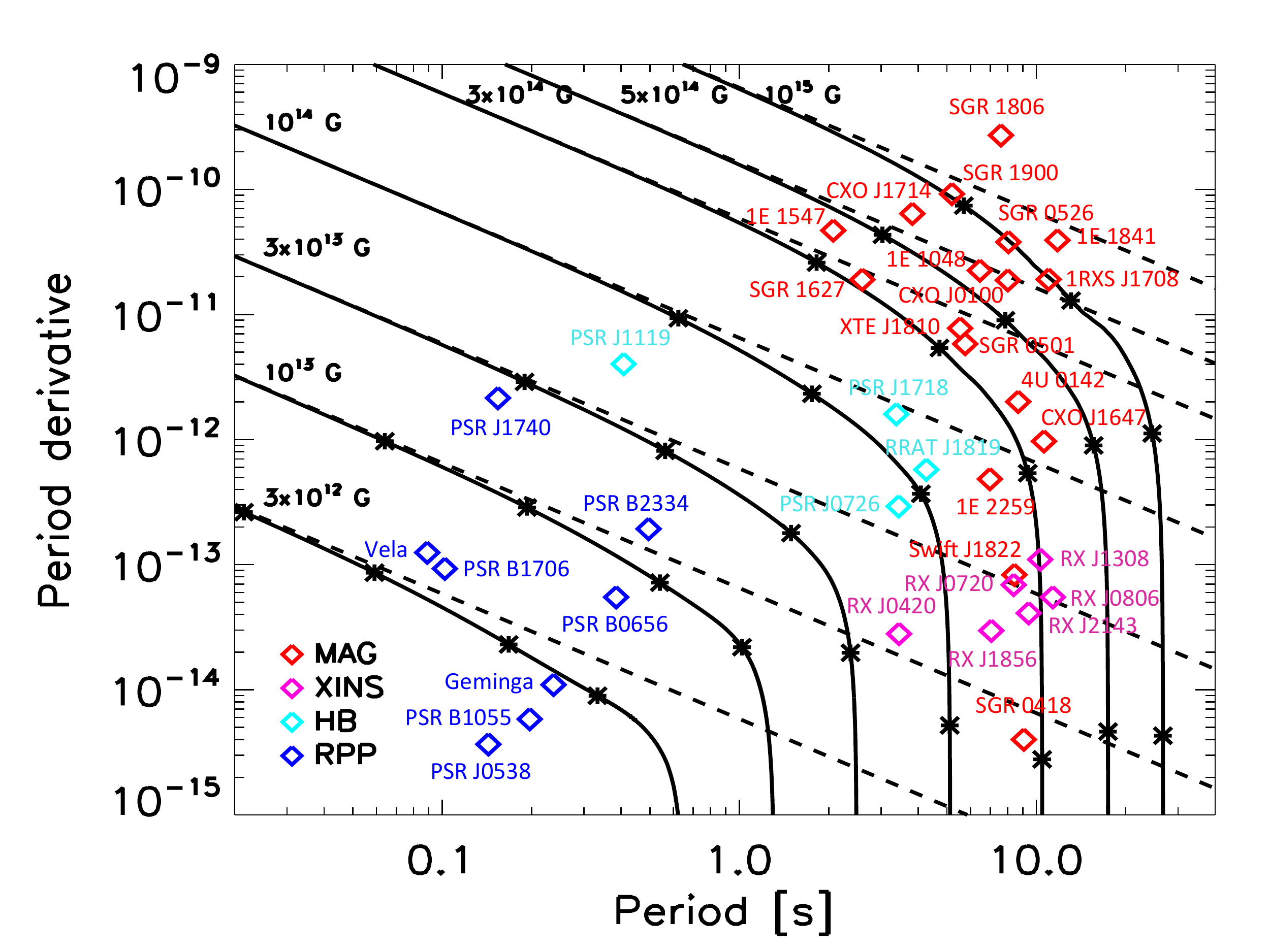}
\caption{Top: Comparison between observational data and theoretical cooling curves. The thermal luminosity was evaluated for each source using the best estimated distance available and ages were estimated from kinematic measurements, when available, or as the characteristic age derived from the timing parameters (in this case, an arrow was used for objects older than 10\,kyr, for which the characteristic age is considered an upper limit on the real age and not a reasonable approximation). Dashed lines are non-magnetic cooling curves, the upper with $M=1.10$\,M$_\odot$ and a light-element envelope, and the lower with $M=1.76$\,M$_\odot$ and an iron envelope. The magneto-thermal evolutionary tracks (solid lines) were computed for magnetic fields in the range $B=3\times 10^{14}$--$3\times 10^{15}$\,G and both iron and light-element (upper) envelopes. Bottom: Evolutionary tracks in the $P$--$\dot{P}$ diagram for a 1.4-M$_\odot$ neutron star with $B_p^0 = 3\times10^{12},~10^{13},~3\times10^{13},~10^{14}, 3\times10^{14},~\mathrm{and}~10^{15}$\,G. Asterisks mark the real ages $t = 10^3,~10^4,~10^5,~5\times10^5$\,yr, while dashed lines show the tracks followed in absence of magnetic field decay.(Adapted from \citealt{vigano13}.)} 
 \label{mt}
%\label{ppdot_evolution}    
 \end{figure*}

During the past decades, the study of the neutron star field decay, thanks to the continual comparison between theoretical modelling and observations, proved to be fundamental for the understanding of the secular evolutions of pulsars. In particular, recent advances in the magneto-thermal evolutionary models and the availability of deep X-ray observations of many thermally emitting isolated neutron stars, allowed a significant improvement towards the unification of the ``bestiary" of different classes of isolated  neutron stars \citep[see e.g.][for an overview of the different  observational manifestations of  neutron stars]{kaspi10,kaspi16}. The sample of detected neutron stars with  thermal emission consist of about 40 sources, ranging from magnetars, X-ray dim isolated neutron stars, and  central compact objects, to rotational-powered pulsars and `high-$B$' pulsars \citep{vigano13}. 
Figure\,\ref{mt} (upper panel) shows the thermal  luminosity of all these neutron stars as a function of the age together with several cooling curves for magnetic fields in the range of $3\times10^{14}$--$3\times10^{15}$\,G and for two envelope compositions, hydrogen and iron, as computed by \citet{vigano13}. Note that for young neutron stars ($t<100$\,kyr, still in the neutrino cooling era), light-elements envelopes are able to maintain a higher luminosity (up to an order of magnitude) than iron envelopes.
%%%%%%%%%%%%%%%%%%%%%%%%%%%%%%%%%%%%%%%%%%%%%%%%%
%%%%%%%%%%%%%%%%%%%%%%%%%%%%%%%%%%%%%%%%%%%%%%%%%
The inspection of a range of theoretical models, as well as observations, has shown that the magnetic field has little effect on the luminosity for `weakly' magnetized neutron stars with $B < 10^{13}$\,G. These objects, of which the radio pulsars are the most notable representatives, have thermal luminosities that are compatible with those predicted by standard non-magnetic cooling models. Overall, the magneto-thermal simulations can broadly reproduce the observed X-ray luminosities for a range of initial magnetic field strengths, envelope compositions, and neutron star masses. As the neutron stars age and become colder, they also spin down, primarily due to dipolar
radiation losses. In the absence of field decay, pulsars should follow linear tracks in the $P$--$\dot{P}$ diagram (see dashed lines in the lower panel of Fig.\,\ref{mt}). However, when magnetic  field dissipation is taken into account, evolutionary tracks in the $P$--$\dot{P}$ diagram bend down (Fig.\,\ref{mt}). 

Comparing observations and theoretical modelling of the neutron star magneto-thermal evolution, considering both the luminosity and the rotational period properties, we can gather that objects like the traditional rotation-powered radio pulsars were born with magnetic fields in the range of a few $10^{12}$--$10^{13}$\,G. When they cool and slow down, they eventually become invisible in both the radio and the X-ray bands, and hence they lack observable counterparts. On the other hand, pulsars born with fields exceeding the $10^{14}$\,G, will be observed now as young magnetars or high-$B$ pulsars (depending on the strength and the configuration of the field at birth), and have as descendants the objects known as X-ray dim isolated neutron stars. These simulations point to evolutionary connections (some of which have been suspected for long) between apparently different groups of pulsars: Most likely, they are all essentially the same kind of objects, but they were born with different magnetic field strength and geometry, and are observed at different evolutionary stages of their life.

\subsection{Low-$B$ magnetars and high-$B$ pulsars}

Recently, the long standing belief that magnetars must posses supercritical magnetic fields\footnote{The electron quantum critical magnetic field $B_{\mathrm{Q}} =m^2_ec^2/(\hbar e) \simeq 4.4\times10^{13}$ G was traditionally considered the threshold above which magnetars could be found.} has been challenged by the discovery of full-fledged magnetars  with a dipole magnetic field well in the range of ordinary radio pulsars: \src, \lowbb, and \3xmm\ (\citealt{rea10,rie12,rea14}; see \citealt{turolla13} for a review). Those three magnetars are in fact not dissimilar from the other members of the class, except for the strength of the dipole magnetic field $B_p$ estimated from the spin parameters, in the range $(0.6$--$4)\times10^{13}$\,G. Furthermore, this results also in a large characteristic age of $>$$10^{6}$\,yr, two or three orders of magnitudes larger than the typical magnetar ages, suggesting that these low-field magnetars might be old objects.
The small number of detected bursts (with comparatively low energetics) and the low persistent luminosity in quiescence have been taken as further hints that these might be worn-out magnetars, approaching the end of their active life \citep{turolla11}. The `old magnetar' scenario sounds appealing since it offers an interpretation of the low-magnetic-field magnetars within an already well established framework, validating the magnetar model also for (surface) field strengths quite far away from those of canonical SGR/AXPs. 

\begin{figure*}[h]
\resizebox{\hsize}{!}{\includegraphics[angle=0]{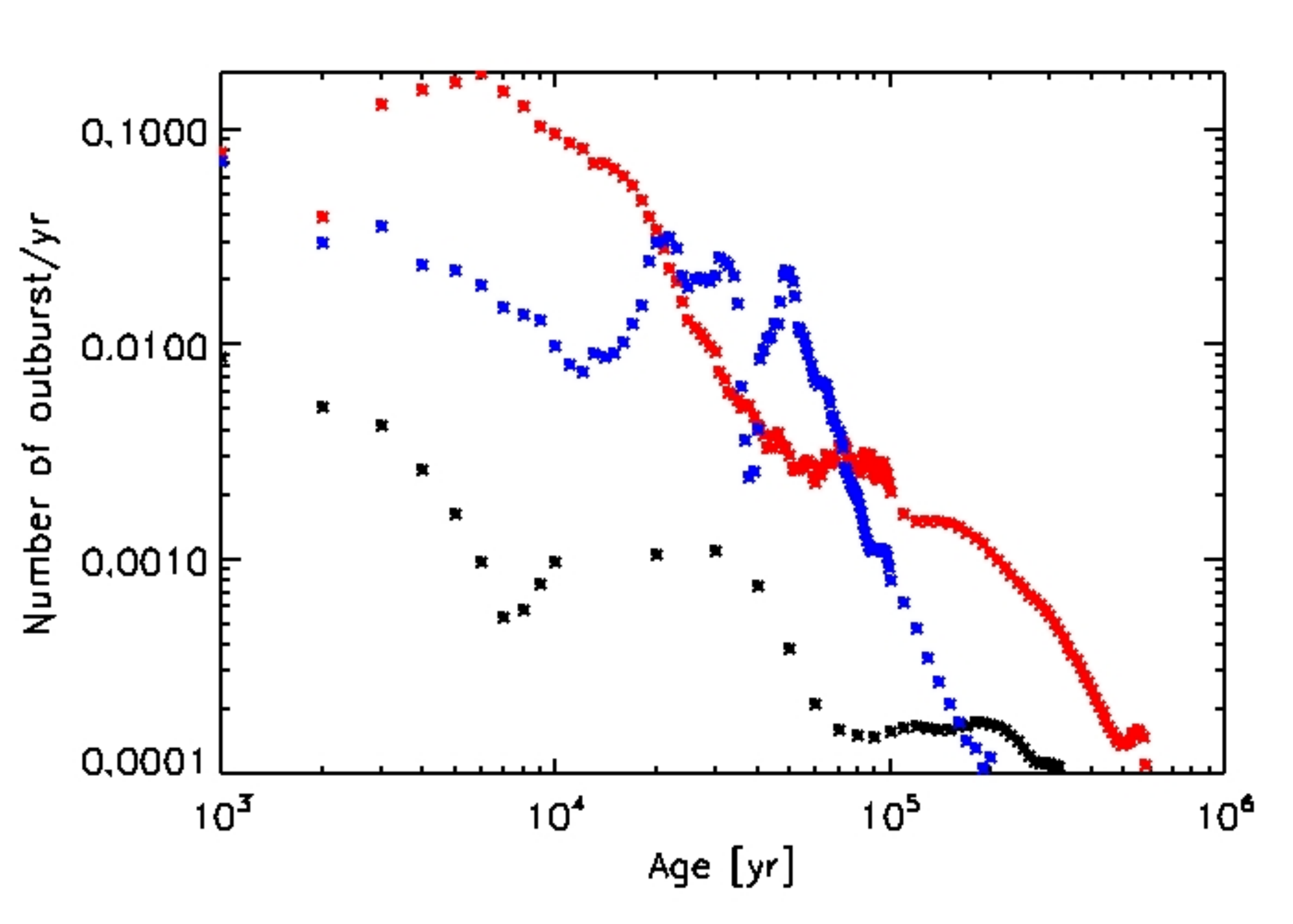}}
\caption{Outburst rate as a function of time for neutron stars with different magnetic field configurations and strength: $B_p^0 = 3\times10^{14}$\,G in black, $B_p^0 = 10^{15}$\,G in red, and in blue $B_p^0 = 10^{14}$\,G but with an initial strong toroidal field of $B_t^0 = 5\times10^{15}$\,G \citep[from][]{vigano13}.} 
\label{burstrates}
 \end{figure*}

\begin{figure*}[h]
\sidecaption
\resizebox{\hsize}{!}{\includegraphics[angle=0]{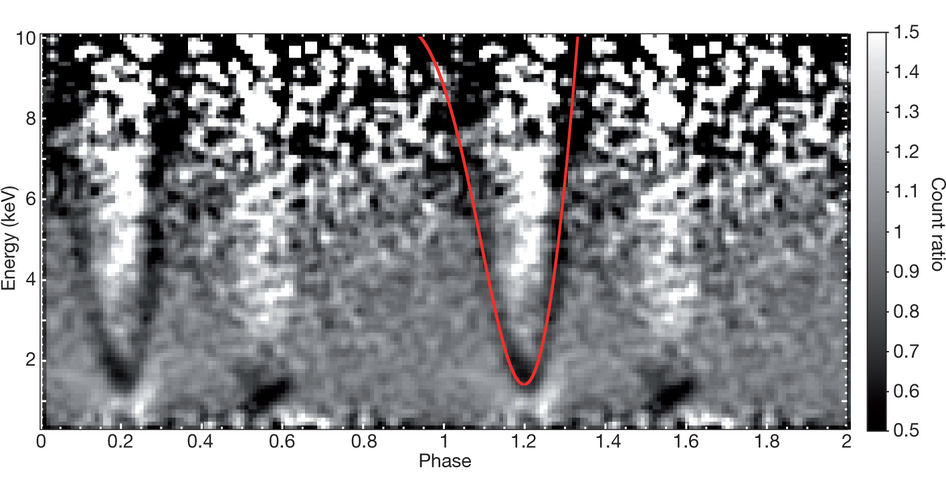}}
\caption{Normalised energy-versus-phase `spectral image' of SGR\,0418+5729. It was obtained from the XMM--Newton data binning the source photons into 100 phase bins and 100-eV-width energy channels and normalising the counts first by the phase-averaged energy spectrum and then by the pulse profile (normalised to the average count rate). The red line shown for one of the cycles represents the the proton cyclotron model of \citet[][from which the image was taken]{tiengo13} and highlights the V-shaped absorption feature.}
\label{sgr0418line}       % Give a unique label
\end{figure*}

The crucial issue is whether a relatively low dipolar field is consistent with the starquake models, in which the primary cause of the outbursts is an internal deposition of energy following a crust failure once the magnetically induced shear stress exceeds a critical value. The magnetic stress needed to break the crust is strongly dependent on the density (it is much easier to break the outer crust than the inner crust); moreover, the crust thickness grows as the temperature drops with age. Detailed calculations show that a local magnetic field of $\approx$$2\times 10^{15}$\,G should be necessary to break the crust, but closer to the surface of the crust,  due to the smaller density, magnetic fields as low as $\sim 10^{14}$\,G may lead to crust fractures \citep{gourgouliatos15,lander15}. At any rate, the minimum requirement seems to be around $10^{14}$\,G. So, can (and how) aged, cold and low-magnetic-field magnetars still produce bursts and outbursts? This depends on the internal toroidal component of its magnetic field. For this reason, objects with similar dipolar magnetic field strength as inferred from their period and period derivative can display very different behaviours.  In general the toroidal component of the magnetic field is unmeasurable in a pulsar (but see the gray box for SGR\,0418+5729), but this reasoning help us understanding and explaining the populations of active magnetars, low-magnetic-field magnetars, and high-$B$ pulsars. A rough prediction of the expected outburst rate for different initial magnetic field configurations and life stages is given for standard assumptions in Fig.\,\ref{burstrates}. For an object similar to \src, a rate of $\approx$$10^{-3}$ starquakes\,yr$^{-1}$ is expected \citep{perna11,vigano13}. Assuming that there are about $10^4$ neutron stars in the Galaxy with similar age, and that a (very approximatively) 10\% of them were born as magnetars, a naive extrapolation of this event rate to the whole neutron star population leads to the occurrence of $\sim$1 low-magnetic-field-magnetar outburst per year. Therefore, we expect that more and more objects of this class will be discovered in the upcoming years. Similarly, we can anticipate some magnetar-like events from only-sporadically-active sources labelled as belonging to other classes of isolated neutron stars, as perhaps shown by the outbursts of PSR\,J1846--0258 and PSR\,J1119--6127 \citep{gavriil08,archibald16,gogus16}.
%%%%%%%%%%%%%%%%%%%%%%%%%%%%%%%%%%%%%%%%%%%%%%%%%
%%%%%%%%%%%%%%%%%%%%%%%%%%%%%%%%%%%%%%%%%%%%%%%%%
\begin{svgraybox}
\textbf{An energy-dependant absorption line in the cornerstone low-magnetic-field magnetar: \src}\\ 

\noindent The most searched-for indicator of the magnetic field strength in neutron stars are cyclotron features in their spectra. The cyclotron energy for a particle of charge and mass $e$ and $m$ is 
$E_{\mathrm{cycl}}=11.6\,(m_e/m)/(1+z)\,\,B_{12}\,\mathrm{keV}$, 
where $z\approx0.8$ is the gravitational redshift, $m_e$ is the mass of the electron, and $B_{12}$ is the magnetic field in units of $10^{12}$\,G. For magnetic fields of $\approx$$10^{14}$\,G, magnetospheric protons can produce cyclotron lines in the soft X-ray range . 

The most convincing of such features in a magnetar was reported by \citet{tiengo13}, who observed a phase-dependent absorption feature in the spectrum of the low-magnetic-field magnetar SGR\,0418+5729 during its 2009 outburst \citep{esposito10,rea13}. The feature was more prominent in a deep XMM--Newton observation but was present also in data collected with RossiXTE and Swift \citep[][see also \citealt{esposito10}]{tiengo13}.
 The line energy was varying between $\sim$1 and 5\,keV in approximately one-fifth of the rotation cycle (Fig.\,\ref{sgr0418line}), corresponding to magnetic field strength values of $10^{14}$ to $10^{15}$\,G if due to protons (\citealt{tiengo13} devised a toy model in which the cyclotron line is due to thermal photons crossing protons localised in a magnetic loop with magnetic field of $10^{14}$--$10^{15}$ G close to the surface of the star). An electron cyclotron feature is indeed implausible, as the electrons, considered the dipole magnetic field of $B_p\simeq6\times10^{12}$\,G, should be confined in a small volume at a few stellar radii from the neutron star surface (while the alternative explanation in terms of atomic transition lines is ruled out by the variability of the line energy with the spin phase). 
If the feature is really a proton cyclotron line, it demonstrates in SGR\,0418+5729 the presence of nondipolar magnetic field components strong enough to break the neutron star crust and give rise to magnetar outbursts.

\citet{rodriguez16} reported the presence of a similar feature in Swift\,J1822.3--1606, the magnetar with the second lower magnetic field ($B_p\sim(1$--$3)\times10^{13}$\,G \citealt{olausen14,scholz12}), again indicating the presence of localised magnetic fields of $10^{14}$--$10^{15}$\,G. Interestingly, spectral features with analogous characteristics have been detected also in two X-ray dim isolated neutron stars, RX\,J0720.4--3125. and RX\,J1308.6+2127 \citep{borghese15,borghese17}; in these cases, the magnetic fields deduced in the hypothesis of a proton cyclotron line are of $\sim$$2\times10^{14}$\,G, in both cases around 5 times the values inferred from the spin parameters.
\end{svgraybox}
%%%%%%%%%%%%%%%%%%%%%%%%%%%%%%%%%%%%%%%%
%%%%%%%%%%%%%%%%%%%%%%%%%%%%%%%%%%%%%%%%

\subsection{Magnetars in binary systems}\label{binaries}

It is not clear whether magnetar exist in binary systems but, although it is possible that magnetars to form have to sacrifice the companion, binarity may be an important ingredient for the production of magnetars for at least two reasons. Firstly, if magnetars descend from particularly massive stars ($\gtrsim$20\,M$_\odot$), the binary interaction may prevent the formation of a black hole instead of a neutron star. Secondly, in the case the magnetic fields of magnetars are formed through dynamo amplification, even if magnetars form from stars with lower masses, a binary system may help the stellar core to maintain the angular momentum necessary for the dynamo mechanisms. On the other hand, in the fossil field scenario, in which the neutron star magnetic field reflects that of the stellar precursor, progenitors belonging to binary systems may be disfavoured, since magnetic hot stars in compact binary systems are very rare \citep{neiner15}.

Both scenarios for the origin of magnetar magnetic field require a very massive progenitor, heavier than $\sim$20\,M$_\odot$. For the fossil field, because there seems to be in main sequence stars a trend of stronger magnetic fields with larger masses (\citealt{ferrario08} and references therein). In the dynamo hypothesis, because to produce neutron stars rotating fast enough to generate a magnetar field via $\alpha$-$\omega$ dynamo,  a star more massive than $\sim$20--35\,M$_\odot$ star is necessary \citep{heger05}. 
There are also observational facts that favour very massive stars as the magnetar precursors. Considering the spatial distribution of the known magnetars, it has been noted that their height on the Galactic plane is smaller than that of OB stars. This suggests that they were produced by the most massive O stars \citep{olausen14}. Furthermore, there are some tentative associations of magnetars with massive star clusters. 

In the most convincing case, CXOU\,J164710.2--455216 in Westerlund~1, the young age of the cluster ($\sim$4\,Myr) implies a progenitor with minimum mass of $\sim$40\,M$_\odot$ \citep{muno06}.\footnote{For the progenitor of SGR\,1900+14, a lower mass of $\sim$17\,M$_\odot$ has been inferred, suggesting that magnetars could form from stars with a wide spectrum of initial masses \citep{clark08,davies09}.}
To allow such a massive star to produce a neutron star, \citet{clark14} suggested a binary with $(41+35)$\,M$_\odot$ stars and an orbital period shorter than 8 days.\footnote{They also found a candidate for the other member of the pre-supernova system: Wd1--5, a $\sim$9-M$_\odot$ runaway star that is escaping the cluster at high velocity and has a peculiar carbon excess that may be due to the binary evolution.} The idea is that the binary interaction drives the primary in a Wolf--Rayet star that through its powerful stellar winds loses mass to the point that the formation of a neutron star is possible. The star also avoids the supergiant stage, during which the core would lose angular momentum because of the core--envelope coupling. 

Observationally, the presence of magnetars is often invoked in high-mass X-ray binaries \citep[e.g.][]{bozzo08} and in particular for some persistent Be systems with long-spin-period ($\gtrsim$1000\,s) neutron stars. The issue is that, according to the standard picture,  after a short propeller phase, the neutron star enters the accretor stage and its spin period quickly settles at an equilibrium value. To have an equilibrium period longer than 1000\,s, the neutron star dipole field must be $\gtrsim$$10^{14}$\,G  \citep{davies81}. This is unless the accretion rate is very low, but it would be orders of magnitude below the rate necessary to account for the luminosity of these persistent Be X-ray binary systems, hence the puzzle. However, in the recent model of quasi-spherical settling accretion in wind-fed high-mass X-ray binaries by \citet{shakura12}, the equilibrium period can be of $\sim$1000\,s even for `ordinary' magnetic fields of $\sim$$10^{12}$--$10^{13}$\,G and acceptable accretion rates. The model has been applied successfully in populations studies and to model samples and singles sources \citep[e.g.][]{chashkina12,li16}, including the slowest known accreting pulsar, in the high-mass X-ray binary system AX\,J1910.7+0917 \citep[spin period of 36.2\,ks;][]{sidoli17}. 

There is however an exception, the Be X-ray binary  SXP\,1062 in the Small Magellanic Cloud, which has a spin period of 1062\,s and is robustly associated to a supernova remnant with kinematic age of $(2$--$4)\times10^4$\,yr (\citealt{henault12}; \citealt{haberl12} propose an even younger age using a temperature--size relationship).
In fact, for typical values of magnetic field and accretion rate, it would have been impossible for the neutron star to enter the propeller stage and become an accretor within the time constraint of a few $\times$$10^4$\,yr dictated by the age of the supernova remnant. After considering several possible scenarios, \citet{popov12} proposed a neutron star born with an initial magnetic field of $\gtrsim$$10^{14}$\,G that then decayed to a present value of $\sim$$10^{13}$\,G (derived in the assumption that the star is rotating close to the equilibrium period). 

Magnetars are increasingly popular also in the field of ultraluminous X-ray sources (ULXs). These sources---a few hundreds of them are known---are observed in off-nucleus regions of nearby galaxies at X-ray luminosities exceeding a few $10^{39}$\,erg s$^{-1}$ \citep{kaaret17}. Since this threshold for isotropic luminosity is larger than the Eddington limit for spherical accretion of fully ionized hydrogen onto a $\sim$10-M$_\odot$ compact object (a scale value of the black holes of stellar origin observed in our Galaxy), ULXs were considered the observational manifestation of massive black holes of stellar origin ($\lesssim$80--100\,M$_\odot$) and, the brightest ones in particular, promising candidates of intermediate-mass black holes of $10^3$--$10^5$\,M$_\odot$ \citep{miller04}. For these reason, the recent discovery in three ULXs (M82\,X--2, NGC\,5907~ULX, and NGC\,7793~P13) of pulsars with spin periods from 0.4 to 1.4\,s has been a blow, showing both that some ULXs (even in the high side of their luminosity distribution) may host neutron stars and that neutron stars can achieve extreme supper-Eddington luminosities \citep{bachetti14,israel17,ipe17}. The most luminous of the bunch, NGC\,5907~ULX, which has a period of $\sim$1.1\,s, was observed at a maximum X-ray luminosity of $\sim$$10^{41}$\,erg s$^{-1}$, more than 500 times the Eddington limit for a 1.4-M$_\odot$ neutron star \citep{israel17}. In principle, a neutron star with a magnetic field of $\gtrsim$$10^{15}$\,G could attain such a super-Eddington luminosity \citep{dallosso15,mushtukov15}, since the magnetic field reduces the electron scattering cross section (see also Sect.\,\ref{giantflares}). However, this explanation is not viable in the cases of NGC\,5907~ULX and NGC\,7793~P13, because such a huge magnetic field coupled with the rapid spinning of the star would inhibit the accretion via the propeller mechanisms. A possible solution proposed by \citet{israel17} is that the magnetic field is indeed of a few $10^{14}$\,G at the neutron star surface, but is dominated ($\sim$90\%) by multipolar components, which vanish rapidly with the distance, so that at the magnetospheric radius ($\approx$100 neutron star radii) the magnetic field is virtually the dipolar one, low enough for the accretion to proceed. It is interesting to note that this hypothesis does not require for the neutron star only a superstrong magnetic field (a purely dipolar magnetic field would not work), but a complex magnetic field configuration similar to that envisaged for magnetars.

Finally, there is a peculiar binary source in which the presence of an ultra-magnetized neutron star has been suggested not as a wildcard to explain its properties, but because it actually showed one of the specific characteristics of magnetars: LS\,I\,+61$^\circ$303. It is one of the few gamma-ray (TeV) binaries \citep{dubus13} and is in a orbit of 27 days with a 10--15-M$_\odot$ Be star. LS\,I\,+61$^\circ$303 is also a periodic (27\,d) and variable radio source (it is often referred to as a \emph{microquasar}) and, since no direct evidence for the presence of a neutron star has been obtained so far (for example, pulsations, mass limits or thermonuclear bursts), the nature of its compact object is still debated. In 2008, Swift triggered on a short, SGR-like burst from LS\,I\,+61$^\circ$303. With a duration of $\sim$0.2--0.3\,s, a blackbody spectrum with temperature of 7.5\,keV, and luminosity of $\sim$$2\times10^{37}$\,erg s$^{-1}$, the event had the characteristics of a magnetar burst \citep{torres12}. A second magnetar-like burst was detected again by Swift in 2012 \citep{burrows12}. \citet{torres12} discussed the implication of the presence of a magnetar in LS\,I\,+61$^\circ$303 and showed that it would be compatible with the properties of the source.

\section{Final remarks}

Until about a decade or little more ago, magnetars were regarded as a sort of astrophysical oddities and only comparatively few small groups of astronomers were interested in them. In recent times however, a number of surprising observational discoveries connected more strictly magnetars to the other classes of neutron stars and made them hard to be ignored by the large community studying pulsars at different wavelengths. In fact, it has been discovered that magnetars can be radio pulsars and also that `ordinary' X-ray and radio pulsars, as well as other sources, such as the peculiar neutron star in the supernova remnant RCW\,103, can behave like magnetars, showing the whole array of magnetar activity: bursts, outbursts, dramatic and abrupt pulse profile changes and other timing anomalies.
We have also learnt that magnetars can populate unexpected regions of the $P$--$\dot{P}$ diagram, with dipole magnetic fields measured from the rotation parameters that are comparable to or lower than those of the radio pulsars, disguising at the same time much stronger nondipolar magnetic field components. On the other hand, there are mounting evidences that magnetar (nondipolar) magnetic fields may be present also in other classes of isolated and binary neutron stars; in the future, they might manifest magnetar behaviour. Summarising, episodes of magnetar activity and their frequency seem to be related to the total magnetic energy stored in the internal field of a neutron star but, while a huge tank of this energy is typically associated to the pulsars in the up--right corner of the $P$--$\dot{P}$ diagram, the external dipolar magnetic field inferred from the spin period and the slow-down rate is not necessary a good proxy for it. For this reason, maybe it would be more appropriate to speak of magnetar-like activity (or `magnetic restlessness') in neutron stars rather than of magnetars.

Perhaps even more importantly, magnetars are proving to be key objects in understanding the puzzling observational diversity among the different classes of isolated neutron stars. After all, neutron stars are relatively simply, collapsed, objects, presumably all governed by the same equation of state: Why do they come in so many flavours? Theoretical progresses, chiefly about the complexity and the evolution of their magnetic field, are paving the way to a unifying view, in which the age, the different magnetic field strength and geometry at birth, and few other pivotal parameters, such as the mass and the chemical composition of the envelope, can explain the neutron star diversity.

Magnetars are being increasingly invoked in a variety of astrophysical sources and phenomena, from high-mass X-ray binaries and ULXs, to gamma-ray bursts, superluminous supernovae, fast radio bursts, sources of gravitational waves, and many others. While in some cases they are used in a lighthearted way as jacks-of-all-trades, because a huge magnetic field offers an easy way to solve an observational or theoretical problem, magnetar are finally receiving  the attention they deserve. This will trigger more and more observational and theoretical efforts which, together with new forthcoming powerful and innovative instruments, such as CTA, SKA, Athena+, X-ray polarimeters, space interferometers, and giant optical telescopes, are bound to deliver many important and surprising discoveries. Magnetars enthusiasts are well positioned to enjoy the next few decades.

\begin{acknowledgement}
PE and NR acknowledge funding in the framework of the NWO Vidi award A.2320.0076. 
\end{acknowledgement}

 \bibliographystyle{natbib}
 \bibliography{biblio}

\begin{thebibliography}{}

\bibitem[{Aguilera} {\em et~al.}(2008){Aguilera}, {Pons}, and
  {Miralles}]{aguilera08}
{Aguilera}, D.~N., {Pons}, J.~A., and {Miralles}, J.~A. (2008).
\newblock {2D Cooling of magnetized neutron stars}.
\newblock {\em \aap\/}, {\bf 486}, 255--271.

\bibitem[{Aleksi{\'c}} {\em et~al.}(2013){Aleksi{\'c}}, {Antonelli},
  {Antoranz}, {Asensio}, {Barres de Almeida}, {Barrio}, {Becerra Gonz{\'a}lez},
  {Bednarek}, {Berger}, {Bernardini}, {Biland}, {Blanch}, {Bock}, {Boller},
  {Bonnoli}, {Borla Tridon}, {Bretz}, {Carmona}, {Carosi}, {Colin}, {Colombo},
  {Contreras}, {Cortina}, {Cossio}, {Covino}, {Da Vela}, {Dazzi}, {De Angelis},
  {De Caneva}, {De Cea del Pozo}, {De Lotto}, {Delgado Mendez}, {Diago Ortega},
  {Doert}, {Dominis Prester}, {Dorner}, {Doro}, {Eisenacher}, {Elsaesser},
  {Ferenc}, {Fonseca}, {Font}, {Fruck}, {Garc{\'{\i}}a L{\'o}pez},
  {Garczarczyk}, {Garrido Terrats}, {Gaug}, {Giavitto}, {Godinovi{\'c}},
  {Gonz{\'a}lez Mu{\~n}oz}, {Gozzini}, {Hadamek}, {Hadasch}, {H{\"a}fner},
  {Herrero}, {Hose}, {Hrupec}, {Huber}, {Jankowski}, {Jogler}, {Kadenius},
  {Klepser}, {Knoetig}, {Kr{\"a}henb{\"u}hl}, {Krause}, {Kushida}, {La
  Barbera}, {Lelas}, {Leonardo}, {Lewandowska}, {Lindfors}, {Lombardi},
  {L{\'o}pez}, {L{\'o}pez-Coto}, {L{\'o}pez-Oramas}, {Lorenz}, {Makariev},
  {Maneva}, {Mankuzhiyil}, {Mannheim}, {Maraschi}, {Marcote}, {Mariotti},
  {Mart{\'{\i}}nez}, {Mazin}, {Meucci}, {Miranda}, {Mirzoyan}, {Mold{\'o}n},
  {Moralejo}, {Munar-Adrover}, {Niedzwiecki}, {Nieto}, {Nilsson}, {Nowak},
  {Orito}, {Paiano}, {Palatiello}, {Paneque}, {Paoletti}, {Paredes}, {Partini},
  {Persic}, {Pilia}, {Pochon}, {Prada}, {Prada Moroni}, {Prandini}, {Puljak},
  {Reichardt}, {Reinthal}, {Rhode}, {Rib{\'o}}, {Rico}, {R{\"u}gamer},
  {Saggion}, {Saito}, {Saito}, {Salvati}, {Satalecka}, {Scalzotto}, {Scapin},
  {Schultz}, {Schweizer}, {Shore}, {Sillanp{\"a}{\"a}}, {Sitarek}, {Snidaric},
  {Sobczynska}, {Spanier}, {Spiro}, {Stamatescu}, {Stamerra}, {Steinke},
  {Storz}, {Sun}, {Suri{\'c}}, {Takalo}, {Takami}, {Tavecchio}, {Temnikov},
  {Terzi{\'c}}, {Tescaro}, {Teshima}, {Tibolla}, {Torres}, {Toyama}, {Treves},
  {Uellenbeck}, {Vogler}, {Wagner}, {Weitzel}, {Zabalza}, {Zandanel}, {Zanin},
  {Rea}, and {Backes}]{aleksic13}
{Aleksi{\'c}}, J., {Antonelli}, L.~A., {Antoranz}, P., {Asensio}, M., {Barres
  de Almeida}, U., {Barrio}, J.~A., {Becerra Gonz{\'a}lez}, J., {Bednarek}, W.,
  {Berger}, K., {Bernardini}, E., {Biland}, A., {Blanch}, O., {Bock}, R.~K.,
  {Boller}, A., {Bonnoli}, G., {Borla Tridon}, D., {Bretz}, T., {Carmona}, E.,
  {Carosi}, A., {Colin}, P., {Colombo}, E., {Contreras}, J.~L., {Cortina}, J.,
  {Cossio}, L., {Covino}, S., {Da Vela}, P., {Dazzi}, F., {De Angelis}, A., {De
  Caneva}, G., {De Cea del Pozo}, E., {De Lotto}, B., {Delgado Mendez}, C.,
  {Diago Ortega}, A., {Doert}, M., {Dominis Prester}, D., {Dorner}, D., {Doro},
  M., {Eisenacher}, D., {Elsaesser}, D., {Ferenc}, D., {Fonseca}, M.~V.,
  {Font}, L., {Fruck}, C., {Garc{\'{\i}}a L{\'o}pez}, R.~J., {Garczarczyk}, M.,
  {Garrido Terrats}, D., {Gaug}, M., {Giavitto}, G., {Godinovi{\'c}}, N.,
  {Gonz{\'a}lez Mu{\~n}oz}, A., {Gozzini}, S.~R., {Hadamek}, A., {Hadasch}, D.,
  {H{\"a}fner}, D., {Herrero}, A., {Hose}, J., {Hrupec}, D., {Huber}, B.,
  {Jankowski}, F., {Jogler}, T., {Kadenius}, V., {Klepser}, S., {Knoetig},
  M.~L., {Kr{\"a}henb{\"u}hl}, T., {Krause}, J., {Kushida}, J., {La Barbera},
  A., {Lelas}, D., {Leonardo}, E., {Lewandowska}, N., {Lindfors}, E.,
  {Lombardi}, S., {L{\'o}pez}, M., {L{\'o}pez-Coto}, R., {L{\'o}pez-Oramas},
  A., {Lorenz}, E., {Makariev}, M., {Maneva}, G., {Mankuzhiyil}, N.,
  {Mannheim}, K., {Maraschi}, L., {Marcote}, B., {Mariotti}, M.,
  {Mart{\'{\i}}nez}, M., {Mazin}, D., {Meucci}, M., {Miranda}, J.~M.,
  {Mirzoyan}, R., {Mold{\'o}n}, J., {Moralejo}, A., {Munar-Adrover}, P.,
  {Niedzwiecki}, A., {Nieto}, D., {Nilsson}, K., {Nowak}, N., {Orito}, R.,
  {Paiano}, S., {Palatiello}, M., {Paneque}, D., {Paoletti}, R., {Paredes},
  J.~M., {Partini}, S., {Persic}, M., {Pilia}, M., {Pochon}, J., {Prada}, F.,
  {Prada Moroni}, P.~G., {Prandini}, E., {Puljak}, I., {Reichardt}, I.,
  {Reinthal}, R., {Rhode}, W., {Rib{\'o}}, M., {Rico}, J., {R{\"u}gamer}, S.,
  {Saggion}, A., {Saito}, K., {Saito}, T.~Y., {Salvati}, M., {Satalecka}, K.,
  {Scalzotto}, V., {Scapin}, V., {Schultz}, C., {Schweizer}, T., {Shore},
  S.~N., {Sillanp{\"a}{\"a}}, A., {Sitarek}, J., {Snidaric}, I., {Sobczynska},
  D., {Spanier}, F., {Spiro}, S., {Stamatescu}, V., {Stamerra}, A., {Steinke},
  B., {Storz}, J., {Sun}, S., {Suri{\'c}}, T., {Takalo}, L., {Takami}, H.,
  {Tavecchio}, F., {Temnikov}, P., {Terzi{\'c}}, T., {Tescaro}, D., {Teshima},
  M., {Tibolla}, O., {Torres}, D.~F., {Toyama}, T., {Treves}, A., {Uellenbeck},
  M., {Vogler}, P., {Wagner}, R.~M., {Weitzel}, Q., {Zabalza}, V., {Zandanel},
  F., {Zanin}, R., {Rea}, N., and {Backes}, M. (2013).
\newblock {Observations of the magnetars 4U 0142+61 and 1E 2259+586 with the
  MAGIC telescopes}.
\newblock {\em \aap\/}, {\bf 549}, A23.

\bibitem[{An} {\em et~al.}(2014){An}, {Kaspi}, {Archibald}, {Bachetti},
  {Bhalerao}, {Bellm}, {Beloborodov}, {Boggs}, {Chakrabarty}, {Christensen},
  {Craig}, {Dufour}, {Forster}, {Gotthelf}, {Grefenstette}, {Hailey},
  {Harrison}, {Hasco{\"e}t}, {Kitaguchi}, {Kouveliotou}, {Madsen}, {Mori},
  {Pivovaroff}, {Rana}, {Stern}, {Tendulkar}, {Tomsick}, {Vogel}, {Zhang}, and
  {NuSTAR Team}]{an14}
{An}, H., {Kaspi}, V.~M., {Archibald}, R., {Bachetti}, M., {Bhalerao}, V.,
  {Bellm}, E.~C., {Beloborodov}, A.~M., {Boggs}, S.~E., {Chakrabarty}, D.,
  {Christensen}, F.~E., {Craig}, W.~W., {Dufour}, F., {Forster}, K.,
  {Gotthelf}, E.~V., {Grefenstette}, B.~W., {Hailey}, C.~J., {Harrison}, F.~A.,
  {Hasco{\"e}t}, R., {Kitaguchi}, T., {Kouveliotou}, C., {Madsen}, K.~K.,
  {Mori}, K., {Pivovaroff}, M.~J., {Rana}, V.~R., {Stern}, D., {Tendulkar}, S.,
  {Tomsick}, J.~A., {Vogel}, J.~K., {Zhang}, W.~W., and {NuSTAR Team} (2014).
\newblock {NuSTAR results and future plans for magnetar and rotation-powered
  pulsar observations}.
\newblock {\em Astronomische Nachrichten\/}, {\bf 335}, 280--284.

\bibitem[{Anderson} {\em et~al.}(2012){Anderson}, {Gaensler}, {Slane}, {Rea},
  {Kaplan}, {Posselt}, {Levin}, {Johnston}, {Murray}, {Brogan}, {Bailes},
  {Bates}, {Benjamin}, {Bhat}, {Burgay}, {Burke-Spolaor}, {Chakrabarty},
  {D'Amico}, {Drake}, {Esposito}, {Grindlay}, {Hong}, {Israel}, {Keith},
  {Kramer}, {Lazio}, {Lee}, {Mauerhan}, {Milia}, {Possenti}, {Stappers}, and
  {Steeghs}]{anderson12}
{Anderson}, G.~E., {Gaensler}, B.~M., {Slane}, P.~O., {Rea}, N., {Kaplan},
  D.~L., {Posselt}, B., {Levin}, L., {Johnston}, S., {Murray}, S.~S., {Brogan},
  C.~L., {Bailes}, M., {Bates}, S., {Benjamin}, R.~A., {Bhat}, N.~D.~R.,
  {Burgay}, M., {Burke-Spolaor}, S., {Chakrabarty}, D., {D'Amico}, N., {Drake},
  J.~J., {Esposito}, P., {Grindlay}, J.~E., {Hong}, J., {Israel}, G.~L.,
  {Keith}, M.~J., {Kramer}, M., {Lazio}, T.~J.~W., {Lee}, J.~C., {Mauerhan},
  J.~C., {Milia}, S., {Possenti}, A., {Stappers}, B., and {Steeghs}, D.~T.~H.
  (2012).
\newblock {Multi-wavelength Observations of the Radio Magnetar PSR J1622-4950
  and Discovery of Its Possibly Associated Supernova Remnant}.
\newblock {\em \apj\/}, {\bf 751}, 53.

\bibitem[{Aptekar} {\em et~al.}(2001){Aptekar}, {Frederiks}, {Golenetskii},
  {Il'inskii}, {Mazets}, {Pal'shin}, {Butterworth}, and {Cline}]{aptekar01}
{Aptekar}, R.~L., {Frederiks}, D.~D., {Golenetskii}, S.~V., {Il'inskii}, V.~N.,
  {Mazets}, E.~P., {Pal'shin}, V.~D., {Butterworth}, P.~S., and {Cline}, T.~L.
  (2001).
\newblock {Konus Catalog of Soft Gamma Repeater Activity: 1978 to 2000}.
\newblock {\em \apjs\/}, {\bf 137}, 227--277.

\bibitem[{Archibald} {\em et~al.}(2013){Archibald}, {Kaspi}, {Ng},
  {Gourgouliatos}, {Tsang}, {Scholz}, {Beardmore}, {Gehrels}, and
  {Kennea}]{archibald13}
{Archibald}, R.~F., {Kaspi}, V.~M., {Ng}, C.-Y., {Gourgouliatos}, K.~N.,
  {Tsang}, D., {Scholz}, P., {Beardmore}, A.~P., {Gehrels}, N., and {Kennea},
  J.~A. (2013).
\newblock {An anti-glitch in a magnetar}.
\newblock {\em \nat\/}, {\bf 497}, 591--593.

\bibitem[{Archibald} {\em et~al.}(2016){Archibald}, {Kaspi}, {Tendulkar}, and
  {Scholz}]{archibald16}
{Archibald}, R.~F., {Kaspi}, V.~M., {Tendulkar}, S.~P., and {Scholz}, P.
  (2016).
\newblock {A Magnetar-like Outburst from a High-B Radio Pulsar}.
\newblock {\em \apjl\/}, {\bf 829}, L21.

\bibitem[{Archibald} {\em et~al.}(2017){Archibald}, {Burgay}, {Lyutikov},
  {Kaspi}, {Esposito}, {Israel}, {Kerr}, {Possenti}, {Rea}, {Sarkissian},
  {Scholz}, and {Tendulkar}]{archibald17}
{Archibald}, R.~F., {Burgay}, M., {Lyutikov}, M., {Kaspi}, V.~M., {Esposito},
  P., {Israel}, G., {Kerr}, M., {Possenti}, A., {Rea}, N., {Sarkissian}, J.,
  {Scholz}, P., and {Tendulkar}, S.~P. (2017).
\newblock {Magnetar-like X-Ray Bursts Suppress Pulsar Radio Emission}.
\newblock {\em \apjl\/}, {\bf 849}, L20.

\bibitem[Aschwanden(2013)Aschwanden]{aschwanden13}
Aschwanden, M., editor (2013).
\newblock {\em {Self-Organized Criticality Systems}\/}.
\newblock Open Academic Press, Berlin, Warsaw, 483pp.

\bibitem[{Bachetti} {\em et~al.}(2014){Bachetti}, {Harrison}, {Walton},
  {Grefenstette}, {Chakrabarty}, {F{\"u}rst}, {Barret}, {Beloborodov}, {Boggs},
  {Christensen}, {Craig}, {Fabian}, {Hailey}, {Hornschemeier}, {Kaspi},
  {Kulkarni}, {Maccarone}, {Miller}, {Rana}, {Stern}, {Tendulkar}, {Tomsick},
  {Webb}, and {Zhang}]{bachetti14}
{Bachetti}, M., {Harrison}, F.~A., {Walton}, D.~J., {Grefenstette}, B.~W.,
  {Chakrabarty}, D., {F{\"u}rst}, F., {Barret}, D., {Beloborodov}, A., {Boggs},
  S.~E., {Christensen}, F.~E., {Craig}, W.~W., {Fabian}, A.~C., {Hailey},
  C.~J., {Hornschemeier}, A., {Kaspi}, V., {Kulkarni}, S.~R., {Maccarone}, T.,
  {Miller}, J.~M., {Rana}, V., {Stern}, D., {Tendulkar}, S.~P., {Tomsick}, J.,
  {Webb}, N.~A., and {Zhang}, W.~W. (2014).
\newblock {An ultraluminous X-ray source powered by an accreting neutron star}.
\newblock {\em \nat\/}, {\bf 514}, 202--204.

\bibitem[{Bak} and {Chen}(1991){Bak} and {Chen}]{bak91}
{Bak}, P. and {Chen}, K. (1991).
\newblock {Self-Organized Criticality}.
\newblock {\em Scientific American\/}, {\bf 264}, 46--53.

\bibitem[{Barat} {\em et~al.}(1983){Barat}, {Hayles}, {Hurley}, {Niel},
  {Vedrenne}, {Desai}, {Kurt}, {Zenchenko}, and {Estulin}]{barat83}
{Barat}, C., {Hayles}, R.~I., {Hurley}, K., {Niel}, M., {Vedrenne}, G.,
  {Desai}, U., {Kurt}, V.~G., {Zenchenko}, V.~M., and {Estulin}, I.~V. (1983).
\newblock {Fine time structure in the 1979 March 5 gamma ray burst}.
\newblock {\em \aap\/}, {\bf 126}, 400--402.

\bibitem[{Baring} and {Harding}(2007){Baring} and {Harding}]{baring07}
{Baring}, M.~G. and {Harding}, A.~K. (2007).
\newblock {Resonant Compton upscattering in anomalous X-ray pulsars}.
\newblock {\em \apss\/}, {\bf 308}, 109--118.

\bibitem[{Baym} {\em et~al.}(1969){Baym}, {Pethick}, and {Pines}]{baym69}
{Baym}, G., {Pethick}, C., and {Pines}, D. (1969).
\newblock {Superfluidity in Neutron Stars}.
\newblock {\em \nat\/}, {\bf 224}, 673--674.

\bibitem[{Beloborodov}(2013a){Beloborodov}]{b13}
{Beloborodov}, A.~M. (2013a).
\newblock {Electron-Positron Flows around Magnetars}.
\newblock {\em \apj\/}, {\bf 777}, 114.

\bibitem[{Beloborodov}(2013b){Beloborodov}]{beloborodov13}
{Beloborodov}, A.~M. (2013b).
\newblock {On the Mechanism of Hard X-Ray Emission from Magnetars}.
\newblock {\em \apj\/}, {\bf 762}, 13.

\bibitem[{Borghese} {\em et~al.}(2015){Borghese}, {Rea}, {Coti Zelati},
  {Tiengo}, and {Turolla}]{borghese15}
{Borghese}, A., {Rea}, N., {Coti Zelati}, F., {Tiengo}, A., and {Turolla}, R.
  (2015).
\newblock {Discovery of a Strongly Phase-variable Spectral Feature in the
  Isolated Neutron Star RX J0720.4-3125}.
\newblock {\em \apjl\/}, {\bf 807}, L20.

\bibitem[{Borghese} {\em et~al.}(2017){Borghese}, {Rea}, {Coti Zelati},
  {Tiengo}, {Turolla}, and {Zane}]{borghese17}
{Borghese}, A., {Rea}, N., {Coti Zelati}, F., {Tiengo}, A., {Turolla}, R., and
  {Zane}, S. (2017).
\newblock {Narrow phase-dependent features in X-ray dim isolated neutron stars:
  a new detection and upper limits}.
\newblock {\em \mnras\/}, {\bf 468}, 2975--2983.

\bibitem[{Bower} {\em et~al.}(2015){Bower}, {Deller}, {Demorest}, {Brunthaler},
  {Falcke}, {Moscibrodzka}, {O'Leary}, {Eatough}, {Kramer}, {Lee}, {Spitler},
  {Desvignes}, {Rushton}, {Doeleman}, and {Reid}]{bower15}
{Bower}, G.~C., {Deller}, A., {Demorest}, P., {Brunthaler}, A., {Falcke}, H.,
  {Moscibrodzka}, M., {O'Leary}, R.~M., {Eatough}, R.~P., {Kramer}, M., {Lee},
  K.~J., {Spitler}, L., {Desvignes}, G., {Rushton}, A.~P., {Doeleman}, S., and
  {Reid}, M.~J. (2015).
\newblock {The Proper Motion of the Galactic Center Pulsar Relative to
  Sagittarius A*}.
\newblock {\em \apj\/}, {\bf 798}, 120.

\bibitem[{Bozzo} {\em et~al.}(2008){Bozzo}, {Falanga}, and {Stella}]{bozzo08}
{Bozzo}, E., {Falanga}, M., and {Stella}, L. (2008).
\newblock {Are There Magnetars in High-Mass X-Ray Binaries? The Case of
  Supergiant Fast X-Ray Transients}.
\newblock {\em \apj\/}, {\bf 683}, 1031--1044.

\bibitem[{Burgay} {\em et~al.}(2009){Burgay}, {Israel}, {Possenti}, {Rea},
  {Esposito}, {Mereghetti}, {Tiengo}, and {G\"otz}]{burgay09}
{Burgay}, M., {Israel}, G.~L., {Possenti}, A., {Rea}, N., {Esposito}, P.,
  {Mereghetti}, S., {Tiengo}, A., and {G\"otz}, D. (2009).
\newblock {Back to radio: Parkes detection of radio pulses from the transient
  AXP 1E1547.0-5408}.
\newblock {\em Astron. Tel.}, {\bf 1913}.

\bibitem[{Burgay} {\em et~al.}(2016){Burgay}, {Possenti}, {Kerr}, {Esposito},
  {Rea}, {Zelati}, {Israel}, and {Johnston}]{burgay16}
{Burgay}, M., {Possenti}, A., {Kerr}, M., {Esposito}, P., {Rea}, N., {Zelati},
  F.~C., {Israel}, G.~L., and {Johnston}, S. (2016).
\newblock {Pulsed Radio Emission from PSR J1119-6127 disappeared}.
\newblock {\em Astron. Tel.}, {\bf 9286}.

\bibitem[{Burrows} {\em et~al.}(2012){Burrows}, {Chester}, {D'Elia}, {Palmer},
  {Romano}, {Saxton}, {Sonbas}, {Stamatikos}, and {Stratta}]{burrows12}
{Burrows}, D.~N., {Chester}, M.~M., {D'Elia}, V., {Palmer}, D.~M., {Romano},
  P., {Saxton}, C.~J., {Sonbas}, E., {Stamatikos}, M., and {Stratta}, G.
  (2012).
\newblock {Swift detection of a burst from LS I +61$^\circ$ 303.}
\newblock {\em GCN Circ.}, {\bf 12914}.

\bibitem[{Camilo} {\em et~al.}(2006){Camilo}, {Ransom}, {Halpern}, {Reynolds},
  {Helfand}, {Zimmerman}, and {Sarkissian}]{camilo06}
{Camilo}, F., {Ransom}, S.~M., {Halpern}, J.~P., {Reynolds}, J., {Helfand},
  D.~J., {Zimmerman}, N., and {Sarkissian}, J. (2006).
\newblock {Transient pulsed radio emission from a magnetar}.
\newblock {\em \nat\/}, {\bf 442}, 892--895.

\bibitem[{Camilo} {\em et~al.}(2007a){Camilo}, {Ransom}, {Halpern}, and
  {Reynolds}]{camilo07}
{Camilo}, F., {Ransom}, S.~M., {Halpern}, J.~P., and {Reynolds}, J. (2007a).
\newblock {1E 1547.0-5408: A Radio-emitting Magnetar with a Rotation Period of
  2 Seconds}.
\newblock {\em \apjl\/}, {\bf 666}, L93--L96.

\bibitem[{Camilo} {\em et~al.}(2007b){Camilo}, {Cognard}, {Ransom}, {Halpern},
  {Reynolds}, {Zimmerman}, {Gotthelf}, {Helfand}, {Demorest}, {Theureau}, and
  {Backer}]{ccr07}
{Camilo}, F., {Cognard}, I., {Ransom}, S.~M., {Halpern}, J.~P., {Reynolds}, J.,
  {Zimmerman}, N., {Gotthelf}, E.~V., {Helfand}, D.~J., {Demorest}, P.,
  {Theureau}, G., and {Backer}, D.~C. (2007b).
\newblock {The Magnetar XTE J1810-197: Variations in Torque, Radio Flux
  Density, and Pulse Profile Morphology}.
\newblock {\em \apj\/}, {\bf 663}, 497--504.

\bibitem[{Camilo} {\em et~al.}(2008){Camilo}, {Reynolds}, {Johnston},
  {Halpern}, and {Ransom}]{camilo08}
{Camilo}, F., {Reynolds}, J., {Johnston}, S., {Halpern}, J.~P., and {Ransom},
  S.~M. (2008).
\newblock {The Magnetar 1E 1547.0-5408: Radio Spectrum, Polarimetry, and
  Timing}.
\newblock {\em \apj\/}, {\bf 679}, 681--686.

\bibitem[{Camilo} {\em et~al.}(2016){Camilo}, {Ransom}, {Halpern}, {Alford},
  {Cognard}, {Reynolds}, {Johnston}, {Sarkissian}, and {van Straten}]{camilo16}
{Camilo}, F., {Ransom}, S.~M., {Halpern}, J.~P., {Alford}, J.~A.~J., {Cognard},
  I., {Reynolds}, J.~E., {Johnston}, S., {Sarkissian}, J., and {van Straten},
  W. (2016).
\newblock {Radio Disappearance of the Magnetar XTE J1810-197 and Continued
  X-ray Timing}.
\newblock {\em \apj\/}, {\bf 820}, 110.

\bibitem[{Cavallo} and {Rees}(1978){Cavallo} and {Rees}]{cavallo78}
{Cavallo}, G. and {Rees}, M.~J. (1978).
\newblock {A qualitative study of cosmic fireballs and gamma-ray bursts}.
\newblock {\em \mnras\/}, {\bf 183}, 359--365.

\bibitem[{Chashkina} and {Popov}(2012){Chashkina} and {Popov}]{chashkina12}
{Chashkina}, A. and {Popov}, S.~B. (2012).
\newblock {Magnetic field estimates for accreting neutron stars in massive
  binary systems and models of magnetic field decay}.
\newblock {\em \na\/}, {\bf 17}, 594--602.

\bibitem[{Chatterjee} {\em et~al.}(2000){Chatterjee}, {Hernquist}, and
  {Narayan}]{chatterjee00}
{Chatterjee}, P., {Hernquist}, L., and {Narayan}, R. (2000).
\newblock {An Accretion Model for Anomalous X-Ray Pulsars}.
\newblock {\em \apj\/}, {\bf 534}, 373--379.

\bibitem[{Clark} {\em et~al.}(2008){Clark}, {Muno}, {Negueruela}, {Dougherty},
  {Crowther}, {Goodwin}, and {de Grijs}]{clark08}
{Clark}, J.~S., {Muno}, M.~P., {Negueruela}, I., {Dougherty}, S.~M.,
  {Crowther}, P.~A., {Goodwin}, S.~P., and {de Grijs}, R. (2008).
\newblock {Unveiling the X-ray point source population of the Young Massive
  Cluster Westerlund 1}.
\newblock {\em \aap\/}, {\bf 477}, 147--163.

\bibitem[{Clark} {\em et~al.}(2014){Clark}, {Ritchie}, {Najarro}, {Langer}, and
  {Negueruela}]{clark14}
{Clark}, J.~S., {Ritchie}, B.~W., {Najarro}, F., {Langer}, N., and
  {Negueruela}, I. (2014).
\newblock {A VLT/FLAMES survey for massive binaries in Westerlund 1. IV. Wd1-5
  - binary product and a pre-supernova companion for the magnetar CXOU
  J1647-45?}
\newblock {\em \aap\/}, {\bf 565}, A90.

\bibitem[{Cline} {\em et~al.}(1980){Cline}, {Desai}, {Pizzichini}, {Teegarden},
  {Evans}, {Klebesadel}, {Laros}, {Hurley}, {Niel}, and {Vedrenne}]{cline80}
{Cline}, T.~L., {Desai}, U.~D., {Pizzichini}, G., {Teegarden}, B.~J., {Evans},
  W.~D., {Klebesadel}, R.~W., {Laros}, J.~G., {Hurley}, K., {Niel}, M., and
  {Vedrenne}, G. (1980).
\newblock {Detection of a fast, intense and unusual gamma-ray transient}.
\newblock {\em \apjl\/}, {\bf 237}, L1--L5.

\bibitem[{Coti Zelati} {\em et~al.}(2015){Coti Zelati}, {Rea}, {Papitto},
  {Vigan{\`o}}, {Pons}, {Turolla}, {Esposito}, {Haggard}, {Baganoff}, {Ponti},
  {Israel}, {Campana}, {Torres}, {Tiengo}, {Mereghetti}, {Perna}, {Zane},
  {Mignani}, {Possenti}, and {Stella}]{cotizelati15}
{Coti Zelati}, F., {Rea}, N., {Papitto}, A., {Vigan{\`o}}, D., {Pons}, J.~A.,
  {Turolla}, R., {Esposito}, P., {Haggard}, D., {Baganoff}, F.~K., {Ponti}, G.,
  {Israel}, G.~L., {Campana}, S., {Torres}, D.~F., {Tiengo}, A., {Mereghetti},
  S., {Perna}, R., {Zane}, S., {Mignani}, R.~P., {Possenti}, A., and {Stella},
  L. (2015).
\newblock {The X-ray outburst of the Galactic Centre magnetar SGR J1745-2900
  during the first 1.5 year}.
\newblock {\em \mnras\/}, {\bf 449}, 2685--2699.

\bibitem[{Coti Zelati} {\em et~al.}(2017){Coti Zelati}, {Rea}, {Turolla},
  {Pons}, {Papitto}, {Esposito}, {Israel}, {Campana}, {Zane}, {Tiengo},
  {Mignani}, {Mereghetti}, {Baganoff}, {Haggard}, {Ponti}, {Torres},
  {Borghese}, and {Elfritz}]{cotizelati17}
{Coti Zelati}, F., {Rea}, N., {Turolla}, R., {Pons}, J.~A., {Papitto}, A.,
  {Esposito}, P., {Israel}, G.~L., {Campana}, S., {Zane}, S., {Tiengo}, A.,
  {Mignani}, R.~P., {Mereghetti}, S., {Baganoff}, F.~K., {Haggard}, D.,
  {Ponti}, G., {Torres}, D.~F., {Borghese}, A., and {Elfritz}, J. (2017).
\newblock {Chandra monitoring of the Galactic Centre magnetar SGR J1745-2900
  during the initial 3.5 years of outburst decay}.
\newblock {\em \mnras\/}, {\bf 471}, 1819--1829.

\bibitem[{Coti Zelati} {\em et~al.}(2018){Coti Zelati}, {Rea}, {Pons},
  {Campana}, and {Esposito}]{cotizelati18}
{Coti Zelati}, F., {Rea}, N., {Pons}, J.~A., {Campana}, S., and {Esposito}, P.
  (2018).
\newblock {Systematic study of magnetar outbursts}.
\newblock {\em \mnras\/}, {\bf 474}, 961--1017.

\bibitem[{D'A{\`i}} {\em et~al.}(2016){D'A{\`i}}, {Evans}, {Burrows}, {Kuin},
  {Kann}, {Campana}, {Maselli}, {Romano}, {Cusumano}, {La Parola}, {Barthelmy},
  {Beardmore}, {Cenko}, {De Pasquale}, {Gehrels}, {Greiner}, {Kennea}, {Klose},
  {Melandri}, {Nousek}, {Osborne}, {Palmer}, {Sbarufatti}, {Schady}, {Siegel},
  {Tagliaferri}, {Yates}, and {Zane}]{dai16}
{D'A{\`i}}, A., {Evans}, P.~A., {Burrows}, D.~N., {Kuin}, N.~P.~M., {Kann},
  D.~A., {Campana}, S., {Maselli}, A., {Romano}, P., {Cusumano}, G., {La
  Parola}, V., {Barthelmy}, S.~D., {Beardmore}, A.~P., {Cenko}, S.~B., {De
  Pasquale}, M., {Gehrels}, N., {Greiner}, J., {Kennea}, J.~A., {Klose}, S.,
  {Melandri}, A., {Nousek}, J.~A., {Osborne}, J.~P., {Palmer}, D.~M.,
  {Sbarufatti}, B., {Schady}, P., {Siegel}, M.~H., {Tagliaferri}, G., {Yates},
  R., and {Zane}, S. (2016).
\newblock {Evidence for the magnetar nature of 1E 161348-5055 in RCW 103}.
\newblock {\em \mnras\/}, {\bf 463}, 2394--2404.

\bibitem[{Dall'Osso} {\em et~al.}(2009){Dall'Osso}, {Shore}, and
  {Stella}]{dallosso09}
{Dall'Osso}, S., {Shore}, S.~N., and {Stella}, L. (2009).
\newblock {Early evolution of newly born magnetars with a strong toroidal
  field}.
\newblock {\em \mnras\/}, {\bf 398}, 1869--1885.

\bibitem[{Dall'Osso} {\em et~al.}(2015){Dall'Osso}, {Perna}, and
  {Stella}]{dallosso15}
{Dall'Osso}, S., {Perna}, R., and {Stella}, L. (2015).
\newblock {NuSTAR J095551+6940.8: a highly magnetized neutron star with
  super-Eddington mass accretion}.
\newblock {\em \mnras\/}, {\bf 449}, 2144--2150.

\bibitem[{Davies} {\em et~al.}(2009){Davies}, {Figer}, {Kudritzki}, {Trombley},
  {Kouveliotou}, and {Wachter}]{davies09}
{Davies}, B., {Figer}, D.~F., {Kudritzki}, R.-P., {Trombley}, C.,
  {Kouveliotou}, C., and {Wachter}, S. (2009).
\newblock {The Progenitor Mass of the Magnetar SGR1900+14}.
\newblock {\em \apj\/}, {\bf 707}, 844--851.

\bibitem[{Davies} and {Pringle}(1981){Davies} and {Pringle}]{davies81}
{Davies}, R.~E. and {Pringle}, J.~E. (1981).
\newblock {Spindown of neutron stars in close binary systems. II}.
\newblock {\em \mnras\/}, {\bf 196}, 209--224.

\bibitem[{De Luca}(2017){De Luca}]{deluca17}
{De Luca}, A. (2017).
\newblock Central compact objects in supernova remnants.
\newblock In G.~G. {Pavlov}, J.~A. {Pons}, P.~S. {Shternin}, and D.~G.
  {Yakovlev}, editors, {\em International Conference Physics of Neutron Stars -
  2017. 50 years after, St. Petersburg, Russian Federation\/}, volume 932, page
  012006. Journal of Physics Conference Series.

\bibitem[{De Luca} {\em et~al.}(2006){De Luca}, {Caraveo}, {Mereghetti},
  {Tiengo}, and {Bignami}]{deluca06}
{De Luca}, A., {Caraveo}, P.~A., {Mereghetti}, S., {Tiengo}, A., and {Bignami},
  G.~F. (2006).
\newblock {A Long-Period, Violently Variable X-ray Source in a Young Supernova
  Remnant}.
\newblock {\em Science\/}, {\bf 313}, 814--817.

\bibitem[{Degenaar} {\em et~al.}(2013){Degenaar}, {Reynolds}, {Miller},
  {Kennea}, and {Wijnands}]{drm13}
{Degenaar}, N., {Reynolds}, M.~T., {Miller}, J.~M., {Kennea}, J.~A., and
  {Wijnands}, R. (2013).
\newblock {Large Flare from Sgr A* Detected by Swift}.
\newblock {\em The Astronomer's Telegram\/}, {\bf 5006}.

\bibitem[{Deller} {\em et~al.}(2012){Deller}, {Camilo}, {Reynolds}, and
  {Halpern}]{deller12}
{Deller}, A.~T., {Camilo}, F., {Reynolds}, J.~E., and {Halpern}, J.~P. (2012).
\newblock {The Proper Motion of PSR J1550-5418 Measured with VLBI: A Second
  Magnetar Velocity Measurement}.
\newblock {\em \apjl\/}, {\bf 748}, L1.

\bibitem[{den Hartog} {\em et~al.}(2008a){den Hartog}, {Kuiper}, and
  {Hermsen}]{dkh08}
{den Hartog}, P.~R., {Kuiper}, L., and {Hermsen}, W. (2008a).
\newblock {Detailed high-energy characteristics of AXP 1RXS J170849-400910.
  Probing the magnetosphere using INTEGRAL, RXTE, and XMM-Newton}.
\newblock {\em \aap\/}, {\bf 489}, 263--279.

\bibitem[{den Hartog} {\em et~al.}(2008b){den Hartog}, {Kuiper}, {Hermsen},
  {Kaspi}, {Dib}, {Kn{\"o}dlseder}, and {Gavriil}]{denhartog08}
{den Hartog}, P.~R., {Kuiper}, L., {Hermsen}, W., {Kaspi}, V.~M., {Dib}, R.,
  {Kn{\"o}dlseder}, J., and {Gavriil}, F.~P. (2008b).
\newblock {Detailed high-energy characteristics of AXP 4U 0142+61. Multi-year
  observations with INTEGRAL, RXTE, XMM-Newton, and ASCA}.
\newblock {\em \aap\/}, {\bf 489}, 245--261.

\bibitem[{Dhillon} {\em et~al.}(2005){Dhillon}, {Marsh}, {Hulleman}, {van
  Kerkwijk}, {Shearer}, {Littlefair}, {Gavriil}, and {Kaspi}]{dhillon05}
{Dhillon}, V.~S., {Marsh}, T.~R., {Hulleman}, F., {van Kerkwijk}, M.~H.,
  {Shearer}, A., {Littlefair}, S.~P., {Gavriil}, F.~P., and {Kaspi}, V.~M.
  (2005).
\newblock {High-speed, multicolour optical photometry of the anomalous X-ray
  pulsar 4U 0142+61 with ULTRACAM}.
\newblock {\em \mnras\/}, {\bf 363}, 609--614.

\bibitem[{Dhillon} {\em et~al.}(2009){Dhillon}, {Marsh}, {Littlefair},
  {Copperwheat}, {Kerry}, {Dib}, {Durant}, {Kaspi}, {Mignani}, and
  {Shearer}]{dhillon09}
{Dhillon}, V.~S., {Marsh}, T.~R., {Littlefair}, S.~P., {Copperwheat}, C.~M.,
  {Kerry}, P., {Dib}, R., {Durant}, M., {Kaspi}, V.~M., {Mignani}, R.~P., and
  {Shearer}, A. (2009).
\newblock {Optical pulsations from the anomalous X-ray pulsar 1E1048.1-5937}.
\newblock {\em \mnras\/}, {\bf 394}, L112--L116.

\bibitem[{Dhillon} {\em et~al.}(2011){Dhillon}, {Marsh}, {Littlefair},
  {Copperwheat}, {Hickman}, {Kerry}, {Levan}, {Rea}, {Savoury}, {Tanvir},
  {Turolla}, and {Wiersema}]{dhillon11}
{Dhillon}, V.~S., {Marsh}, T.~R., {Littlefair}, S.~P., {Copperwheat}, C.~M.,
  {Hickman}, R.~D.~G., {Kerry}, P., {Levan}, A.~J., {Rea}, N., {Savoury},
  C.~D.~J., {Tanvir}, N.~R., {Turolla}, R., and {Wiersema}, K. (2011).
\newblock {The first observation of optical pulsations from a soft gamma
  repeater: SGR 0501+4516}.
\newblock {\em \mnras\/}, {\bf 416}, L16--L20.

\bibitem[{Dib} and {Kaspi}(2014){Dib} and {Kaspi}]{dib14}
{Dib}, R. and {Kaspi}, V.~M. (2014).
\newblock {16 yr of RXTE Monitoring of Five Anomalous X-Ray Pulsars}.
\newblock {\em \apj\/}, {\bf 784}, 37.

\bibitem[{Dib} {\em et~al.}(2008){Dib}, {Kaspi}, and {Gavriil}]{dib08}
{Dib}, R., {Kaspi}, V.~M., and {Gavriil}, F.~P. (2008).
\newblock {Glitches in Anomalous X-Ray Pulsars}.
\newblock {\em \apj\/}, {\bf 673}, 1044--1061.

\bibitem[{Dubus}(2013){Dubus}]{dubus13}
{Dubus}, G. (2013).
\newblock {Gamma-ray binaries and related systems}.
\newblock {\em \aapr\/}, {\bf 21}, 64.

\bibitem[{Duncan}(1998){Duncan}]{duncan98}
{Duncan}, R.~C. (1998).
\newblock {Global Seismic Oscillations in Soft Gamma Repeaters}.
\newblock {\em \apjl\/}, {\bf 498}, L45--L49.

\bibitem[{Duncan} and {Thompson}(1992){Duncan} and {Thompson}]{duncan92}
{Duncan}, R.~C. and {Thompson}, C. (1992).
\newblock {Formation of very strongly magnetized neutron stars - Implications
  for gamma-ray bursts}.
\newblock {\em \apjl\/}, {\bf 392}, L9--L13.

\bibitem[{Durant} and {van Kerkwijk}(2005){Durant} and {van Kerkwijk}]{dk05}
{Durant}, M. and {van Kerkwijk}, M.~H. (2005).
\newblock {The Broadband Spectrum and Infrared Variability of the Magnetar AXP
  1E 1048.1-5937}.
\newblock {\em \apj\/}, {\bf 627}, 376--382.

\bibitem[{Eatough} {\em et~al.}(2013){Eatough}, {Falcke}, {Karuppusamy}, {Lee},
  {Champion}, {Keane}, {Desvignes}, {Schnitzeler}, {Spitler}, {Kramer},
  {Klein}, {Bassa}, {Bower}, {Brunthaler}, {Cognard}, {Deller}, {Demorest},
  {Freire}, {Kraus}, {Lyne}, {Noutsos}, {Stappers}, and {Wex}]{eatough13}
{Eatough}, R.~P., {Falcke}, H., {Karuppusamy}, R., {Lee}, K.~J., {Champion},
  D.~J., {Keane}, E.~F., {Desvignes}, G., {Schnitzeler}, D.~H.~F.~M.,
  {Spitler}, L.~G., {Kramer}, M., {Klein}, B., {Bassa}, C., {Bower}, G.~C.,
  {Brunthaler}, A., {Cognard}, I., {Deller}, A.~T., {Demorest}, P.~B.,
  {Freire}, P.~C.~C., {Kraus}, A., {Lyne}, A.~G., {Noutsos}, A., {Stappers},
  B., and {Wex}, N. (2013).
\newblock {A strong magnetic field around the supermassive black hole at the
  centre of the Galaxy}.
\newblock {\em \nat\/}, {\bf 501}, 391--394.

\bibitem[{Elfritz} {\em et~al.}(2016){Elfritz}, {Pons}, {Rea}, {Glampedakis},
  and {Vigan{\`o}}]{elfritz16}
{Elfritz}, J.~G., {Pons}, J.~A., {Rea}, N., {Glampedakis}, K., and
  {Vigan{\`o}}, D. (2016).
\newblock {Simulated magnetic field expulsion in neutron star cores}.
\newblock {\em \mnras\/}, {\bf 456}, 4461--4474.

\bibitem[{Enoto} {\em et~al.}(2017){Enoto}, {Shibata}, {Kitaguchi}, {Suwa},
  {Uchide}, {Nishioka}, {Kisaka}, {Nakano}, {Murakami}, and
  {Makishima}]{enoto17}
{Enoto}, T., {Shibata}, S., {Kitaguchi}, T., {Suwa}, Y., {Uchide}, T.,
  {Nishioka}, H., {Kisaka}, S., {Nakano}, T., {Murakami}, H., and {Makishima},
  K. (2017).
\newblock {Magnetar Broadband X-Ray Spectra Correlated with Magnetic Fields:
  Suzaku Archive of SGRs and AXPs Combined with NuSTAR, Swift, and RXTE}.
\newblock {\em \apjs\/}, {\bf 231}, 8.

\bibitem[{Esposito} {\em et~al.}(2007){Esposito}, {Mereghetti}, {Tiengo},
  {Sidoli}, {Feroci}, and {Woods}]{esposito07}
{Esposito}, P., {Mereghetti}, S., {Tiengo}, A., {Sidoli}, L., {Feroci}, M., and
  {Woods}, P. (2007).
\newblock {Five years of SGR 1900+14 observations with BeppoSAX}.
\newblock {\em \aap\/}, {\bf 461}, 605--612.

\bibitem[{Esposito} {\em et~al.}(2008){Esposito}, {Israel}, {Zane}, {Senziani},
  {Starling}, {Rea}, {Palmer}, {Gehrels}, {Tiengo}, {De Luca}, {G{\"o}tz},
  {Mereghetti}, {Romano}, {Sakamoto}, {Barthelmy}, {Stella}, {Turolla},
  {Feroci}, and {Mangano}]{eiz08}
{Esposito}, P., {Israel}, G.~L., {Zane}, S., {Senziani}, F., {Starling},
  R.~L.~C., {Rea}, N., {Palmer}, D.~M., {Gehrels}, N., {Tiengo}, A., {De Luca},
  A., {G{\"o}tz}, D., {Mereghetti}, S., {Romano}, P., {Sakamoto}, T.,
  {Barthelmy}, S.~D., {Stella}, L., {Turolla}, R., {Feroci}, M., and {Mangano},
  V. (2008).
\newblock {The 2008 May burst activation of SGR1627-41}.
\newblock {\em \mnras\/}, {\bf 390}, L34--L38.

\bibitem[{Esposito} {\em et~al.}(2010){Esposito}, {Israel}, {Turolla},
  {Tiengo}, {G{\"o}tz}, {De Luca}, {Mignani}, {Zane}, {Rea}, {Testa},
  {Caraveo}, {Chaty}, {Mattana}, {Mereghetti}, {Pellizzoni}, and
  {Romano}]{esposito10}
{Esposito}, P., {Israel}, G.~L., {Turolla}, R., {Tiengo}, A., {G{\"o}tz}, D.,
  {De Luca}, A., {Mignani}, R.~P., {Zane}, S., {Rea}, N., {Testa}, V.,
  {Caraveo}, P.~A., {Chaty}, S., {Mattana}, F., {Mereghetti}, S., {Pellizzoni},
  A., and {Romano}, P. (2010).
\newblock {Early X-ray and optical observations of the soft gamma-ray repeater
  SGR 0418+5729}.
\newblock {\em \mnras\/}, {\bf 405}, 1787--1795.

\bibitem[{Esposito} {\em et~al.}(2011a){Esposito}, {Israel}, {Turolla},
  {Mattana}, {Tiengo}, {Possenti}, {Zane}, {Rea}, {Burgay}, {G{\"o}tz},
  {Mereghetti}, {Stella}, {Wieringa}, {Sarkissian}, {Enoto}, {Romano},
  {Sakamoto}, {Nakagawa}, {Makishima}, {Nakazawa}, {Nishioka}, and {Fran{\c
  c}ois-Martin}]{esposito11}
{Esposito}, P., {Israel}, G.~L., {Turolla}, R., {Mattana}, F., {Tiengo}, A.,
  {Possenti}, A., {Zane}, S., {Rea}, N., {Burgay}, M., {G{\"o}tz}, D.,
  {Mereghetti}, S., {Stella}, L., {Wieringa}, M.~H., {Sarkissian}, J.~M.,
  {Enoto}, T., {Romano}, P., {Sakamoto}, T., {Nakagawa}, Y.~E., {Makishima},
  K., {Nakazawa}, K., {Nishioka}, H., and {Fran{\c c}ois-Martin}, C. (2011a).
\newblock {Long-term spectral and timing properties of the soft gamma-ray
  repeater SGR 1833-0832 and detection of extended X-ray emission around the
  radio pulsar PSR B1830-08}.
\newblock {\em \mnras\/}, {\bf 416}, 205--215.

\bibitem[{Esposito} {\em et~al.}(2011b){Esposito}, {Turolla}, {De Luca},
  {Israel}, {Possenti}, and {Burrows}]{etdl11}
{Esposito}, P., {Turolla}, R., {De Luca}, A., {Israel}, G.~L., {Possenti}, A.,
  and {Burrows}, D.~N. (2011b).
\newblock {Swift monitoring of the central X-ray source in RCW 103}.
\newblock {\em \mnras\/}, {\bf 418}, 170--175.

\bibitem[{Fabian}(1979){Fabian}]{fabian79}
{Fabian}, A.~C. (1979).
\newblock {Theories of the nuclei of active galaxies}.
\newblock {\em Proceedings of the Royal Society of London Series A\/}, {\bf
  366}, 449--459.

\bibitem[{Fern{\'a}ndez} and {Thompson}(2007){Fern{\'a}ndez} and
  {Thompson}]{fernandez07}
{Fern{\'a}ndez}, R. and {Thompson}, C. (2007).
\newblock {Resonant Cyclotron Scattering in Three Dimensions and the Quiescent
  Nonthermal X-ray Emission of Magnetars}.
\newblock {\em \apj\/}, {\bf 660}, 615--640.

\bibitem[{Feroci} {\em et~al.}(2016){Feroci}, {Bozzo}, {Brandt}, {Hernanz},
  {van der Klis}, {Liu}, {Orleanski}, {Pohl}, {Santangelo}, {Schanne}, and
  et~al.]{feroci16}
{Feroci}, M., {Bozzo}, E., {Brandt}, S., {Hernanz}, M., {van der Klis}, M.,
  {Liu}, L.-P., {Orleanski}, P., {Pohl}, M., {Santangelo}, A., {Schanne}, S.,
  and et~al. (2016).
\newblock {The LOFT mission concept: a status update}.
\newblock In {\em Space Telescopes and Instrumentation 2016: Ultraviolet to
  Gamma Ray\/}, volume 9905 of {\em \procspie\/}, page 99051R.

\bibitem[{Ferrario} and {Wickramasinghe}(2006){Ferrario} and
  {Wickramasinghe}]{ferrario06}
{Ferrario}, L. and {Wickramasinghe}, D. (2006).
\newblock {Modelling of isolated radio pulsars and magnetars on the fossil
  field hypothesis}.
\newblock {\em \mnras\/}, {\bf 367}, 1323--1328.

\bibitem[{Ferrario} and {Wickramasinghe}(2008){Ferrario} and
  {Wickramasinghe}]{ferrario08}
{Ferrario}, L. and {Wickramasinghe}, D. (2008).
\newblock {Origin and evolution of magnetars}.
\newblock {\em \mnras\/}, {\bf 389}, L66--L70.

\bibitem[{Frail} {\em et~al.}(1999){Frail}, {Kulkarni}, and {Bloom}]{frail99}
{Frail}, D.~A., {Kulkarni}, S.~R., and {Bloom}, J.~S. (1999).
\newblock {An outburst of relativistic particles from the soft gamma-ray
  repeater SGR 1900+14.}
\newblock {\em \nat\/}, {\bf 398}, 127--129.

\bibitem[{Gaensler} {\em et~al.}(2005){Gaensler}, {Kouveliotou}, {Gelfand},
  {Taylor}, {Eichler}, {Wijers}, {Granot}, {Ramirez-Ruiz}, {Lyubarsky},
  {Hunstead}, {Campbell-Wilson}, {van der Horst}, {McLaughlin}, {Fender},
  {Garrett}, {Newton-McGee}, {Palmer}, {Gehrels}, and {Woods}]{gaensler05}
{Gaensler}, B.~M., {Kouveliotou}, C., {Gelfand}, J.~D., {Taylor}, G.~B.,
  {Eichler}, D., {Wijers}, R.~A.~M.~J., {Granot}, J., {Ramirez-Ruiz}, E.,
  {Lyubarsky}, Y.~E., {Hunstead}, R.~W., {Campbell-Wilson}, D., {van der
  Horst}, A.~J., {McLaughlin}, M.~A., {Fender}, R.~P., {Garrett}, M.~A.,
  {Newton-McGee}, K.~J., {Palmer}, D.~M., {Gehrels}, N., and {Woods}, P.~M.
  (2005).
\newblock {An expanding radio nebula produced by a giant flare from the
  magnetar SGR 1806-20}.
\newblock {\em \nat\/}, {\bf 434}, 1104--1106.

\bibitem[{Gavriil} {\em et~al.}(2002){Gavriil}, {Kaspi}, and
  {Woods}]{gavriil02}
{Gavriil}, F.~P., {Kaspi}, V.~M., and {Woods}, P.~M. (2002).
\newblock {Magnetar-like X-ray bursts from an anomalous X-ray pulsar}.
\newblock {\em \nat\/}, {\bf 419}, 142--144.

\bibitem[{Gavriil} {\em et~al.}(2008){Gavriil}, {Gonzalez}, {Gotthelf},
  {Kaspi}, {Livingstone}, and {Woods}]{gavriil08}
{Gavriil}, F.~P., {Gonzalez}, M.~E., {Gotthelf}, E.~V., {Kaspi}, V.~M.,
  {Livingstone}, M.~A., and {Woods}, P.~M. (2008).
\newblock {Magnetar-Like Emission from the Young Pulsar in Kes 75}.
\newblock {\em Science\/}, {\bf 319}, 1802--.

\bibitem[{Gehrels} {\em et~al.}(2004){Gehrels}, {Chincarini}, {Giommi},
  {Mason}, {Nousek}, {Wells}, {White}, {Barthelmy}, {Burrows}, {Cominsky},
  {Hurley}, {Marshall}, {M{\'e}sz{\'a}ros}, {Roming}, {Angelini}, {Barbier},
  {Belloni}, {Campana}, {Caraveo}, {Chester}, {Citterio}, {Cline}, {Cropper},
  {Cummings}, {Dean}, {Feigelson}, {Fenimore}, {Frail}, {Fruchter}, {Garmire},
  {Gendreau}, {Ghisellini}, {Greiner}, {Hill}, {Hunsberger}, {Krimm},
  {Kulkarni}, {Kumar}, {Lebrun}, {Lloyd-Ronning}, {Markwardt}, {Mattson},
  {Mushotzky}, {Norris}, {Osborne}, {Paczynski}, {Palmer}, {Park}, {Parsons},
  {Paul}, {Rees}, {Reynolds}, {Rhoads}, {Sasseen}, {Schaefer}, {Short},
  {Smale}, {Smith}, {Stella}, {Tagliaferri}, {Takahashi}, {Tashiro},
  {Townsley}, {Tueller}, {Turner}, {Vietri}, {Voges}, {Ward}, {Willingale},
  {Zerbi}, and {Zhang}]{gehrels04}
{Gehrels}, N., {Chincarini}, G., {Giommi}, P., {Mason}, K.~O., {Nousek}, J.~A.,
  {Wells}, A.~A., {White}, N.~E., {Barthelmy}, S.~D., {Burrows}, D.~N.,
  {Cominsky}, L.~R., {Hurley}, K.~C., {Marshall}, F.~E., {M{\'e}sz{\'a}ros},
  P., {Roming}, P.~W.~A., {Angelini}, L., {Barbier}, L.~M., {Belloni}, T.,
  {Campana}, S., {Caraveo}, P.~A., {Chester}, M.~M., {Citterio}, O., {Cline},
  T.~L., {Cropper}, M.~S., {Cummings}, J.~R., {Dean}, A.~J., {Feigelson},
  E.~D., {Fenimore}, E.~E., {Frail}, D.~A., {Fruchter}, A.~S., {Garmire},
  G.~P., {Gendreau}, K., {Ghisellini}, G., {Greiner}, J., {Hill}, J.~E.,
  {Hunsberger}, S.~D., {Krimm}, H.~A., {Kulkarni}, S.~R., {Kumar}, P.,
  {Lebrun}, F., {Lloyd-Ronning}, N.~M., {Markwardt}, C.~B., {Mattson}, B.~J.,
  {Mushotzky}, R.~F., {Norris}, J.~P., {Osborne}, J., {Paczynski}, B.,
  {Palmer}, D.~M., {Park}, H.-S., {Parsons}, A.~M., {Paul}, J., {Rees}, M.~J.,
  {Reynolds}, C.~S., {Rhoads}, J.~E., {Sasseen}, T.~P., {Schaefer}, B.~E.,
  {Short}, A.~T., {Smale}, A.~P., {Smith}, I.~A., {Stella}, L., {Tagliaferri},
  G., {Takahashi}, T., {Tashiro}, M., {Townsley}, L.~K., {Tueller}, J.,
  {Turner}, M.~J.~L., {Vietri}, M., {Voges}, W., {Ward}, M.~J., {Willingale},
  R., {Zerbi}, F.~M., and {Zhang}, W.~W. (2004).
\newblock {The Swift Gamma-Ray Burst Mission}.
\newblock {\em \apj\/}, {\bf 611}, 1005--1020.

\bibitem[{Gelfand} {\em et~al.}(2005){Gelfand}, {Lyubarsky}, {Eichler},
  {Gaensler}, {Taylor}, {Granot}, {Newton-McGee}, {Ramirez-Ruiz},
  {Kouveliotou}, and {Wijers}]{gelfand05}
{Gelfand}, J.~D., {Lyubarsky}, Y.~E., {Eichler}, D., {Gaensler}, B.~M.,
  {Taylor}, G.~B., {Granot}, J., {Newton-McGee}, K.~J., {Ramirez-Ruiz}, E.,
  {Kouveliotou}, C., and {Wijers}, R.~A.~M.~J. (2005).
\newblock {A Rebrightening of the Radio Nebula Associated with the 2004
  December 27 Giant Flare from SGR 1806-20}.
\newblock {\em \apjl\/}, {\bf 634}, L89--L92.

\bibitem[{Geppert} and {Rheinhardt}(2002){Geppert} and {Rheinhardt}]{geppert02}
{Geppert}, U. and {Rheinhardt}, M. (2002).
\newblock {Non-linear magnetic field decay in neutron stars. Theory and
  observations}.
\newblock {\em \aap\/}, {\bf 392}, 1015--1024.

\bibitem[{Geppert} and {Rheinhardt}(2006){Geppert} and {Rheinhardt}]{geppert06}
{Geppert}, U. and {Rheinhardt}, M. (2006).
\newblock {Magnetars versus radio pulsars. MHD stability in newborn highly
  magnetized neutron stars}.
\newblock {\em \aap\/}, {\bf 456}, 639--649.

\bibitem[{Gillessen} {\em et~al.}(2012){Gillessen}, {Genzel}, {Fritz},
  {Quataert}, {Alig}, {Burkert}, {Cuadra}, {Eisenhauer}, {Pfuhl}, {Dodds-Eden},
  {Gammie}, and {Ott}]{gillessen12}
{Gillessen}, S., {Genzel}, R., {Fritz}, T.~K., {Quataert}, E., {Alig}, C.,
  {Burkert}, A., {Cuadra}, J., {Eisenhauer}, F., {Pfuhl}, O., {Dodds-Eden}, K.,
  {Gammie}, C.~F., and {Ott}, T. (2012).
\newblock {A gas cloud on its way towards the supermassive black hole at the
  Galactic Centre}.
\newblock {\em \nat\/}, {\bf 481}, 51--54.

\bibitem[{Gillessen} {\em et~al.}(2013){Gillessen}, {Genzel}, {Fritz},
  {Eisenhauer}, {Pfuhl}, {Ott}, {Cuadra}, {Schartmann}, and
  {Burkert}]{gillessen13}
{Gillessen}, S., {Genzel}, R., {Fritz}, T.~K., {Eisenhauer}, F., {Pfuhl}, O.,
  {Ott}, T., {Cuadra}, J., {Schartmann}, M., and {Burkert}, A. (2013).
\newblock {New Observations of the Gas Cloud G2 in the Galactic Center}.
\newblock {\em \apj\/}, {\bf 763}, 78.

\bibitem[{Glampedakis} and {Jones}(2014){Glampedakis} and
  {Jones}]{glampedakis14}
{Glampedakis}, K. and {Jones}, D.~I. (2014).
\newblock {Three evolutionary paths for magnetar oscillations}.
\newblock {\em \mnras\/}, {\bf 439}, 1522--1535.

\bibitem[{Glampedakis} {\em et~al.}(2006){Glampedakis}, {Samuelsson}, and
  {Andersson}]{glampedakis06}
{Glampedakis}, K., {Samuelsson}, L., and {Andersson}, N. (2006).
\newblock {Elastic or magnetic? A toy model for global magnetar oscillations
  with implications for quasi-periodic oscillations during flares}.
\newblock {\em \mnras\/}, {\bf 371}, L74--L77.

\bibitem[{Goldreich} and {Reisenegger}(1992){Goldreich} and
  {Reisenegger}]{goldreich92}
{Goldreich}, P. and {Reisenegger}, A. (1992).
\newblock {Magnetic field decay in isolated neutron stars}.
\newblock {\em \apj\/}, {\bf 395}, 250--258.

\bibitem[{Gonzalez} and {Reisenegger}(2010){Gonzalez} and
  {Reisenegger}]{gonzalez10}
{Gonzalez}, D. and {Reisenegger}, A. (2010).
\newblock {Internal heating of old neutron stars: contrasting different
  mechanisms}.
\newblock {\em \aap\/}, {\bf 522}, A16.

\bibitem[{Gotthelf} {\em et~al.}(2013){Gotthelf}, {Halpern}, and
  {Alford}]{gotthelf13}
{Gotthelf}, E.~V., {Halpern}, J.~P., and {Alford}, J. (2013).
\newblock {The Spin-down of PSR J0821-4300 and PSR J1210-5226: Confirmation of
  Central Compact Objects as Anti-magnetars}.
\newblock {\em \apj\/}, {\bf 765}, 58.

\bibitem[{G{\"o}tz} {\em et~al.}(2004){G{\"o}tz}, {Mereghetti}, {Mirabel}, and
  {Hurley}]{gotz04}
{G{\"o}tz}, D., {Mereghetti}, S., {Mirabel}, I.~F., and {Hurley}, K. (2004).
\newblock {Spectral evolution of weak bursts from SGR 1806-20 observed with
  INTEGRAL}.
\newblock {\em \aap\/}, {\bf 417}, L45--L48.

\bibitem[{G{\"o}tz} {\em et~al.}(2006a){G{\"o}tz}, {Mereghetti}, {Tiengo}, and
  {Esposito}]{gotz06}
{G{\"o}tz}, D., {Mereghetti}, S., {Tiengo}, A., and {Esposito}, P. (2006a).
\newblock {Magnetars as persistent hard X-ray sources: INTEGRAL discovery of a
  hard tail in SGR 1900+14}.
\newblock {\em \aap\/}, {\bf 449}, L31--L34.

\bibitem[{G{\"o}tz} {\em et~al.}(2006b){G{\"o}tz}, {Mereghetti}, {Molkov},
  {Hurley}, {Mirabel}, {Sunyaev}, {Weidenspointner}, {Brandt}, {del Santo},
  {Feroci}, {G{\"o}{\u g}{\"u}{\c s}}, {von Kienlin}, {van der Klis},
  {Kouveliotou}, {Lund}, {Pizzichini}, {Ubertini}, {Winkler}, and
  {Woods}]{gmm06}
{G{\"o}tz}, D., {Mereghetti}, S., {Molkov}, S., {Hurley}, K., {Mirabel}, I.~F.,
  {Sunyaev}, R., {Weidenspointner}, G., {Brandt}, S., {del Santo}, M.,
  {Feroci}, M., {G{\"o}{\u g}{\"u}{\c s}}, E., {von Kienlin}, A., {van der
  Klis}, M., {Kouveliotou}, C., {Lund}, N., {Pizzichini}, G., {Ubertini}, P.,
  {Winkler}, C., and {Woods}, P.~M. (2006b).
\newblock {Two years of INTEGRAL monitoring of the soft gamma-ray repeater SGR
  1806-20: from quiescence to frenzy}.
\newblock {\em \aap\/}, {\bf 445}, 313--321.

\bibitem[{G{\"o}tz} {\em et~al.}(2007){G{\"o}tz}, {Rea}, {Israel}, {Zane},
  {Esposito}, {Gotthelf}, {Mereghetti}, {Tiengo}, and {Turolla}]{gri07}
{G{\"o}tz}, D., {Rea}, N., {Israel}, G.~L., {Zane}, S., {Esposito}, P.,
  {Gotthelf}, E.~V., {Mereghetti}, S., {Tiengo}, A., and {Turolla}, R. (2007).
\newblock {Long term hard X-ray variability of the anomalous X-ray pulsar 1RXS
  J170849.0-400910 discovered with INTEGRAL}.
\newblock {\em \aap\/}, {\bf 475}, 317--321.

\bibitem[{G{\"o}{\u g}{\"u}{\c s}} {\em et~al.}(2001){G{\"o}{\u g}{\"u}{\c s}},
  {Kouveliotou}, {Woods}, {Thompson}, {Duncan}, and {Briggs}]{gogus01}
{G{\"o}{\u g}{\"u}{\c s}}, E., {Kouveliotou}, C., {Woods}, P.~M., {Thompson},
  C., {Duncan}, R.~C., and {Briggs}, M.~S. (2001).
\newblock {Temporal and Spectral Characteristics of Short Bursts from the Soft
  Gamma Repeaters 1806-20 and 1900+14}.
\newblock {\em \apj\/}, {\bf 558}, 228--236.

\bibitem[{G{\"o}{\u g}{\"u}{\c s}} {\em et~al.}(2016){G{\"o}{\u g}{\"u}{\c s}},
  {Lin}, {Kaneko}, {Kouveliotou}, {Watts}, {Chakraborty}, {Alpar},
  {Huppenkothen}, {Roberts}, {Younes}, and {van der Horst}]{gogus16}
{G{\"o}{\u g}{\"u}{\c s}}, E., {Lin}, L., {Kaneko}, Y., {Kouveliotou}, C.,
  {Watts}, A.~L., {Chakraborty}, M., {Alpar}, M.~A., {Huppenkothen}, D.,
  {Roberts}, O.~J., {Younes}, G., and {van der Horst}, A.~J. (2016).
\newblock {Magnetar-like X-Ray Bursts from a Rotation-powered Pulsar, PSR
  J1119-6127}.
\newblock {\em \apjl\/}, {\bf 829}, L25.

\bibitem[{Gourgouliatos} and {Cumming}(2015){Gourgouliatos} and
  {Cumming}]{gourgouliatos15}
{Gourgouliatos}, K.~N. and {Cumming}, A. (2015).
\newblock {Hall drift and the braking indices of young pulsars}.
\newblock {\em \mnras\/}, {\bf 446}, 1121--1128.

\bibitem[{G{\"o}{\v g}{\"u}{\c s} } {\em et~al.}(1999){G{\"o}{\v g}{\"u}{\c s}
  }, {Woods}, {Kouveliotou}, {van Paradijs}, {Briggs}, {Duncan}, and
  {Thompson}]{gogus99}
{G{\"o}{\v g}{\"u}{\c s} }, E., {Woods}, P.~M., {Kouveliotou}, C., {van
  Paradijs}, J., {Briggs}, M.~S., {Duncan}, R.~C., and {Thompson}, C. (1999).
\newblock {Statistical Properties of SGR 1900+14 Bursts}.
\newblock {\em \apjl\/}, {\bf 526}, L93--L96.

\bibitem[{G{\"o}{\v g}{\"u}{\c s}} {\em et~al.}(2000){G{\"o}{\v g}{\"u}{\c s}},
  {Woods}, {Kouveliotou}, {van Paradijs}, {Briggs}, {Duncan}, and
  {Thompson}]{gogus00}
{G{\"o}{\v g}{\"u}{\c s}}, E., {Woods}, P.~M., {Kouveliotou}, C., {van
  Paradijs}, J., {Briggs}, M.~S., {Duncan}, R.~C., and {Thompson}, C. (2000).
\newblock {Statistical Properties of SGR 1806-20 Bursts}.
\newblock {\em \apjl\/}, {\bf 532}, L121--L124.

\bibitem[{G{\"o}{\v g}{\"u}{\c s}} {\em et~al.}(2011){G{\"o}{\v g}{\"u}{\c s}},
  {Woods}, {Kouveliotou}, {Finger}, {Pal'shin}, {Kaneko}, {Golenetskii},
  {Frederiks}, and {Airhart}]{gwk11}
{G{\"o}{\v g}{\"u}{\c s}}, E., {Woods}, P.~M., {Kouveliotou}, C., {Finger},
  M.~H., {Pal'shin}, V., {Kaneko}, Y., {Golenetskii}, S., {Frederiks}, D., and
  {Airhart}, C. (2011).
\newblock {Extended Tails from SGR 1806-20 Bursts}.
\newblock {\em \apj\/}, {\bf 740}, 55.

\bibitem[{Granot} {\em et~al.}(2006){Granot}, {Ramirez-Ruiz}, {Taylor},
  {Eichler}, {Lyubarsky}, {Wijers}, {Gaensler}, {Gelfand}, and
  {Kouveliotou}]{granot06}
{Granot}, J., {Ramirez-Ruiz}, E., {Taylor}, G.~B., {Eichler}, D., {Lyubarsky},
  Y.~E., {Wijers}, R.~A.~M.~J., {Gaensler}, B.~M., {Gelfand}, J.~D., and
  {Kouveliotou}, C. (2006).
\newblock {Diagnosing the Outflow from the SGR 1806-20 Giant Flare with Radio
  Observations}.
\newblock {\em \apj\/}, {\bf 638}, 391--396.

\bibitem[{Haberl} {\em et~al.}(2012){Haberl}, {Sturm}, {Filipovi{\'c}},
  {Pietsch}, and {Crawford}]{haberl12}
{Haberl}, F., {Sturm}, R., {Filipovi{\'c}}, M.~D., {Pietsch}, W., and
  {Crawford}, E.~J. (2012).
\newblock {SXP 1062, a young Be X-ray binary pulsar with long spin period.
  Implications for the neutron star birth spin}.
\newblock {\em \aap\/}, {\bf 537}, L1.

\bibitem[{Heger} {\em et~al.}(2005){Heger}, {Woosley}, and {Spruit}]{heger05}
{Heger}, A., {Woosley}, S.~E., and {Spruit}, H.~C. (2005).
\newblock {Presupernova Evolution of Differentially Rotating Massive Stars
  Including Magnetic Fields}.
\newblock {\em \apj\/}, {\bf 626}, 350--363.

\bibitem[{Helfand} {\em et~al.}(2007){Helfand}, {Chatterjee}, {Brisken},
  {Camilo}, {Reynolds}, {van Kerkwijk}, {Halpern}, and {Ransom}]{helfand07}
{Helfand}, D.~J., {Chatterjee}, S., {Brisken}, W.~F., {Camilo}, F., {Reynolds},
  J., {van Kerkwijk}, M.~H., {Halpern}, J.~P., and {Ransom}, S.~M. (2007).
\newblock {VLBA Measurement of the Transverse Velocity of the Magnetar XTE
  J1810-197}.
\newblock {\em \apj\/}, {\bf 662}, 1198--1203.

\bibitem[{H{\'e}nault-Brunet} {\em et~al.}(2012){H{\'e}nault-Brunet},
  {Oskinova}, {Guerrero}, {Sun}, {Chu}, {Evans}, {Gallagher}, {Gruendl}, and
  {Reyes-Iturbide}]{henault12}
{H{\'e}nault-Brunet}, V., {Oskinova}, L.~M., {Guerrero}, M.~A., {Sun}, W.,
  {Chu}, Y.-H., {Evans}, C.~J., {Gallagher}, III, J.~S., {Gruendl}, R.~A., and
  {Reyes-Iturbide}, J. (2012).
\newblock {Discovery of a Be/X-ray pulsar binary and associated supernova
  remnant in the Wing of the Small Magellanic Cloud}.
\newblock {\em \mnras\/}, {\bf 420}, L13--L17.

\bibitem[{Herold}(1979){Herold}]{herold79}
{Herold}, H. (1979).
\newblock {Compton and Thomson scattering in strong magnetic fields}.
\newblock {\em \prd\/}, {\bf 19}, 2868--2875.

\bibitem[{Ho}(2011){Ho}]{ho11}
{Ho}, W.~C.~G. (2011).
\newblock {Evolution of a buried magnetic field in the central compact object
  neutron stars}.
\newblock {\em \mnras\/}, {\bf 414}, 2567--2575.

\bibitem[{Ho} and {Andersson}(2017){Ho} and {Andersson}]{ho17}
{Ho}, W.~C.~G. and {Andersson}, N. (2017).
\newblock {Ejector and propeller spin-down: how might a superluminous supernova
  millisecond magnetar become the 6.67 h pulsar in RCW 103}.
\newblock {\em \mnras\/}, {\bf 464}, L65--L69.

\bibitem[{Hobbs} {\em et~al.}(2010){Hobbs}, {Lyne}, and {Kramer}]{hobbs10}
{Hobbs}, G., {Lyne}, A.~G., and {Kramer}, M. (2010).
\newblock {An analysis of the timing irregularities for 366 pulsars}.
\newblock {\em \mnras\/}, {\bf 402}, 1027--1048.

\bibitem[{Hulleman} {\em et~al.}(2000){Hulleman}, {van Kerkwijk}, and
  {Kulkarni}]{hulleman00}
{Hulleman}, F., {van Kerkwijk}, M.~H., and {Kulkarni}, S.~R. (2000).
\newblock {An optical counterpart to the anomalous X-ray pulsar 4U0142+61}.
\newblock {\em \nat\/}, {\bf 408}, 689--692.

\bibitem[{Hulleman} {\em et~al.}(2004){Hulleman}, {van Kerkwijk}, and
  {Kulkarni}]{hulleman04}
{Hulleman}, F., {van Kerkwijk}, M.~H., and {Kulkarni}, S.~R. (2004).
\newblock {The Anomalous X-ray Pulsar 4U 0142+61: Variability in the infrared
  and a spectral break in the optical}.
\newblock {\em \aap\/}, {\bf 416}, 1037--1045.

\bibitem[{Huppenkothen} {\em et~al.}(2013){Huppenkothen}, {Watts}, {Uttley},
  {van der Horst}, {van der Klis}, {Kouveliotou}, {G{\"o}{\v g}{\"u}{\c s}},
  {Granot}, {Vaughan}, and {Finger}]{huppenkothen13}
{Huppenkothen}, D., {Watts}, A.~L., {Uttley}, P., {van der Horst}, A.~J., {van
  der Klis}, M., {Kouveliotou}, C., {G{\"o}{\v g}{\"u}{\c s}}, E., {Granot},
  J., {Vaughan}, S., and {Finger}, M.~H. (2013).
\newblock {Quasi-periodic Oscillations and Broadband Variability in Short
  Magnetar Bursts}.
\newblock {\em \apj\/}, {\bf 768}, 87.

\bibitem[{Huppenkothen} {\em et~al.}(2014a){Huppenkothen}, {Heil}, {Watts}, and
  {G{\"o}{\u g}{\"u}{\c s}}]{hhw14}
{Huppenkothen}, D., {Heil}, L.~M., {Watts}, A.~L., and {G{\"o}{\u g}{\"u}{\c
  s}}, E. (2014a).
\newblock {Quasi-periodic Oscillations in Short Recurring Bursts of Magnetars
  SGR 1806-20 and SGR 1900+14 Observed with RXTE}.
\newblock {\em \apj\/}, {\bf 795}, 114.

\bibitem[{Huppenkothen} {\em et~al.}(2014b){Huppenkothen}, {D'Angelo}, {Watts},
  {Heil}, {van der Klis}, {van der Horst}, {Kouveliotou}, {Baring}, {G{\"o}{\u
  g}{\"u}{\c s}}, {Granot}, {Kaneko}, {Lin}, {von Kienlin}, and
  {Younes}]{huppenkothen14}
{Huppenkothen}, D., {D'Angelo}, C., {Watts}, A.~L., {Heil}, L., {van der Klis},
  M., {van der Horst}, A.~J., {Kouveliotou}, C., {Baring}, M.~G., {G{\"o}{\u
  g}{\"u}{\c s}}, E., {Granot}, J., {Kaneko}, Y., {Lin}, L., {von Kienlin}, A.,
  and {Younes}, G. (2014b).
\newblock {Quasi-periodic Oscillations in Short Recurring Bursts of the Soft
  Gamma Repeater J1550-5418}.
\newblock {\em \apj\/}, {\bf 787}, 128.

\bibitem[{Huppenkothen} {\em et~al.}(2015){Huppenkothen}, {Brewer}, {Hogg},
  {Murray}, {Frean}, {Elenbaas}, {Watts}, {Levin}, {van der Horst}, and
  {Kouveliotou}]{huppenkothen15}
{Huppenkothen}, D., {Brewer}, B.~J., {Hogg}, D.~W., {Murray}, I., {Frean}, M.,
  {Elenbaas}, C., {Watts}, A.~L., {Levin}, Y., {van der Horst}, A.~J., and
  {Kouveliotou}, C. (2015).
\newblock {Dissecting Magnetar Variability with Bayesian Hierarchical Models}.
\newblock {\em \apj\/}, {\bf 810}, 66.

\bibitem[{Hurley} {\em et~al.}(1999){Hurley}, {Cline}, {Mazets}, {Barthelmy},
  {Butterworth}, {Marshall}, {Palmer}, {Aptekar}, {Golenetskii}, {Il'Inskii},
  {Frederiks}, {McTiernan}, {Gold}, and {Trombka}]{hurley99}
{Hurley}, K., {Cline}, T., {Mazets}, E., {Barthelmy}, S., {Butterworth}, P.,
  {Marshall}, F., {Palmer}, D., {Aptekar}, R., {Golenetskii}, S., {Il'Inskii},
  V., {Frederiks}, D., {McTiernan}, J., {Gold}, R., and {Trombka}, J. (1999).
\newblock {A giant periodic flare from the soft gamma-ray repeater SGR
  1900+14.}
\newblock {\em \nat\/}, {\bf 397}, 41--43.

\bibitem[{Hurley} {\em et~al.}(2005){Hurley}, {Boggs}, {Smith}, {Duncan},
  {Lin}, {Zoglauer}, {Krucker}, {Hurford}, {Hudson}, {Wigger}, {Hajdas},
  {Thompson}, {Mitrofanov}, {Sanin}, {Boynton}, {Fellows}, {von Kienlin},
  {Lichti}, {Rau}, and {Cline}]{hurley05}
{Hurley}, K., {Boggs}, S.~E., {Smith}, D.~M., {Duncan}, R.~C., {Lin}, R.,
  {Zoglauer}, A., {Krucker}, S., {Hurford}, G., {Hudson}, H., {Wigger}, C.,
  {Hajdas}, W., {Thompson}, C., {Mitrofanov}, I., {Sanin}, A., {Boynton}, W.,
  {Fellows}, C., {von Kienlin}, A., {Lichti}, G., {Rau}, A., and {Cline}, T.
  (2005).
\newblock {An exceptionally bright flare from SGR 1806-20 and the origins of
  short-duration {$\gamma$}-ray bursts}.
\newblock {\em \nat\/}, {\bf 434}, 1098--1103.

\bibitem[{Inan} {\em et~al.}(1999){Inan}, {Lehtinen}, {Lev-Tov}, {Johnson},
  {Bell}, and {Hurley}]{inan99}
{Inan}, U.~S., {Lehtinen}, N.~G., {Lev-Tov}, S.~J., {Johnson}, M.~P., {Bell},
  T.~F., and {Hurley}, K. (1999).
\newblock {Ionization of the lower ionosphere by {$\gamma$}-rays from a
  magnetar: Detection of a low energy (3-10 keV) component}.
\newblock {\em Geophys. Res. Lett.}, {\bf 26}, 3357--3360.

\bibitem[{Inan} {\em et~al.}(2007){Inan}, {Lehtinen}, {Moore}, {Hurley},
  {Boggs}, {Smith}, and {Fishman}]{inan07}
{Inan}, U.~S., {Lehtinen}, N.~G., {Moore}, R.~C., {Hurley}, K., {Boggs}, S.,
  {Smith}, D.~M., and {Fishman}, G.~J. (2007).
\newblock {Massive disturbance of the daytime lower ionosphere by the giant
  {$\gamma$}-ray flare from magnetar SGR 1806-20}.
\newblock {\em Geophys. Res. Lett.}, {\bf 34}, L08103.

\bibitem[{Israel} and {Dall'Osso}(2011){Israel} and {Dall'Osso}]{israel11}
{Israel}, G. and {Dall'Osso}, S. (2011).
\newblock Bursts and flares from highly magnetic pulsars.
\newblock In D.~F. Torres and N.~Rea, editors, {\em High-Energy Emission from
  Pulsars and their Systems. Proceedings of the First Session of the Sant Cugat
  Forum on Astrophysics\/}, Astrophysics and Space Science Proceedings, pages
  279--298. Springer, Heidelberg.

\bibitem[{Israel} {\em et~al.}(2004){Israel}, {Stella}, {Covino}, {Campana},
  {Angelini}, {Mignani}, {Mereghetti}, {Marconi}, and {Perna}]{israel04}
{Israel}, G., {Stella}, L., {Covino}, S., {Campana}, S., {Angelini}, L.,
  {Mignani}, R., {Mereghetti}, S., {Marconi}, G., and {Perna}, R. (2004).
\newblock {Unveiling the Multi-wavelength Phenomenology of Anomalous X-ray
  Pulsars}.
\newblock In F.~{Camilo} and B.~M. {Gaensler}, editors, {\em Young Neutron
  Stars and Their Environments, IAU Symposium no. 218\/}, page 247.
  Astronomical Society of the Pacific (san Francisco).

\bibitem[{Israel} {\em et~al.}(2005){Israel}, {Belloni}, {Stella}, {Rephaeli},
  {Gruber}, {Casella}, {Dall'Osso}, {Rea}, {Persic}, and {Rothschild}]{ibs05}
{Israel}, G.~L., {Belloni}, T., {Stella}, L., {Rephaeli}, Y., {Gruber}, D.~E.,
  {Casella}, P., {Dall'Osso}, S., {Rea}, N., {Persic}, M., and {Rothschild},
  R.~E. (2005).
\newblock {The Discovery of Rapid X-Ray Oscillations in the Tail of the SGR
  1806-20 Hyperflare}.
\newblock {\em \apjl\/}, {\bf 628}, L53--L56.

\bibitem[{Israel} {\em et~al.}(2007){Israel}, {G{\"o}tz}, {Zane}, {Dall'Osso},
  {Rea}, and {Stella}]{israel07}
{Israel}, G.~L., {G{\"o}tz}, D., {Zane}, S., {Dall'Osso}, S., {Rea}, N., and
  {Stella}, L. (2007).
\newblock {Linking the X-ray timing and spectral properties of the glitching
  AXP 1RXS J170849-400910}.
\newblock {\em \aap\/}, {\bf 476}, L9--L12.

\bibitem[{Israel} {\em et~al.}(2008){Israel}, {Romano}, {Mangano}, {Dall'Osso},
  {Chincarini}, {Stella}, {Campana}, {Belloni}, {Tagliaferri}, {Blustin},
  {Sakamoto}, {Hurley}, {Zane}, {Moretti}, {Palmer}, {Guidorzi}, {Burrows},
  {Gehrels}, and {Krimm}]{israel08}
{Israel}, G.~L., {Romano}, P., {Mangano}, V., {Dall'Osso}, S., {Chincarini},
  G., {Stella}, L., {Campana}, S., {Belloni}, T., {Tagliaferri}, G., {Blustin},
  A.~J., {Sakamoto}, T., {Hurley}, K., {Zane}, S., {Moretti}, A., {Palmer}, D.,
  {Guidorzi}, C., {Burrows}, D.~N., {Gehrels}, N., and {Krimm}, H.~A. (2008).
\newblock {A Swift Gaze into the 2006 March 29 Burst Forest of SGR 1900+14}.
\newblock {\em \apj\/}, {\bf 685}, 1114--1128.

\bibitem[{Israel} {\em et~al.}(2010){Israel}, {Esposito}, {Rea}, {Dall'Osso},
  {Senziani}, {Romano}, {Mangano}, {G{\"o}tz}, {Zane}, {Tiengo}, {Palmer},
  {Krimm}, {Gehrels}, {Mereghetti}, {Stella}, {Turolla}, {Campana}, {Perna},
  {Angelini}, and {De Luca}]{israel10}
{Israel}, G.~L., {Esposito}, P., {Rea}, N., {Dall'Osso}, S., {Senziani}, F.,
  {Romano}, P., {Mangano}, V., {G{\"o}tz}, D., {Zane}, S., {Tiengo}, A.,
  {Palmer}, D.~M., {Krimm}, H., {Gehrels}, N., {Mereghetti}, S., {Stella}, L.,
  {Turolla}, R., {Campana}, S., {Perna}, R., {Angelini}, L., and {De Luca}, A.
  (2010).
\newblock {The 2008 October Swift detection of X-ray bursts/outburst from the
  transient SGR-like AXP 1E1547.0-5408}.
\newblock {\em \mnras\/}, {\bf 408}, 1387--1395.

\bibitem[{Israel} {\em et~al.}(2017a){Israel}, {Belfiore}, {Stella},
  {Esposito}, {Casella}, {De Luca}, {Marelli}, {Papitto}, {Perri}, {Puccetti},
  {Castillo}, {Salvetti}, {Tiengo}, {Zampieri}, {D'Agostino}, {Greiner},
  {Haberl}, {Novara}, {Salvaterra}, {Turolla}, {Watson}, {Wilms}, and
  {Wolter}]{israel17}
{Israel}, G.~L., {Belfiore}, A., {Stella}, L., {Esposito}, P., {Casella}, P.,
  {De Luca}, A., {Marelli}, M., {Papitto}, A., {Perri}, M., {Puccetti}, S.,
  {Castillo}, G.~A.~R., {Salvetti}, D., {Tiengo}, A., {Zampieri}, L.,
  {D'Agostino}, D., {Greiner}, J., {Haberl}, F., {Novara}, G., {Salvaterra},
  R., {Turolla}, R., {Watson}, M., {Wilms}, J., and {Wolter}, A. (2017a).
\newblock {An accreting pulsar with extreme properties drives an ultraluminous
  x-ray source in NGC 5907}.
\newblock {\em Science\/}, {\bf 355}, 817--819.

\bibitem[{Israel} {\em et~al.}(2017b){Israel}, {Papitto}, {Esposito}, {Stella},
  {Zampieri}, {Belfiore}, {Rodr{\'{\i}}guez Castillo}, {De Luca}, {Tiengo},
  {Haberl}, {Greiner}, {Salvaterra}, {Sandrelli}, and {Lisini}]{ipe17}
{Israel}, G.~L., {Papitto}, A., {Esposito}, P., {Stella}, L., {Zampieri}, L.,
  {Belfiore}, A., {Rodr{\'{\i}}guez Castillo}, G.~A., {De Luca}, A., {Tiengo},
  A., {Haberl}, F., {Greiner}, J., {Salvaterra}, R., {Sandrelli}, S., and
  {Lisini}, G. (2017b).
\newblock {Discovery of a 0.42-s pulsar in the ultraluminous X-ray source NGC
  7793 P13}.
\newblock {\em \mnras\/}, {\bf 466}, L48--L52.

\bibitem[{Kaaret} {\em et~al.}(2017){Kaaret}, {Feng}, and {Roberts}]{kaaret17}
{Kaaret}, P., {Feng}, H., and {Roberts}, T.~P. (2017).
\newblock {Ultraluminous X-Ray Sources}.
\newblock {\em \araa\/}, {\bf 55}, 303--341.

\bibitem[{Kaspi}(2010){Kaspi}]{kaspi10}
{Kaspi}, V.~M. (2010).
\newblock {Grand unification of neutron stars}.
\newblock {\em Proc.~of the National Academy of Science\/}, {\bf 107},
  7147--7152.

\bibitem[{Kaspi} and {Beloborodov}(2017){Kaspi} and {Beloborodov}]{kaspi17}
{Kaspi}, V.~M. and {Beloborodov}, A.~M. (2017).
\newblock {Magnetars}.
\newblock {\em \araa\/}, {\bf 55}, 261--301.

\bibitem[{Kaspi} and {Boydstun}(2010){Kaspi} and {Boydstun}]{kb10}
{Kaspi}, V.~M. and {Boydstun}, K. (2010).
\newblock {On the X-Ray Spectra of Anomalous X-Ray Pulsars and Soft Gamma
  Repeaters}.
\newblock {\em \apjl\/}, {\bf 710}, L115--L120.

\bibitem[{Kaspi} and {Kramer}(2016){Kaspi} and {Kramer}]{kaspi16}
{Kaspi}, V.~M. and {Kramer}, M. (2016).
\newblock {Radio Pulsars: The Neutron Star Population {\&} Fundamental
  Physics}.
\newblock In R.~{Blandford}, D.~{Gross}, and A.~{Sevrin}, editors, {\em
  Astrophysics and Cosmology\/}, Proceedings of the 26th Solvay Conference on
  Physics, pages 22--61. World Scientific, Singapore.

\bibitem[{Kennea} {\em et~al.}(2013){Kennea}, {Burrows}, {Kouveliotou},
  {Palmer}, {G{\"o}{\u g}{\"u}{\c s}}, {Kaneko}, {Evans}, {Degenaar},
  {Reynolds}, {Miller}, {Wijnands}, {Mori}, and {Gehrels}]{kennea13}
{Kennea}, J.~A., {Burrows}, D.~N., {Kouveliotou}, C., {Palmer}, D.~M.,
  {G{\"o}{\u g}{\"u}{\c s}}, E., {Kaneko}, Y., {Evans}, P.~A., {Degenaar}, N.,
  {Reynolds}, M.~T., {Miller}, J.~M., {Wijnands}, R., {Mori}, K., and
  {Gehrels}, N. (2013).
\newblock {Swift Discovery of a New Soft Gamma Repeater, SGR J1745--29, near
  Sagittarius A*}.
\newblock {\em \apjl\/}, {\bf 770}, L24.

\bibitem[{Kern} and {Martin}(2002){Kern} and {Martin}]{kern02}
{Kern}, B. and {Martin}, C. (2002).
\newblock {Optical pulsations from the anomalous X-ray pulsar 4U0142+61}.
\newblock {\em \nat\/}, {\bf 417}, 527--529.

\bibitem[{Kouveliotou} {\em et~al.}(1998){Kouveliotou}, {Dieters},
  {Strohmayer}, {van Paradijs}, {Fishman}, {Meegan}, {Hurley}, {Kommers},
  {Smith}, {Frail}, and {Murakami}]{kouveliotou98}
{Kouveliotou}, C., {Dieters}, S., {Strohmayer}, T., {van Paradijs}, J.,
  {Fishman}, G.~J., {Meegan}, C.~A., {Hurley}, K., {Kommers}, J., {Smith}, I.,
  {Frail}, D., and {Murakami}, T. (1998).
\newblock {An X-ray pulsar with a superstrong magnetic field in the soft
  gamma-ray repeater SGR 1806-20.}
\newblock {\em \nat\/}, {\bf 393}, 235--237.

\bibitem[{Kramer} {\em et~al.}(2006){Kramer}, {Stairs}, {Manchester},
  {McLaughlin}, {Lyne}, {Ferdman}, {Burgay}, {Lorimer}, {Possenti}, {D'Amico},
  {Sarkissian}, {Hobbs}, {Reynolds}, {Freire}, and {Camilo}]{kramer06}
{Kramer}, M., {Stairs}, I.~H., {Manchester}, R.~N., {McLaughlin}, M.~A.,
  {Lyne}, A.~G., {Ferdman}, R.~D., {Burgay}, M., {Lorimer}, D.~R., {Possenti},
  A., {D'Amico}, N., {Sarkissian}, J.~M., {Hobbs}, G.~B., {Reynolds}, J.~E.,
  {Freire}, P.~C.~C., and {Camilo}, F. (2006).
\newblock {Tests of General Relativity from Timing the Double Pulsar}.
\newblock {\em Science\/}, {\bf 314}, 97--102.

\bibitem[{Kramer} {\em et~al.}(2007){Kramer}, {Stappers}, {Jessner}, {Lyne},
  and {Jordan}]{kramer07}
{Kramer}, M., {Stappers}, B.~W., {Jessner}, A., {Lyne}, A.~G., and {Jordan},
  C.~A. (2007).
\newblock {Polarized radio emission from a magnetar}.
\newblock {\em \mnras\/}, {\bf 377}, 107--119.

\bibitem[{Kuiper} {\em et~al.}(2004){Kuiper}, {Hermsen}, and
  {Mendez}]{kuiper04}
{Kuiper}, L., {Hermsen}, W., and {Mendez}, M. (2004).
\newblock {Discovery of Hard Nonthermal Pulsed X-Ray Emission from the
  Anomalous X-Ray Pulsar 1E 1841-045}.
\newblock {\em \apj\/}, {\bf 613}, 1173--1178.

\bibitem[{Kuiper} {\em et~al.}(2006){Kuiper}, {Hermsen}, {den Hartog}, and
  {Collmar}]{kuiper06}
{Kuiper}, L., {Hermsen}, W., {den Hartog}, P.~R., and {Collmar}, W. (2006).
\newblock {Discovery of Luminous Pulsed Hard X-Ray Emission from Anomalous
  X-Ray Pulsars 1RXS J1708-4009, 4U 0142+61, and 1E 2259+586 by INTEGRAL and
  RXTE}.
\newblock {\em \apj\/}, {\bf 645}, 556--575.

\bibitem[{Lander} {\em et~al.}(2015){Lander}, {Andersson}, {Antonopoulou}, and
  {Watts}]{lander15}
{Lander}, S.~K., {Andersson}, N., {Antonopoulou}, D., and {Watts}, A.~L.
  (2015).
\newblock {Magnetically driven crustquakes in neutron stars}.
\newblock {\em \mnras\/}, {\bf 449}, 2047--2058.

\bibitem[{Langer}(2012){Langer}]{langer12}
{Langer}, N. (2012).
\newblock {Presupernova Evolution of Massive Single and Binary Stars}.
\newblock {\em \araa\/}, {\bf 50}, 107--164.

\bibitem[{Lenters} {\em et~al.}(2003){Lenters}, {Woods}, {Goupell},
  {Kouveliotou}, {G{\"o}{\u g}{\"u}{\c s}}, {Hurley}, {Frederiks},
  {Golenetskii}, and {Swank}]{lenters03}
{Lenters}, G.~T., {Woods}, P.~M., {Goupell}, J.~E., {Kouveliotou}, C.,
  {G{\"o}{\u g}{\"u}{\c s}}, E., {Hurley}, K., {Frederiks}, D., {Golenetskii},
  S., and {Swank}, J. (2003).
\newblock {An Extended Burst Tail from SGR 1900+14 with a Thermal X-Ray
  Spectrum}.
\newblock {\em \apj\/}, {\bf 587}, 761--770.

\bibitem[{Levin} {\em et~al.}(2010){Levin}, {Bailes}, {Bates}, {Bhat},
  {Burgay}, {Burke-Spolaor}, {D'Amico}, {Johnston}, {Keith}, {Kramer}, {Milia},
  {Possenti}, {Rea}, {Stappers}, and {van Straten}]{levin10}
{Levin}, L., {Bailes}, M., {Bates}, S., {Bhat}, N.~D.~R., {Burgay}, M.,
  {Burke-Spolaor}, S., {D'Amico}, N., {Johnston}, S., {Keith}, M., {Kramer},
  M., {Milia}, S., {Possenti}, A., {Rea}, N., {Stappers}, B., and {van
  Straten}, W. (2010).
\newblock {A Radio-loud Magnetar in X-ray Quiescence}.
\newblock {\em \apjl\/}, {\bf 721}, L33--L37.

\bibitem[{Levin} {\em et~al.}(2012){Levin}, {Bailes}, {Bates}, {Bhat},
  {Burgay}, {Burke-Spolaor}, {D'Amico}, {Johnston}, {Keith}, {Kramer}, {Milia},
  {Possenti}, {Stappers}, and {van Straten}]{levin12}
{Levin}, L., {Bailes}, M., {Bates}, S.~D., {Bhat}, N.~D.~R., {Burgay}, M.,
  {Burke-Spolaor}, S., {D'Amico}, N., {Johnston}, S., {Keith}, M.~J., {Kramer},
  M., {Milia}, S., {Possenti}, A., {Stappers}, B., and {van Straten}, W.
  (2012).
\newblock {Radio emission evolution, polarimetry and multifrequency single
  pulse analysis of the radio magnetar PSR J1622-4950}.
\newblock {\em \mnras\/}, {\bf 422}, 2489--2500.

\bibitem[{Levin}(2006){Levin}]{levin06}
{Levin}, Y. (2006).
\newblock {QPOs during magnetar flares are not driven by mechanical normal
  modes of the crust}.
\newblock {\em \mnras\/}, {\bf 368}, L35--L38.

\bibitem[{Levin}(2007){Levin}]{levin07}
{Levin}, Y. (2007).
\newblock {On the theory of magnetar QPOs}.
\newblock {\em \mnras\/}, {\bf 377}, 159--167.

\bibitem[{Levin} and {van Hoven}(2011){Levin} and {van Hoven}]{levin11}
{Levin}, Y. and {van Hoven}, M. (2011).
\newblock {On the excitation of f modes and torsional modes by magnetar giant
  flares}.
\newblock {\em \mnras\/}, {\bf 418}, 659--663.

\bibitem[{Li} {\em et~al.}(2017){Li}, {Rea}, {Torres}, and {de
  O{\~n}a-Wilhelmi}]{li17}
{Li}, J., {Rea}, N., {Torres}, D.~F., and {de O{\~n}a-Wilhelmi}, E. (2017).
\newblock {Gamma-ray Upper Limits on Magnetars with Six Years of Fermi-LAT
  Observations}.
\newblock {\em \apj\/}, {\bf 835}, 30.

\bibitem[{Li} {\em et~al.}(2016){Li}, {Shao}, and {Li}]{li16}
{Li}, T., {Shao}, Y., and {Li}, X.-D. (2016).
\newblock {Can the Subsonic Accretion Model Explain the Spin Period
  Distribution of Wind-fed X-Ray Pulsars?}
\newblock {\em \apj\/}, {\bf 824}, 143.

\bibitem[{Lynch} {\em et~al.}(2015){Lynch}, {Archibald}, {Kaspi}, and
  {Scholz}]{lynch15}
{Lynch}, R.~S., {Archibald}, R.~F., {Kaspi}, V.~M., and {Scholz}, P. (2015).
\newblock {Green Bank Telescope and Swift X-Ray Telescope Observations of the
  Galactic Center Radio Magnetar SGR J1745-2900}.
\newblock {\em \apj\/}, {\bf 806}, 266.

\bibitem[{Lyne} {\em et~al.}(2010){Lyne}, {Hobbs}, {Kramer}, {Stairs}, and
  {Stappers}]{lyne10}
{Lyne}, A., {Hobbs}, G., {Kramer}, M., {Stairs}, I., and {Stappers}, B. (2010).
\newblock {Switched Magnetospheric Regulation of Pulsar Spin-Down}.
\newblock {\em Science\/}, {\bf 329}, 408.

\bibitem[{Mandea} and {Balasis}(2006){Mandea} and {Balasis}]{mandea06}
{Mandea}, M. and {Balasis}, G. (2006).
\newblock {FAST TRACK PAPER: The SGR 1806-20 magnetar signature on the Earth's
  magnetic field}.
\newblock {\em Geophysical Journal International\/}, {\bf 167}, 586--591.

\bibitem[{Martin} {\em et~al.}(2014){Martin}, {Rea}, {Torres}, and
  {Papitto}]{martin14}
{Martin}, J., {Rea}, N., {Torres}, D.~F., and {Papitto}, A. (2014).
\newblock {Comparing supernova remnants around strongly magnetized and
  canonical pulsars}.
\newblock {\em \mnras\/}, {\bf 444}, 2910--2924.

\bibitem[{Mazets} {\em et~al.}(1979){Mazets}, {Golentskii}, {Ilinskii},
  {Aptekar}, and {Guryan}]{mazets79}
{Mazets}, E.~P., {Golentskii}, S.~V., {Ilinskii}, V.~N., {Aptekar}, R.~L., and
  {Guryan}, I.~A. (1979).
\newblock {Observations of a flaring X-ray pulsar in Dorado}.
\newblock {\em \nat\/}, {\bf 282}, 587--589.

\bibitem[{Mereghetti} and {Stella}(1995){Mereghetti} and
  {Stella}]{mereghetti95}
{Mereghetti}, S. and {Stella}, L. (1995).
\newblock {The very low mass X-ray binary pulsars: A new class of sources?}
\newblock {\em \apjl\/}, {\bf 442}, L17--L20.

\bibitem[{Mereghetti} {\em et~al.}(2002){Mereghetti}, {Chiarlone}, {Israel},
  and {Stella}]{mci02}
{Mereghetti}, S., {Chiarlone}, L., {Israel}, G.~L., and {Stella}, L. (2002).
\newblock {The Anomalous X-ray Pulsars}.
\newblock In W.~{Becker}, H.~{Lesch}, and J.~{Tr{\"u}mper}, editors, {\em
  Neutron Stars, Pulsars, and Supernova Remnants\/}, pages 29--43. MPE Report
  278.

\bibitem[{Mereghetti} {\em et~al.}(2005){Mereghetti}, {Tiengo}, {Esposito},
  {G{\"o}tz}, {Stella}, {Israel}, {Rea}, {Feroci}, {Turolla}, and
  {Zane}]{mte05}
{Mereghetti}, S., {Tiengo}, A., {Esposito}, P., {G{\"o}tz}, D., {Stella}, L.,
  {Israel}, G.~L., {Rea}, N., {Feroci}, M., {Turolla}, R., and {Zane}, S.
  (2005).
\newblock {An XMM-Newton View of the Soft Gamma Repeater SGR 1806-20: Long-Term
  Variability in the Pre-Giant Flare Epoch}.
\newblock {\em \apj\/}, {\bf 628}, 938--945.

\bibitem[{Mereghetti} {\em et~al.}(2015){Mereghetti}, {Pons}, and
  {Melatos}]{mereghetti15}
{Mereghetti}, S., {Pons}, J.~A., and {Melatos}, A. (2015).
\newblock {Magnetars: Properties, Origin and Evolution}.
\newblock {\em \ssr\/}, {\bf 191}, 315--338.

\bibitem[{Metzger} {\em et~al.}(2011){Metzger}, {Giannios}, {Thompson},
  {Bucciantini}, and {Quataert}]{metzger11}
{Metzger}, B.~D., {Giannios}, D., {Thompson}, T.~A., {Bucciantini}, N., and
  {Quataert}, E. (2011).
\newblock {The protomagnetar model for gamma-ray bursts}.
\newblock {\em \mnras\/}, {\bf 413}, 2031--2056.

\bibitem[{Mignani}(2011){Mignani}]{mignani11}
{Mignani}, R.~P. (2011).
\newblock {Optical, ultraviolet, and infrared observations of isolated neutron
  stars}.
\newblock {\em Advances in Space Research\/}, {\bf 47}, 1281--1293.

\bibitem[{Miller} and {Colbert}(2004){Miller} and {Colbert}]{miller04}
{Miller}, M.~C. and {Colbert}, E.~J.~M. (2004).
\newblock {Intermediate-Mass Black Holes}.
\newblock {\em International Journal of Modern Physics D\/}, {\bf 13}, 1--64.

\bibitem[{Mori} {\em et~al.}(2013){Mori}, {Gotthelf}, {Zhang}, {An},
  {Baganoff}, {Barri{\`e}re}, {Beloborodov}, {Boggs}, {Christensen}, {Craig},
  {Dufour}, {Grefenstette}, {Hailey}, {Harrison}, {Hong}, {Kaspi}, {Kennea},
  {Madsen}, {Markwardt}, {Nynka}, {Stern}, {Tomsick}, and {Zhang}]{mori13}
{Mori}, K., {Gotthelf}, E.~V., {Zhang}, S., {An}, H., {Baganoff}, F.~K.,
  {Barri{\`e}re}, N.~M., {Beloborodov}, A.~M., {Boggs}, S.~E., {Christensen},
  F.~E., {Craig}, W.~W., {Dufour}, F., {Grefenstette}, B.~W., {Hailey}, C.~J.,
  {Harrison}, F.~A., {Hong}, J., {Kaspi}, V.~M., {Kennea}, J.~A., {Madsen},
  K.~K., {Markwardt}, C.~B., {Nynka}, M., {Stern}, D., {Tomsick}, J.~A., and
  {Zhang}, W.~W. (2013).
\newblock {NuSTAR Discovery of a 3.76 s Transient Magnetar Near Sagittarius
  A*}.
\newblock {\em \apjl\/}, {\bf 770}, L23.

\bibitem[{Muno} {\em et~al.}(2006){Muno}, {Clark}, {Crowther}, {Dougherty}, {de
  Grijs}, {Law}, {McMillan}, {Morris}, {Negueruela}, {Pooley}, {Portegies
  Zwart}, and {Yusef-Zadeh}]{muno06}
{Muno}, M.~P., {Clark}, J.~S., {Crowther}, P.~A., {Dougherty}, S.~M., {de
  Grijs}, R., {Law}, C., {McMillan}, S.~L.~W., {Morris}, M.~R., {Negueruela},
  I., {Pooley}, D., {Portegies Zwart}, S., and {Yusef-Zadeh}, F. (2006).
\newblock {A Neutron Star with a Massive Progenitor in Westerlund 1}.
\newblock {\em \apjl\/}, {\bf 636}, L41--L44.

\bibitem[{Murakami} {\em et~al.}(1994){Murakami}, {Tanaka}, {Kulkarni},
  {Ogasaka}, {Sonobe}, {Ogawara}, {Aoki}, and {Yoshida}]{murakami94}
{Murakami}, T., {Tanaka}, Y., {Kulkarni}, S.~R., {Ogasaka}, Y., {Sonobe}, T.,
  {Ogawara}, Y., {Aoki}, T., and {Yoshida}, A. (1994).
\newblock {X-Ray Identification of the Soft Gamma-Ray Repeater 1806-20}.
\newblock {\em \nat\/}, {\bf 368}, 127.

\bibitem[{Mushtukov} {\em et~al.}(2015){Mushtukov}, {Suleimanov}, {Tsygankov},
  and {Poutanen}]{mushtukov15}
{Mushtukov}, A.~A., {Suleimanov}, V.~F., {Tsygankov}, S.~S., and {Poutanen}, J.
  (2015).
\newblock {On the maximum accretion luminosity of magnetized neutron stars:
  connecting X-ray pulsars and ultraluminous X-ray sources}.
\newblock {\em \mnras\/}, {\bf 454}, 2539--2548.

\bibitem[{Neiner} {\em et~al.}(2015){Neiner}, {Morin}, and {Alecian}]{neiner15}
{Neiner}, C., {Morin}, J., and {Alecian}, E. (2015).
\newblock {The ``Binarity and Magnetic Interactions in various classes of
  stars'' (BinaMIcS) project}.
\newblock In F.~{Martins}, S.~{Boissier}, V.~{Buat}, L.~{Cambr{\'e}sy}, and
  P.~{Petit}, editors, {\em SF2A-2015: Proceedings of the Annual meeting of the
  French Society of Astronomy and Astrophysics\/}, pages 213--216.

\bibitem[{Newman}(2005){Newman}]{newman05}
{Newman}, M.~E.~J. (2005).
\newblock {Power laws, Pareto distributions and Zipf's law}.
\newblock {\em Contemporary Physics\/}, {\bf 46}, 323--351.

\bibitem[{Norris} {\em et~al.}(1991){Norris}, {Hertz}, {Wood}, and
  {Kouveliotou}]{norris91}
{Norris}, J.~P., {Hertz}, P., {Wood}, K.~S., and {Kouveliotou}, C. (1991).
\newblock {On the nature of soft gamma repeaters}.
\newblock {\em \apj\/}, {\bf 366}, 240--252.

\bibitem[{Olausen} and {Kaspi}(2014){Olausen} and {Kaspi}]{olausen14}
{Olausen}, S.~A. and {Kaspi}, V.~M. (2014).
\newblock {The McGill Magnetar Catalog}.
\newblock {\em \apjs\/}, {\bf 212}, 6.

\bibitem[{Paczynski}(1992){Paczynski}]{paczynski92}
{Paczynski}, B. (1992).
\newblock {GB 790305 as a very strongly magnetized neutron star}.
\newblock {\em Acta Astronomica\/}, {\bf 42}, 145--153.

\bibitem[{Paizis} and {Sidoli}(2014){Paizis} and {Sidoli}]{paizis14}
{Paizis}, A. and {Sidoli}, L. (2014).
\newblock {Cumulative luminosity distributions of supergiant fast X-ray
  transients in hard X-rays}.
\newblock {\em \mnras\/}, {\bf 439}, 3439--3452.

\bibitem[{Palmer} {\em et~al.}(2005){Palmer}, {Barthelmy}, {Gehrels}, {Kippen},
  {Cayton}, {Kouveliotou}, {Eichler}, {Wijers}, {Woods}, {Granot}, {Lyubarsky},
  {Ramirez-Ruiz}, {Barbier}, {Chester}, {Cummings}, {Fenimore}, {Finger},
  {Gaensler}, {Hullinger}, {Krimm}, {Markwardt}, {Nousek}, {Parsons}, {Patel},
  {Sakamoto}, {Sato}, {Suzuki}, and {Tueller}]{palmer05}
{Palmer}, D.~M., {Barthelmy}, S., {Gehrels}, N., {Kippen}, R.~M., {Cayton}, T.,
  {Kouveliotou}, C., {Eichler}, D., {Wijers}, R.~A.~M.~J., {Woods}, P.~M.,
  {Granot}, J., {Lyubarsky}, Y.~E., {Ramirez-Ruiz}, E., {Barbier}, L.,
  {Chester}, M., {Cummings}, J., {Fenimore}, E.~E., {Finger}, M.~H.,
  {Gaensler}, B.~M., {Hullinger}, D., {Krimm}, H., {Markwardt}, C.~B.,
  {Nousek}, J.~A., {Parsons}, A., {Patel}, S., {Sakamoto}, T., {Sato}, G.,
  {Suzuki}, M., and {Tueller}, J. (2005).
\newblock {A giant {$\gamma$}-ray flare from the magnetar SGR 1806 - 20}.
\newblock {\em \nat\/}, {\bf 434}, 1107--1109.

\bibitem[{Pennucci} {\em et~al.}(2015){Pennucci}, {Possenti}, {Esposito},
  {Rea}, {Haggard}, {Baganoff}, {Burgay}, {Coti Zelati}, {Israel}, and
  {Minter}]{pennucci15}
{Pennucci}, T.~T., {Possenti}, A., {Esposito}, P., {Rea}, N., {Haggard}, D.,
  {Baganoff}, F.~K., {Burgay}, M., {Coti Zelati}, F., {Israel}, G.~L., and
  {Minter}, A. (2015).
\newblock {Simultaneous Multi-band Radio and X-Ray Observations of the Galactic
  Center Magnetar SGR 1745-2900}.
\newblock {\em \apj\/}, {\bf 808}, 81.

\bibitem[{Perna} and {Pons}(2011){Perna} and {Pons}]{perna11}
{Perna}, R. and {Pons}, J.~A. (2011).
\newblock {A Unified Model of the Magnetar and Radio Pulsar Bursting
  Phenomenology}.
\newblock {\em \apjl\/}, {\bf 727}, L51.

\bibitem[{Perna} {\em et~al.}(2014){Perna}, {Duffell}, {Cantiello}, and
  {MacFadyen}]{perna14}
{Perna}, R., {Duffell}, P., {Cantiello}, M., and {MacFadyen}, A.~I. (2014).
\newblock {The Fate of Fallback Matter around Newly Born Compact Objects}.
\newblock {\em \apj\/}, {\bf 781}, 119.

\bibitem[{Pfuhl} {\em et~al.}(2015){Pfuhl}, {Gillessen}, {Eisenhauer},
  {Genzel}, {Plewa}, {Ott}, {Ballone}, {Schartmann}, {Burkert}, {Fritz},
  {Sari}, {Steinberg}, and {Madigan}]{pfuhl15}
{Pfuhl}, O., {Gillessen}, S., {Eisenhauer}, F., {Genzel}, R., {Plewa}, P.~M.,
  {Ott}, T., {Ballone}, A., {Schartmann}, M., {Burkert}, A., {Fritz}, T.~K.,
  {Sari}, R., {Steinberg}, E., and {Madigan}, A.-M. (2015).
\newblock {The Galactic Center Cloud G2 and its Gas Streamer}.
\newblock {\em \apj\/}, {\bf 798}, 111.

\bibitem[{Phifer} {\em et~al.}(2013){Phifer}, {Do}, {Meyer}, {Ghez}, {Witzel},
  {Yelda}, {Boehle}, {Lu}, {Morris}, {Becklin}, and {Matthews}]{phifer13}
{Phifer}, K., {Do}, T., {Meyer}, L., {Ghez}, A.~M., {Witzel}, G., {Yelda}, S.,
  {Boehle}, A., {Lu}, J.~R., {Morris}, M.~R., {Becklin}, E.~E., and {Matthews},
  K. (2013).
\newblock {Keck Observations of the Galactic Center Source G2: Gas Cloud or
  Star?}
\newblock {\em \apjl\/}, {\bf 773}, L13.

\bibitem[{Pintore} {\em et~al.}(2017){Pintore}, {Mereghetti}, {Tiengo},
  {Vianello}, {Costantini}, and {Esposito}]{pintore17}
{Pintore}, F., {Mereghetti}, S., {Tiengo}, A., {Vianello}, G., {Costantini},
  E., and {Esposito}, P. (2017).
\newblock {The effect of X-ray dust scattering on a bright burst from the
  magnetar 1E 1547.0-5408}.
\newblock {\em \mnras\/}, {\bf 467}, 3467--3474.

\bibitem[{Plewa} {\em et~al.}(2017){Plewa}, {Gillessen}, {Pfuhl}, {Eisenhauer},
  {Genzel}, {Burkert}, {Dexter}, {Habibi}, {George}, {Ott}, {Waisberg}, and
  {von Fellenberg}]{plewa17}
{Plewa}, P.~M., {Gillessen}, S., {Pfuhl}, O., {Eisenhauer}, F., {Genzel}, R.,
  {Burkert}, A., {Dexter}, J., {Habibi}, M., {George}, E., {Ott}, T.,
  {Waisberg}, I., and {von Fellenberg}, S. (2017).
\newblock {The Post-pericenter Evolution of the Galactic Center Source G2}.
\newblock {\em \apj\/}, {\bf 840}, 50.

\bibitem[{Pons} and {Geppert}(2007){Pons} and {Geppert}]{pg07}
{Pons}, J.~A. and {Geppert}, U. (2007).
\newblock {Magnetic field dissipation in neutron star crusts: from magnetars to
  isolated neutron stars}.
\newblock {\em \aap\/}, {\bf 470}, 303--315.

\bibitem[{Pons} and {Rea}(2012){Pons} and {Rea}]{pons12}
{Pons}, J.~A. and {Rea}, N. (2012).
\newblock {Modeling Magnetar Outbursts: Flux Enhancements and the Connection
  with Short Bursts and Glitches}.
\newblock {\em \apjl\/}, {\bf 750}, L6.

\bibitem[{Pons} {\em et~al.}(2007){Pons}, {Link}, {Miralles}, and
  {Geppert}]{pons07}
{Pons}, J.~A., {Link}, B., {Miralles}, J.~A., and {Geppert}, U. (2007).
\newblock {Evidence for Heating of Neutron Stars by Magnetic Field Decay}.
\newblock {\em \prl\/}, {\bf 98}, 071101.

\bibitem[{Pons} {\em et~al.}(2009){Pons}, {Miralles}, and {Geppert}]{pons09}
{Pons}, J.~A., {Miralles}, J.~A., and {Geppert}, U. (2009).
\newblock {Magneto-thermal evolution of neutron stars}.
\newblock {\em \aap\/}, {\bf 496}, 207--216.

\bibitem[{Ponti} {\em et~al.}(2015){Ponti}, {De Marco}, {Morris}, {Merloni},
  {Mu{\~n}oz-Darias}, {Clavel}, {Haggard}, {Zhang}, {Nandra}, {Gillessen},
  {Mori}, {Neilsen}, {Rea}, {Degenaar}, {Terrier}, and {Goldwurm}]{ponti15}
{Ponti}, G., {De Marco}, B., {Morris}, M.~R., {Merloni}, A.,
  {Mu{\~n}oz-Darias}, T., {Clavel}, M., {Haggard}, D., {Zhang}, S., {Nandra},
  K., {Gillessen}, S., {Mori}, K., {Neilsen}, J., {Rea}, N., {Degenaar}, N.,
  {Terrier}, R., and {Goldwurm}, A. (2015).
\newblock {Fifteen years of XMM-Newton and Chandra monitoring of Sgr A$^{★}$:
  evidence for a recent increase in the bright flaring rate}.
\newblock {\em \mnras\/}, {\bf 454}, 1525--1544.

\bibitem[{Popov} and {Turolla}(2012){Popov} and {Turolla}]{popov12}
{Popov}, S.~B. and {Turolla}, R. (2012).
\newblock {Probing the neutron star spin evolution in the young Small
  Magellanic Cloud Be/X-ray binary SXP 1062}.
\newblock {\em \mnras\/}, {\bf 421}, L127--L131.

\bibitem[Rea and Esposito(2011)Rea and Esposito]{rea11}
Rea, N. and Esposito, P. (2011).
\newblock Magnetar outbursts: an observational review.
\newblock In D.~F. Torres and N.~Rea, editors, {\em High-Energy Emission from
  Pulsars and their Systems. Proceedings of the First Session of the Sant Cugat
  Forum on Astrophysics\/}, Astrophysics and Space Science Proceedings, pages
  247--273. Springer, Heidelberg.

\bibitem[{Rea} {\em et~al.}(2010){Rea}, {Esposito}, {Turolla}, {Israel},
  {Zane}, {Stella}, {Mereghetti}, {Tiengo}, {G{\"o}tz}, {G{\"o}{\u g}{\"u}{\c
  s}}, and {Kouveliotou}]{rea10}
{Rea}, N., {Esposito}, P., {Turolla}, R., {Israel}, G.~L., {Zane}, S.,
  {Stella}, L., {Mereghetti}, S., {Tiengo}, A., {G{\"o}tz}, D., {G{\"o}{\u
  g}{\"u}{\c s}}, E., and {Kouveliotou}, C. (2010).
\newblock {A Low-Magnetic-Field Soft Gamma Repeater}.
\newblock {\em Science\/}, {\bf 330}, 944.

\bibitem[{Rea} {\em et~al.}(2012){Rea}, {Israel}, {Esposito}, {Pons},
  {Camero-Arranz}, {Mignani}, {Turolla}, {Zane}, {Burgay}, {Possenti},
  {Campana}, {Enoto}, {Gehrels}, {G{\"o}{\u g}{\"u}{\c s}}, {G{\"o}tz},
  {Kouveliotou}, {Makishima}, {Mereghetti}, {Oates}, {Palmer}, {Perna},
  {Stella}, and {Tiengo}]{rie12}
{Rea}, N., {Israel}, G.~L., {Esposito}, P., {Pons}, J.~A., {Camero-Arranz}, A.,
  {Mignani}, R.~P., {Turolla}, R., {Zane}, S., {Burgay}, M., {Possenti}, A.,
  {Campana}, S., {Enoto}, T., {Gehrels}, N., {G{\"o}{\u g}{\"u}{\c s}}, E.,
  {G{\"o}tz}, D., {Kouveliotou}, C., {Makishima}, K., {Mereghetti}, S.,
  {Oates}, S.~R., {Palmer}, D.~M., {Perna}, R., {Stella}, L., and {Tiengo}, A.
  (2012).
\newblock {A New Low Magnetic Field Magnetar: The 2011 Outburst of Swift
  J1822.3-1606}.
\newblock {\em \apj\/}, {\bf 754}, 27.

\bibitem[{Rea} {\em et~al.}(2013a){Rea}, {Esposito}, {Pons}, {Turolla},
  {Torres}, {Israel}, {Possenti}, {Burgay}, {Vigan{\`o}}, {Papitto}, {Perna},
  {Stella}, {Ponti}, {Baganoff}, {Haggard}, {Camero-Arranz}, {Zane}, {Minter},
  {Mereghetti}, {Tiengo}, {Sch{\"o}del}, {Feroci}, {Mignani}, and
  {G{\"o}tz}]{rep13}
{Rea}, N., {Esposito}, P., {Pons}, J.~A., {Turolla}, R., {Torres}, D.~F.,
  {Israel}, G.~L., {Possenti}, A., {Burgay}, M., {Vigan{\`o}}, D., {Papitto},
  A., {Perna}, R., {Stella}, L., {Ponti}, G., {Baganoff}, F.~K., {Haggard}, D.,
  {Camero-Arranz}, A., {Zane}, S., {Minter}, A., {Mereghetti}, S., {Tiengo},
  A., {Sch{\"o}del}, R., {Feroci}, M., {Mignani}, R., and {G{\"o}tz}, D.
  (2013a).
\newblock {A Strongly Magnetized Pulsar within the Grasp of the Milky Way's
  Supermassive Black Hole}.
\newblock {\em \apjl\/}, {\bf 775}, L34.

\bibitem[{Rea} {\em et~al.}(2013b){Rea}, {Israel}, {Pons}, {Turolla},
  {Vigan{\`o}}, {Zane}, {Esposito}, {Perna}, {Papitto}, {Terreran}, {Tiengo},
  {Salvetti}, {Girart}, {Palau}, {Possenti}, {Burgay}, {G{\"o}{\u g}{\"u}{\c
  s}}, {Caliandro}, {Kouveliotou}, {G{\"o}tz}, {Mignani}, {Ratti}, and
  {Stella}]{rea13}
{Rea}, N., {Israel}, G.~L., {Pons}, J.~A., {Turolla}, R., {Vigan{\`o}}, D.,
  {Zane}, S., {Esposito}, P., {Perna}, R., {Papitto}, A., {Terreran}, G.,
  {Tiengo}, A., {Salvetti}, D., {Girart}, J.~M., {Palau}, A., {Possenti}, A.,
  {Burgay}, M., {G{\"o}{\u g}{\"u}{\c s}}, E., {Caliandro}, G.~A.,
  {Kouveliotou}, C., {G{\"o}tz}, D., {Mignani}, R.~P., {Ratti}, E., and
  {Stella}, L. (2013b).
\newblock {The Outburst Decay of the Low Magnetic Field Magnetar SGR
  0418+5729}.
\newblock {\em \apj\/}, {\bf 770}, 65.

\bibitem[{Rea} {\em et~al.}(2014){Rea}, {Vigan{\`o}}, {Israel}, {Pons}, and
  {Torres}]{rea14}
{Rea}, N., {Vigan{\`o}}, D., {Israel}, G.~L., {Pons}, J.~A., and {Torres},
  D.~F. (2014).
\newblock {3XMM J185246.6+003317: Another Low Magnetic Field Magnetar}.
\newblock {\em \apjl\/}, {\bf 781}, L17.

\bibitem[{Rea} {\em et~al.}(2016){Rea}, {Borghese}, {Esposito}, {Coti Zelati},
  {Bachetti}, {Israel}, and {De Luca}]{rea16}
{Rea}, N., {Borghese}, A., {Esposito}, P., {Coti Zelati}, F., {Bachetti}, M.,
  {Israel}, G.~L., and {De Luca}, A. (2016).
\newblock {Magnetar-like Activity from the Central Compact Object in the SNR
  RCW103}.
\newblock {\em \apjl\/}, {\bf 828}, L13.

\bibitem[{Rodr{\'{\i}}guez Castillo} {\em et~al.}(2014){Rodr{\'{\i}}guez
  Castillo}, {Israel}, {Esposito}, {Pons}, {Rea}, {Turolla}, {Vigan{\`o}}, and
  {Zane}]{rodriguez14}
{Rodr{\'{\i}}guez Castillo}, G.~A., {Israel}, G.~L., {Esposito}, P., {Pons},
  J.~A., {Rea}, N., {Turolla}, R., {Vigan{\`o}}, D., and {Zane}, S. (2014).
\newblock {Pulse phase-coherent timing and spectroscopy of CXOU J164710.2-45521
  outbursts}.
\newblock {\em \mnras\/}, {\bf 441}, 1305--1316.

\bibitem[{Rodr{\'{\i}}guez Castillo} {\em et~al.}(2016){Rodr{\'{\i}}guez
  Castillo}, {Israel}, {Tiengo}, {Salvetti}, {Turolla}, {Zane}, {Rea},
  {Esposito}, {Mereghetti}, {Perna}, {Stella}, {Pons}, {Campana}, {G{\"o}tz},
  and {Motta}]{rodriguez16}
{Rodr{\'{\i}}guez Castillo}, G.~A., {Israel}, G.~L., {Tiengo}, A., {Salvetti},
  D., {Turolla}, R., {Zane}, S., {Rea}, N., {Esposito}, P., {Mereghetti}, S.,
  {Perna}, R., {Stella}, L., {Pons}, J.~A., {Campana}, S., {G{\"o}tz}, D., and
  {Motta}, S. (2016).
\newblock {The outburst decay of the low magnetic field magnetar SWIFT
  J1822.3-1606: phase-resolved analysis and evidence for a variable cyclotron
  feature}.
\newblock {\em \mnras\/}, {\bf 456}, 4145--4155.

\bibitem[{Rothschild} {\em et~al.}(1994){Rothschild}, {Kulkarni}, and
  {Lingenfelter}]{rothschild94}
{Rothschild}, R.~E., {Kulkarni}, S.~R., and {Lingenfelter}, R.~E. (1994).
\newblock {Discovery of an X-Ray Source Coincident with the Soft Gamma-ray
  Repeater 0525-66}.
\newblock {\em \nat\/}, {\bf 368}, 432.

\bibitem[{Sana} {\em et~al.}(2006){Sana}, {Rauw}, {Naz{\'e}}, {Gosset}, and
  {Vreux}]{sana06}
{Sana}, H., {Rauw}, G., {Naz{\'e}}, Y., {Gosset}, E., and {Vreux}, J.-M.
  (2006).
\newblock {An XMM-Newton view of the young open cluster NGC 6231 - II. The OB
  star population}.
\newblock {\em \mnras\/}, {\bf 372}, 661--678.

\bibitem[{Scholz} {\em et~al.}(2012){Scholz}, {Ng}, {Livingstone}, {Kaspi},
  {Cumming}, and {Archibald}]{scholz12}
{Scholz}, P., {Ng}, C.-Y., {Livingstone}, M.~A., {Kaspi}, V.~M., {Cumming}, A.,
  and {Archibald}, R.~F. (2012).
\newblock {Post-outburst X-Ray Flux and Timing Evolution of Swift
  J1822.3-1606}.
\newblock {\em \apj\/}, {\bf 761}, 66.

\bibitem[{Scholz} {\em et~al.}(2017){Scholz}, {Camilo}, {Sarkissian},
  {Reynolds}, {Levin}, {Bailes}, {Burgay}, {Johnston}, {Kramer}, and
  {Possenti}]{scholz17}
{Scholz}, P., {Camilo}, F., {Sarkissian}, J., {Reynolds}, J.~E., {Levin}, L.,
  {Bailes}, M., {Burgay}, M., {Johnston}, S., {Kramer}, M., and {Possenti}, A.
  (2017).
\newblock {Spin-down Evolution and Radio Disappearance of the Magnetar PSR
  J1622-4950}.
\newblock {\em \apj\/}, {\bf 841}, 126.

\bibitem[{Seiradakis} and {Wielebinski}(2004){Seiradakis} and
  {Wielebinski}]{seiradakis04}
{Seiradakis}, J.~H. and {Wielebinski}, R. (2004).
\newblock {Morphology and characteristics of radio pulsars}.
\newblock {\em \aapr\/}, {\bf 12}, 239--271.

\bibitem[{Shakura} {\em et~al.}(2012){Shakura}, {Postnov}, {Kochetkova}, and
  {Hjalmarsdotter}]{shakura12}
{Shakura}, N., {Postnov}, K., {Kochetkova}, A., and {Hjalmarsdotter}, L.
  (2012).
\newblock {Theory of quasi-spherical accretion in X-ray pulsars}.
\newblock {\em \mnras\/}, {\bf 420}, 216--236.

\bibitem[{Shannon} and {Johnston}(2013){Shannon} and {Johnston}]{shannon13}
{Shannon}, R.~M. and {Johnston}, S. (2013).
\newblock {Radio properties of the magnetar near Sagittarius A* from
  observations with the Australia Telescope Compact Array}.
\newblock {\em \mnras\/}, {\bf 435}, L29--L32.

\bibitem[{Sidoli} {\em et~al.}(2017){Sidoli}, {Israel}, {Esposito},
  {Rodr{\'{\i}}guez Castillo}, and {Postnov}]{sidoli17}
{Sidoli}, L., {Israel}, G.~L., {Esposito}, P., {Rodr{\'{\i}}guez Castillo},
  G.~A., and {Postnov}, K. (2017).
\newblock {AX J1910.7+0917: the slowest X-ray pulsar}.
\newblock {\em \mnras\/}, {\bf 469}, 3056--3061.

\bibitem[{Spruit}(2008){Spruit}]{spruit08}
{Spruit}, H.~C. (2008).
\newblock {Origin of neutron star magnetic fields}.
\newblock In C.~{Bassa}, Z.~{Wang}, A.~{Cumming}, and V.~M. {Kaspi}, editors,
  {\em 40 Years of Pulsars: Millisecond Pulsars, Magnetars and More\/}, volume
  983 of {\em American Institute of Physics Conference Series\/}, pages
  391--398.

\bibitem[{Strohmayer} and {Watts}(2005){Strohmayer} and {Watts}]{sw05}
{Strohmayer}, T.~E. and {Watts}, A.~L. (2005).
\newblock {Discovery of Fast X-Ray Oscillations during the 1998 Giant Flare
  from SGR 1900+14}.
\newblock {\em \apjl\/}, {\bf 632}, L111--L114.

\bibitem[{Strohmayer} and {Watts}(2006){Strohmayer} and {Watts}]{strohmayer06}
{Strohmayer}, T.~E. and {Watts}, A.~L. (2006).
\newblock {The 2004 Hyperflare from SGR 1806-20: Further Evidence for Global
  Torsional Vibrations}.
\newblock {\em \apj\/}, {\bf 653}, 593--601.

\bibitem[{Tam} {\em et~al.}(2004){Tam}, {Kaspi}, {van Kerkwijk}, and
  {Durant}]{tam04}
{Tam}, C.~R., {Kaspi}, V.~M., {van Kerkwijk}, M.~H., and {Durant}, M. (2004).
\newblock {Correlated Infrared and X-Ray Flux Changes Following the 2002 June
  Outburst of the Anomalous X-Ray Pulsar 1E 2259+586}.
\newblock {\em \apjl\/}, {\bf 617}, L53--L56.

\bibitem[{Tendulkar} {\em et~al.}(2012){Tendulkar}, {Cameron}, and
  {Kulkarni}]{tendulkar12}
{Tendulkar}, S.~P., {Cameron}, P.~B., and {Kulkarni}, S.~R. (2012).
\newblock {Proper Motions and Origins of SGR 1806-20 and SGR 1900+14}.
\newblock {\em \apj\/}, {\bf 761}, 76.

\bibitem[{Tendulkar} {\em et~al.}(2013){Tendulkar}, {Cameron}, and
  {Kulkarni}]{tendulkar13}
{Tendulkar}, S.~P., {Cameron}, P.~B., and {Kulkarni}, S.~R. (2013).
\newblock {Proper Motions and Origins of AXP 1E 2259+586 and AXP 4U 0142+61}.
\newblock {\em \apj\/}, {\bf 772}, 31.

\bibitem[{Tendulkar} {\em et~al.}(2015){Tendulkar}, {Hasc{\"o}et}, {Yang},
  {Kaspi}, {Beloborodov}, {An}, {Bachetti}, {Boggs}, {Christensen}, {Craig},
  {Guillot}, {Hailey}, {Harrison}, {Stern}, and {Zhang}]{tendulkar15}
{Tendulkar}, S.~P., {Hasc{\"o}et}, R., {Yang}, C., {Kaspi}, V.~M.,
  {Beloborodov}, A.~M., {An}, H., {Bachetti}, M., {Boggs}, S.~E.,
  {Christensen}, F.~E., {Craig}, W.~W., {Guillot}, S., {Hailey}, C.~A.,
  {Harrison}, F.~A., {Stern}, D., and {Zhang}, W. (2015).
\newblock {Phase-resolved NuSTAR and Swift-XRT Observations of Magnetar 4U
  0142+61}.
\newblock {\em \apj\/}, {\bf 808}, 32.

\bibitem[{Tendulkar} {\em et~al.}(2017){Tendulkar}, {Kaspi}, {Archibald}, and
  {Scholz}]{tendulkar17}
{Tendulkar}, S.~P., {Kaspi}, V.~M., {Archibald}, R.~F., and {Scholz}, P.
  (2017).
\newblock {A Near-infrared Counterpart of 2E1613.5--5053: The Central Source in
  Supernova Remnant RCW103}.
\newblock {\em \apj\/}, {\bf 841}, 11.

\bibitem[{Testa} {\em et~al.}(2008){Testa}, {Rea}, {Mignani}, {Israel},
  {Perna}, {Chaty}, {Stella}, {Covino}, {Turolla}, {Zane}, {Lo Curto},
  {Campana}, {Marconi}, and {Mereghetti}]{testa08}
{Testa}, V., {Rea}, N., {Mignani}, R.~P., {Israel}, G.~L., {Perna}, R.,
  {Chaty}, S., {Stella}, L., {Covino}, S., {Turolla}, R., {Zane}, S., {Lo
  Curto}, G., {Campana}, S., {Marconi}, G., and {Mereghetti}, S. (2008).
\newblock {Adaptive optics, near-infrared observations of magnetars}.
\newblock {\em \aap\/}, {\bf 482}, 607--615.

\bibitem[{Thompson} and {Duncan}(1995){Thompson} and {Duncan}]{thompson95}
{Thompson}, C. and {Duncan}, R.~C. (1995).
\newblock {The soft gamma repeaters as very strongly magnetized neutron stars -
  I. Radiative mechanism for outbursts}.
\newblock {\em \mnras\/}, {\bf 275}, 255--300.

\bibitem[{Thompson} and {Duncan}(1996){Thompson} and {Duncan}]{thompson96}
{Thompson}, C. and {Duncan}, R.~C. (1996).
\newblock {The Soft Gamma Repeaters as Very Strongly Magnetized Neutron Stars.
  II. Quiescent Neutrino, X-Ray, and Alfven Wave Emission}.
\newblock {\em \apj\/}, {\bf 473}, 322--342.

\bibitem[{Tiengo} {\em et~al.}(2008){Tiengo}, {Esposito}, and
  {Mereghetti}]{tiengo08}
{Tiengo}, A., {Esposito}, P., and {Mereghetti}, S. (2008).
\newblock {XMM-Newton Observations of CXOU J010043.1-721134: The First Deep
  Look at the Soft X-Ray Emission of a Magnetar}.
\newblock {\em \apjl\/}, {\bf 680}, L133--L136.

\bibitem[{Tiengo} {\em et~al.}(2010){Tiengo}, {Vianello}, {Esposito},
  {Mereghetti}, {Giuliani}, {Costantini}, {Israel}, {Stella}, {Turolla},
  {Zane}, {Rea}, {G{\"o}tz}, {Bernardini}, {Moretti}, {Romano}, {Ehle}, and
  {Gehrels}]{tiengo10}
{Tiengo}, A., {Vianello}, G., {Esposito}, P., {Mereghetti}, S., {Giuliani}, A.,
  {Costantini}, E., {Israel}, G.~L., {Stella}, L., {Turolla}, R., {Zane}, S.,
  {Rea}, N., {G{\"o}tz}, D., {Bernardini}, F., {Moretti}, A., {Romano}, P.,
  {Ehle}, M., and {Gehrels}, N. (2010).
\newblock {The Dust-scattering X-ray Rings of the Anomalous X-ray Pulsar 1E
  1547.0-5408}.
\newblock {\em \apj\/}, {\bf 710}, 227--235.

\bibitem[{Tiengo} {\em et~al.}(2013){Tiengo}, {Esposito}, {Mereghetti},
  {Turolla}, {Nobili}, {Gastaldello}, {G{\"o}tz}, {Israel}, {Rea}, {Stella},
  {Zane}, and {Bignami}]{tiengo13}
{Tiengo}, A., {Esposito}, P., {Mereghetti}, S., {Turolla}, R., {Nobili}, L.,
  {Gastaldello}, F., {G{\"o}tz}, D., {Israel}, G.~L., {Rea}, N., {Stella}, L.,
  {Zane}, S., and {Bignami}, G.~F. (2013).
\newblock {A variable absorption feature in the X-ray spectrum of a magnetar}.
\newblock {\em \nat\/}, {\bf 500}, 312--314.

\bibitem[{Tong} {\em et~al.}(2016){Tong}, {Wang}, {Liu}, and {Xu}]{tong16}
{Tong}, H., {Wang}, W., {Liu}, X.~W., and {Xu}, R.~X. (2016).
\newblock {Rotational Evolution of Magnetars in the Presence of a Fallback
  Disk}.
\newblock {\em \apj\/}, {\bf 833}, 265.

\bibitem[{Torne} {\em et~al.}(2015){Torne}, {Eatough}, {Karuppusamy}, {Kramer},
  {Paubert}, {Klein}, {Desvignes}, {Champion}, {Wiesemeyer}, {Kramer},
  {Spitler}, {Thum}, {G{\"u}sten}, {Schuster}, and {Cognard}]{torne15}
{Torne}, P., {Eatough}, R.~P., {Karuppusamy}, R., {Kramer}, M., {Paubert}, G.,
  {Klein}, B., {Desvignes}, G., {Champion}, D.~J., {Wiesemeyer}, H., {Kramer},
  C., {Spitler}, L.~G., {Thum}, C., {G{\"u}sten}, R., {Schuster}, K.~F., and
  {Cognard}, I. (2015).
\newblock {Simultaneous multifrequency radio observations of the Galactic
  Centre magnetar SGR J1745-2900}.
\newblock {\em \mnras\/}, {\bf 451}, L50--L54.

\bibitem[{Torne} {\em et~al.}(2017){Torne}, {Desvignes}, {Eatough},
  {Karuppusamy}, {Paubert}, {Kramer}, {Cognard}, {Champion}, and
  {Spitler}]{torne17}
{Torne}, P., {Desvignes}, G., {Eatough}, R.~P., {Karuppusamy}, R., {Paubert},
  G., {Kramer}, M., {Cognard}, I., {Champion}, D.~J., and {Spitler}, L.~G.
  (2017).
\newblock {Detection of the magnetar SGR J1745-2900 up to 291 GHz with evidence
  of polarized millimetre emission}.
\newblock {\em \mnras\/}, {\bf 465}, 242--247.

\bibitem[{Torres} {\em et~al.}(2012){Torres}, {Rea}, {Esposito}, {Li}, {Chen},
  and {Zhang}]{torres12}
{Torres}, D.~F., {Rea}, N., {Esposito}, P., {Li}, J., {Chen}, Y., and {Zhang},
  S. (2012).
\newblock {A Magnetar-like Event from LS I +61{$^\circ$}303 and Its Nature as a
  Gamma-Ray Binary}.
\newblock {\em \apj\/}, {\bf 744}, 106.

\bibitem[{Turolla} and {Esposito}(2013){Turolla} and {Esposito}]{turolla13}
{Turolla}, R. and {Esposito}, P. (2013).
\newblock {Low-Magnetic Magnetars}.
\newblock {\em International Journal of Modern Physics D\/}, {\bf 22},
  1330024--163.

\bibitem[{Turolla} {\em et~al.}(2011){Turolla}, {Zane}, {Pons}, {Esposito}, and
  {Rea}]{turolla11}
{Turolla}, R., {Zane}, S., {Pons}, J.~A., {Esposito}, P., and {Rea}, N. (2011).
\newblock {Is SGR 0418+5729 Indeed a Waning Magnetar?}
\newblock {\em \apj\/}, {\bf 740}, 105.

\bibitem[{Turolla} {\em et~al.}(2015){Turolla}, {Zane}, and {Watts}]{turolla15}
{Turolla}, R., {Zane}, S., and {Watts}, A.~L. (2015).
\newblock Magnetars: the physics behind observations. a review.
\newblock {\em Reports on Progress in Physics\/}, {\bf 78}, 116901.

\bibitem[{van der Horst} {\em et~al.}(2012){van der Horst}, {Kouveliotou},
  {Gorgone}, {Kaneko}, {Baring}, {Guiriec}, {G{\"o}{\v g}{\"u}{\c s}},
  {Granot}, {Watts}, {Lin}, {Bhat}, {Bissaldi}, {Chaplin}, {Finger}, {Gehrels},
  {Gibby}, {Giles}, {Goldstein}, {Gruber}, {Harding}, {Kaper}, {von Kienlin},
  {van der Klis}, {McBreen}, {Mcenery}, {Meegan}, {Paciesas}, {Pe'er},
  {Preece}, {Ramirez-Ruiz}, {Rau}, {Wachter}, {Wilson-Hodge}, {Woods}, and
  {Wijers}]{vanderhorst12}
{van der Horst}, A.~J., {Kouveliotou}, C., {Gorgone}, N.~M., {Kaneko}, Y.,
  {Baring}, M.~G., {Guiriec}, S., {G{\"o}{\v g}{\"u}{\c s}}, E., {Granot}, J.,
  {Watts}, A.~L., {Lin}, L., {Bhat}, P.~N., {Bissaldi}, E., {Chaplin}, V.~L.,
  {Finger}, M.~H., {Gehrels}, N., {Gibby}, M.~H., {Giles}, M.~M., {Goldstein},
  A., {Gruber}, D., {Harding}, A.~K., {Kaper}, L., {von Kienlin}, A., {van der
  Klis}, M., {McBreen}, S., {Mcenery}, J., {Meegan}, C.~A., {Paciesas}, W.~S.,
  {Pe'er}, A., {Preece}, R.~D., {Ramirez-Ruiz}, E., {Rau}, A., {Wachter}, S.,
  {Wilson-Hodge}, C., {Woods}, P.~M., and {Wijers}, R.~A.~M.~J. (2012).
\newblock {SGR J1550-5418 Bursts Detected with the Fermi Gamma-Ray Burst
  Monitor during its Most Prolific Activity}.
\newblock {\em \apj\/}, {\bf 749}, 122.

\bibitem[{van Paradijs} {\em et~al.}(1995){van Paradijs}, {Taam}, and {van den
  Heuvel}]{vanparadijs95}
{van Paradijs}, J., {Taam}, R.~E., and {van den Heuvel}, E.~P.~J. (1995).
\newblock {On the nature of the 'anomalous' 6-s X-ray pulsars}.
\newblock {\em \aap\/}, {\bf 299}, L41.

\bibitem[{van Putten} {\em et~al.}(2013){van Putten}, {Watts}, {D'Angelo},
  {Baring}, and {Kouveliotou}]{vanputten13}
{van Putten}, T., {Watts}, A.~L., {D'Angelo}, C.~R., {Baring}, M.~G., and
  {Kouveliotou}, C. (2013).
\newblock {Models of hydrostatic magnetar atmospheres at high luminosities}.
\newblock {\em \mnras\/}, {\bf 434}, 1398--1410.

\bibitem[{Vasisht} {\em et~al.}(1994){Vasisht}, {Kulkarni}, {Frail}, and
  {Greiner}]{vasisht94}
{Vasisht}, G., {Kulkarni}, S.~R., {Frail}, D.~A., and {Greiner}, J. (1994).
\newblock {Supernova remnant candidates for the soft gamma-ray repeater
  1900+14}.
\newblock {\em \apjl\/}, {\bf 431}, L35--L38.

\bibitem[{Vietri} {\em et~al.}(2007){Vietri}, {Stella}, and {Israel}]{vietri07}
{Vietri}, M., {Stella}, L., and {Israel}, G.~L. (2007).
\newblock {SGR 1806-20: Evidence for a Superstrong Magnetic Field from
  Quasi-Periodic Oscillations}.
\newblock {\em \apj\/}, {\bf 661}, 1089--1093.

\bibitem[{Vigan{\`o}} and {Pons}(2012){Vigan{\`o}} and {Pons}]{vigano12}
{Vigan{\`o}}, D. and {Pons}, J.~A. (2012).
\newblock {Central compact objects and the hidden magnetic field scenario}.
\newblock {\em \mnras\/}, {\bf 425}, 2487--2492.

\bibitem[{Vigan{\`o}} {\em et~al.}(2013){Vigan{\`o}}, {Rea}, {Pons}, {Perna},
  {Aguilera}, and {Miralles}]{vigano13}
{Vigan{\`o}}, D., {Rea}, N., {Pons}, J.~A., {Perna}, R., {Aguilera}, D.~N., and
  {Miralles}, J.~A. (2013).
\newblock {Unifying the observational diversity of isolated neutron stars via
  magneto-thermal evolution models}.
\newblock {\em \mnras\/}, {\bf 434}, 123--141.

\bibitem[{Vink} and {Kuiper}(2006){Vink} and {Kuiper}]{vink06}
{Vink}, J. and {Kuiper}, L. (2006).
\newblock {Supernova remnant energetics and magnetars: no evidence in favour of
  millisecond proto-neutron stars}.
\newblock {\em \mnras\/}, {\bf 370}, L14--L18.

\bibitem[{Vogel} {\em et~al.}(2014){Vogel}, {Hasco{\"e}t}, {Kaspi}, {An},
  {Archibald}, {Beloborodov}, {Boggs}, {Christensen}, {Craig}, {Gotthelf},
  {Grefenstette}, {Hailey}, {Harrison}, {Kennea}, {Madsen}, {Pivovaroff},
  {Stern}, and {Zhang}]{vogel14}
{Vogel}, J.~K., {Hasco{\"e}t}, R., {Kaspi}, V.~M., {An}, H., {Archibald}, R.,
  {Beloborodov}, A.~M., {Boggs}, S.~E., {Christensen}, F.~E., {Craig}, W.~W.,
  {Gotthelf}, E.~V., {Grefenstette}, B.~W., {Hailey}, C.~J., {Harrison}, F.~A.,
  {Kennea}, J.~A., {Madsen}, K.~K., {Pivovaroff}, M.~J., {Stern}, D., and
  {Zhang}, W.~W. (2014).
\newblock {NuSTAR Observations of the Magnetar 1E 2259+586}.
\newblock {\em \apj\/}, {\bf 789}, 75.

\bibitem[{Wadiasingh} {\em et~al.}(2018){Wadiasingh}, {Baring}, {Gonthier}, and
  {Harding}]{wadiasingh18}
{Wadiasingh}, Z., {Baring}, M.~G., {Gonthier}, P.~L., and {Harding}, A.~K.
  (2018).
\newblock {Resonant Inverse Compton Scattering Spectra from Highly-magnetized
  Neutron Stars}.
\newblock {\em \apj, in press (preprint: astro-ph.HE/1712.09643\/}.

\bibitem[{Wang} {\em et~al.}(2006){Wang}, {Chakrabarty}, and {Kaplan}]{wang06}
{Wang}, Z., {Chakrabarty}, D., and {Kaplan}, D.~L. (2006).
\newblock {A debris disk around an isolated young neutron star}.
\newblock {\em \nat\/}, {\bf 440}, 772--775.

\bibitem[{Watts} and {Strohmayer}(2006){Watts} and {Strohmayer}]{watts06}
{Watts}, A.~L. and {Strohmayer}, T.~E. (2006).
\newblock {Detection with RHESSI of High-Frequency X-Ray Oscillations in the
  Tailof the 2004 Hyperflare from SGR 1806-20}.
\newblock {\em \apjl\/}, {\bf 637}, L117--L120.

\bibitem[{Woods} and {Thompson}(2006){Woods} and {Thompson}]{woods06}
{Woods}, P.~M. and {Thompson}, C. (2006).
\newblock {\em {in Compact stellar X-ray sources, ed. W. H. G. Levin and M. van
  der Klis}\/}.
\newblock Cambridge: Cambridge University Press, p.~547.

\bibitem[{Woods} {\em et~al.}(1999){Woods}, {Kouveliotou}, {van Paradijs},
  {Finger}, {Thompson}, {Duncan}, {Hurley}, {Strohmayer}, {Swank}, and
  {Murakami}]{wkvp99_3}
{Woods}, P.~M., {Kouveliotou}, C., {van Paradijs}, J., {Finger}, M.~H.,
  {Thompson}, C., {Duncan}, R.~C., {Hurley}, K., {Strohmayer}, T., {Swank}, J.,
  and {Murakami}, T. (1999).
\newblock {Variable Spin-Down in the Soft Gamma Repeater SGR 1900+14 and
  Correlations with Burst Activity}.
\newblock {\em \apjl\/}, {\bf 524}, L55--L58.

\bibitem[{Woods} {\em et~al.}(2001){Woods}, {Kouveliotou}, {G{\"o}{\u
  g}{\"u}{\c s}}, {Finger}, {Swank}, {Smith}, {Hurley}, and
  {Thompson}]{woods01}
{Woods}, P.~M., {Kouveliotou}, C., {G{\"o}{\u g}{\"u}{\c s}}, E., {Finger},
  M.~H., {Swank}, J., {Smith}, D.~A., {Hurley}, K., and {Thompson}, C. (2001).
\newblock {Evidence for a Sudden Magnetic Field Reconfiguration in Soft Gamma
  Repeater 1900+14}.
\newblock {\em \apj\/}, {\bf 552}, 748--755.

\bibitem[{Woods} {\em et~al.}(2004){Woods}, {Kaspi}, {Thompson}, {Gavriil},
  {Marshall}, {Chakrabarty}, {Flanagan}, {Heyl}, and {Hernquist}]{woods04}
{Woods}, P.~M., {Kaspi}, V.~M., {Thompson}, C., {Gavriil}, F.~P., {Marshall},
  H.~L., {Chakrabarty}, D., {Flanagan}, K., {Heyl}, J., and {Hernquist}, L.
  (2004).
\newblock {Changes in the X-Ray Emission from the Magnetar Candidate 1E
  2259+586 during Its 2002 Outburst}.
\newblock {\em \apj\/}, {\bf 605}, 378--399.

\bibitem[{Woods} {\em et~al.}(2007){Woods}, {Kouveliotou}, {Finger}, {G{\"o}{\u
  g}{\"u}{\c s}}, {Wilson}, {Patel}, {Hurley}, and {Swank}]{woods07}
{Woods}, P.~M., {Kouveliotou}, C., {Finger}, M.~H., {G{\"o}{\u g}{\"u}{\c s}},
  E., {Wilson}, C.~A., {Patel}, S.~K., {Hurley}, K., and {Swank}, J.~H. (2007).
\newblock {The Prelude to and Aftermath of the Giant Flare of 2004 December 27:
  Persistent and Pulsed X-Ray Properties of SGR 1806-20 from 1993 to 2005}.
\newblock {\em \apj\/}, {\bf 654}, 470--486.

\bibitem[{Younes} {\em et~al.}(2017a){Younes}, {Baring}, {Kouveliotou},
  {Harding}, {Donovan}, {G{\"o}{\u g}{\"u}{\c s}}, {Kaspi}, and
  {Granot}]{ybk17}
{Younes}, G., {Baring}, M.~G., {Kouveliotou}, C., {Harding}, A., {Donovan}, S.,
  {G{\"o}{\u g}{\"u}{\c s}}, E., {Kaspi}, V., and {Granot}, J. (2017a).
\newblock {The Sleeping Monster: NuSTAR Observations of SGR 1806-20, 11 Years
  After the Giant Flare}.
\newblock {\em \apj\/}, {\bf 851}, 17.

\bibitem[{Younes} {\em et~al.}(2017b){Younes}, {Kouveliotou}, {Jaodand},
  {Baring}, {van der Horst}, {Harding}, {Hessels}, {Gehrels}, {Gill},
  {Huppenkothen}, {Granot}, {G{\"o}{\u g}{\"u}{\c s}}, and {Lin}]{younes17}
{Younes}, G., {Kouveliotou}, C., {Jaodand}, A., {Baring}, M.~G., {van der
  Horst}, A.~J., {Harding}, A.~K., {Hessels}, J.~W.~T., {Gehrels}, N., {Gill},
  R., {Huppenkothen}, D., {Granot}, J., {G{\"o}{\u g}{\"u}{\c s}}, E., and
  {Lin}, L. (2017b).
\newblock {X-Ray and Radio Observations of the Magnetar SGR J1935+2154 during
  Its 2014, 2015, and 2016 Outbursts}.
\newblock {\em \apj\/}, {\bf 847}, 85.

\bibitem[{Zane} {\em et~al.}(2011){Zane}, {Nobili}, and {Turolla}]{zane11}
{Zane}, S., {Nobili}, L., and {Turolla}, R. (2011).
\newblock {The magnetar emission in the IR band: the role of magnetospheric
  currents}.
\newblock In D.~F. Torres and N.~Rea, editors, {\em High-Energy Emission from
  Pulsars and their Systems. Proceedings of the First Session of the Sant Cugat
  Forum on Astrophysics\/}, Astrophysics and Space Science Proceedings, pages
  329--335. Springer, Heidelberg.

\end{thebibliography}

\end{document}